\documentclass[twocolumn,times]{aastex61}
\usepackage{amsmath}
\defcitealias{MacDonald:2014aa}{MM14}
\defcitealias{Stevenson:1979aa}{S79}

\submitjournal{ApJ}

\shorttitle{Effects of Rotation and Magnetism in Depth-dependent MLT}
\shortauthors{Ireland and Browning}

\begin{document}

\title{The Radius and Entropy of a Magnetized, Rotating Fully-convective Star: Analysis With Depth-dependent Mixing Length Theories}

\author[0000-0002-8833-1204]{Lewis G. Ireland}
\affil{Department of Physics and Astronomy, University of Exeter, Stocker Road, Exeter, EX4 4QL, UK}

\author[0000-0002-8634-1003]{Matthew K. Browning}
\affil{Department of Physics and Astronomy, University of Exeter, Stocker Road, Exeter, EX4 4QL, UK}

\correspondingauthor{Lewis G. Ireland}
\email{lireland@astro.ex.ac.uk}



\begin{abstract}


Some low-mass stars appear to have larger radii than predicted by standard 1D structure models; prior work has suggested that inefficient convective heat transport, due to rotation and/or magnetism, may ultimately be responsible. We examine this issue using 1D stellar models constructed using Modules for Experiments in Stellar Astrophysics (MESA). First, we consider standard models that do not explicitly include rotational/magnetic effects, with convective inhibition modeled by decreasing a depth-independent mixing length theory (MLT) parameter $\alpha_{\text{MLT}}$ (following \citealt{1981ApJ...245L..37C,Chabrier:2007aa}). We provide formulae linking changes in $\alpha_{\text{MLT}}$ to changes in the interior specific entropy, and hence to the stellar radius. Next, we modify the MLT formulation in MESA to mimic explicitly the influence of rotation and magnetism, using formulations suggested by \citet{Stevenson:1979aa} and \citet{MacDonald:2014aa} respectively. We find rapid rotation in these models has a negligible impact on stellar structure, primarily because a star's adiabat, and hence its radius, is predominantly affected by layers near the surface; convection is rapid and largely uninfluenced by rotation there. Magnetic fields, if they influenced convective transport in the manner described by \citet{MacDonald:2014aa}, could lead to more noticeable radius inflation. Finally, we show that these non-standard effects on stellar structure can be fabricated using a depth-dependent $\alpha_{\text{MLT}}$: a non-magnetic, non-rotating model can be produced that is virtually indistinguishable from one that explicitly parameterizes rotation and/or magnetism using the two formulations above. We provide formulae linking the radially-variable $\alpha_{\text{MLT}}$ to these putative MLT reformulations.

\end{abstract}




\section{Introduction} \label{sec:intro}

All main-sequence stars are convective somewhere in their interior: low-density, high-temperature fluid parcels rise or fall through the stratified medium, transporting heat by their motion. In high-mass stars this convective transport occurs primarily in the innermost regions, whereas low-mass stars like the Sun have convection occurring in an envelope; stars of sufficiently low mass ($\lesssim 0.35 \, \text{M}_\odot$) are convective throughout their interiors \citep[e.g.,][]{1997A&A...327.1039C}. Pre-main-sequence stars on the Hayashi track are likewise fully-convective \citep[e.g.,][]{1961PASJ...13..450H}. Variations in the opacity, energy generation rate, or adiabatic index determine where and when this convection occurs: broadly, it happens whenever the temperature gradient required to carry a star's flux by radiative processes alone is too steep \citep[e.g.,][]{Bohm-Vitense:1992aa}. This may be encapsulated via the Schwarzschild criterion, which states that convection occurs whenever the dimensionless temperature gradient ${\nabla = d \ln{T} / d \ln{p}}$ is greater than the adiabatic gradient ${\nabla_{\text{ad}} = (d \ln{T} / d \ln{p})_{\text{ad}}}$.

In the interior of a star, modest convective velocities and temperature gradients very close to the adiabatic value are usually sufficient to carry a star's flux outwards \citep[e.g.,][]{kippenhahn2012stellar}, owing mainly to the high density and heat capacity of these regions. For fully-convective stars, this implies that most of the interior lies at nearly-constant specific entropy. However, larger entropy gradients are established near the surface (as discussed in \S~\ref{sec:entropy} below). The interaction between convection and radiative transfer in the region of the surface layer thus creates a specific entropy jump $\Delta s$ between the nearly-constant specific entropy in the deep interior, conventionally labeled $s_{\text{ad}}$, and the specific entropy at the stellar photosphere $s_{\text{ph}}$ \citep[e.g.,][]{Trampedach:2014aa}.

The gross structure of a star is linked to its entropy \citep[see, e.g., discussions in][]{1988PASP..100.1474S,2004sipp.book.....H}. In particular, for isentropic stars, knowledge of $s_{\text{ad}}$---i.e., knowledge of which adiabat the star is on---is enough to specify the entire structure. As emphasized by \citet{Gough:1976aa}, a complete theory of convection would specify the adiabat, but in practice this is typically calibrated by comparison to observations. In standard 1D stellar models employing the mixing length theory (MLT) of convection, fluid parcels are assumed to travel some characteristic mixing length $\ell_{\text{MLT}} = \alpha_{\text{MLT}} H_p$ before transferring their heat to their surroundings, where $\alpha_{\text{MLT}}$ is conventionally a depth-independent dimensionless parameter and $H_p$ is the pressure scale height \citep{1958ZA.....46..108B}. In typical models of fully-convective stars, $\alpha_{\text{MLT}}$ effectively specifies the entropy contrast $\Delta s$, and so fixes the adiabat and the overall stellar structure. 

Observations have suggested that some low-mass stars have radii that are $5-15 \%$ larger than standard 1D models would predict \citep[e.g.,][]{Torres:2002aa,2006Ap&SS.304...89R,2008A&A...478..507M,2010A&ARv..18...67T,2012ApJ...760L...9T}. These ``inflated" radii could in turn lead to erroneous age estimates of stars on the pre-main-sequence \citep[see, e.g.,][]{Feiden:2016aa}. Several authors have argued that the inhibition of convection by some mechanism could explain these modifications to the structure, with rotation and/or magnetic fields both invoked as possible culprits {\citep[e.g.,][]{1981ApJ...245L..37C,Chabrier:2007aa}.}

Rotation is well known to influence convection. For example, in classic linear stability analysis, the onset of convection is impeded by the presence of rotation: the critical Rayleigh number for convective instability (measuring, roughly, how great buoyancy driving must be relative to viscous and thermal dissipation) increases with rotation rate $\Omega$ \citep{1961hhs..book.....C}, scaling as $\Omega^{4/3}$ in appropriate circumstances. The horizontal scale of the most unstable modes likewise diminishes with more rapid rotation. The non-linear effects of rotation on the convection are less clear. Broadly, the reduction of horizontal lengthscales and convective speeds in rapidly-rotating systems is expected to inhibit the heat transport somewhat, leading to higher values of the temperature (or in a stratified system, entropy) gradient \citep{Stevenson:1979aa,2012PhRvL.109y4503J,Barker:2014aa}. Rotation also breaks the spherical symmetry, with motions increasingly aligned with the rotation axis at rapid rotation rates, in keeping with the Taylor-Proudman constraint \citep{1916RSPSA..92..408P,1917RSPSA..93...99T}. Other aspects of the non-linear impact of rotation, such as its effect on heat transport and on the establishment of zonal flows, have also been extensively explored using theory and simulation \citep[e.g.,][]{1994GApFD..76..223B,julien_knobloch_1998,sprague_julien_knobloch_werne_2006,2012A&A...546A..19G,king_stellmach_aurnou_2012,2012PhRvL.109y4503J,2014PhRvL.113y4501S,2015PEPI..246...52A,2015JFM...780..143C,2015GApFD.109..145G,2016JFM...808..690G,2016JFM...798...50J,2017JFM...813..558A}.

A reformulation of MLT to treat rapidly-rotating cases was proposed for example by \citet{Stevenson:1979aa}, who argued following \citet{1954RSPSA.225..196M} that the non-linear state was likely to be dominated by the modes that transport the most heat. \citet{2012PhRvL.109y4503J} also examined the transport in rapidly-rotating systems, by scaling to the state of marginal stability; they argue that in contrast to classical non-rotating convection, in which heat transport is ``throttled" in narrow boundary layers, the heat transport of rapidly-rotating systems is limited by the efficiency of turbulent motion in the bulk of the fluid. Recently, \citet{Barker:2014aa} derived a version of rotating MLT equivalent to \citet{Stevenson:1979aa} in a different way, and tested it using 3D simulations in Cartesian domains. Broadly, several methods of analysis suggest that the temperature gradient in the middle of the rotating convective layer ($dT/dz$) increases with rotation rate $\Omega$. In particular, \citet{Barker:2014aa} have argued specifically that $dT/dz \propto \Omega^{4/5}$ in the rapidly rotating limit. Their simulations support this scaling, though it must be noted that their models encompass only a single latitude (namely the pole); extensions to other latitudes are under way (L. Currie, A. Barker, and Y. Lithwick, {personal} communication). 

Magnetic fields are likewise known to influence convection in some manner, but it is not clear how this affects the heat transport in the stellar context. Magnetic fields can inhibit convection in the stellar interior via the Lorentz force, hindering fluid flow perpendicular to the field \citep[e.g.,][]{Stein:2012aa}. Like rotation, magnetic fields influence the linear stability of the fluid to convective motions: in the absence of rotation, magnetism is stabilizing \citep{1961hhs..book.....C,Gough:1966aa}. When rotation is present, the linear stability is more complex, and in fact the critical Rayleigh number for convection with \emph{both} rotation and magnetism can be lower than in the presence of either rotation or magnetism alone \citep{1961hhs..book.....C,Stevenson:1979aa}. Again, the non-linear impact of the magnetism is much less clear. \citet{Stevenson:1979aa} also fashioned a ``magnetic" version of MLT, but (to our knowledge) this has not been incorporated into 1D stellar structure models. \citet{Mullan:2001ab}, drawing on the linear stability analysis of \citet{Gough:1966aa}, argued that the effects of magnetism in a 1D stellar model could be mimicked simply by modifying the adiabatic gradient $\nabla_{\text{ad}}$ (wherever it appears in the MLT prescription) to include a perturbation term proportional to the magnetic pressure (relative to the gas pressure). Physically, this amounts to asserting that the end-state of magnetized convection is to approach a state of marginal stability---where this stability now depends on the strength of the magnetism---in much the same way that non-magnetic convection might be taken to approach an isentropic state. 

\citet{Chabrier:2007aa}, noting that even fairly modest magnetic fields might strongly feed back on the flows through Lorentz forces, modeled rotational and magnetic effects simply by varying the depth-independent $\alpha_{\text{MLT}}$; they also briefly considered the effects of near-surface spots, taken to be regions of cool effective temperature covering some fraction of the surface. \citet{Feiden:2012aa}, drawing on \citet{Lydon:1995aa}, have implemented a more complex magnetic MLT model into the Dartmouth stellar evolution code, with properties of the resulting structure dependent on the strength and (imposed) spatial distribution of the magnetism. Broadly, these authors have argued that magnetic fields can affect the radius of a star, either by inhibiting convection or through the effects of near-surface spots {\citep[e.g.,][]{1981ApJ...245L..37C,Mullan:2001ab,Chabrier:2007aa,MacDonald:2012aa,MacDonald:2013aa,MacDonald:2014aa,MacDonald:2017aa,Feiden:2012aa,Feiden:2014aa,Feiden:2016aa}.}

In this paper, we examine the effects of rotation and magnetic fields on the structure of fully-convective stars via 1D stellar structure models, using the Modules for Experiments in Stellar Astrophysics (MESA) code \citep{Paxton:2011aa,Paxton:2013aa,Paxton:2015aa}. All the reformulations of MLT noted above can modify the adiabat of the star, by changing the efficiency of convective heat transport in the stellar interior. Thus, in \S~\ref{sec:entropy}, we begin by giving an overview of the role of entropy in standard 1D stellar structure models; in particular, we recall how the stellar radius is sensitive to changes in the specific entropy, which is itself sensitive to differing levels of convective inhibition via changes in $\alpha_{\text{MLT}}$. We give an explicit relationship between specific entropy, stellar radius, and $\alpha_{\text{MLT}}$ for these ``standard" models with a depth-independent $\alpha_{\text{MLT}}$.

We then examine the ``rotating" and ``magnetic" MLT reformulations by \citet{Stevenson:1979aa} and \citet{MacDonald:2014aa} respectively in \S~\ref{sec:rot_inhibition_conv} and \S~\ref{sec:mag_inhibition_conv} to determine how these mechanisms inhibit convection, and so influence the stellar radius, compared to solely changing $\alpha_{\text{MLT}}$. We set aside for now the question of whether these formulations correctly capture the complex interaction between rotation, convection, and magnetism in a star; here, we simply examine the consequences of these prescriptions for the entropy and radius of the star. We also investigate the influence on stellar structure as a result of combining these ``rotating" and ``magnetic" MLT reformulations in \S~\ref{sec:rot_and_mag}.

In \S~\ref{sec:depth_dep_alpha}, we show that these reformulations to MLT may be precisely duplicated in a standard (non-magnetic, non-rotating) 1D model by the introduction of a depth-dependent $\alpha_{\text{MLT}}$. We provide formulae for depth-dependent $\alpha_{\text{MLT}}$ profiles that can be used to mimic the effects of rotation or magnetism on the stellar superadiabaticity, and hence on the stellar radius (assuming these are captured by the \citet{Stevenson:1979aa} and \citet{MacDonald:2014aa} formulations respectively), providing a simple way for users to model these non-standard effects. Finally, we discuss our results in \S~\ref{sec:discussion_conclusion}.
 
\section{Entropy, convection, and the radii of standard 1D stellar structure models} \label{sec:entropy}

\begin{figure*}
\gridline{\fig{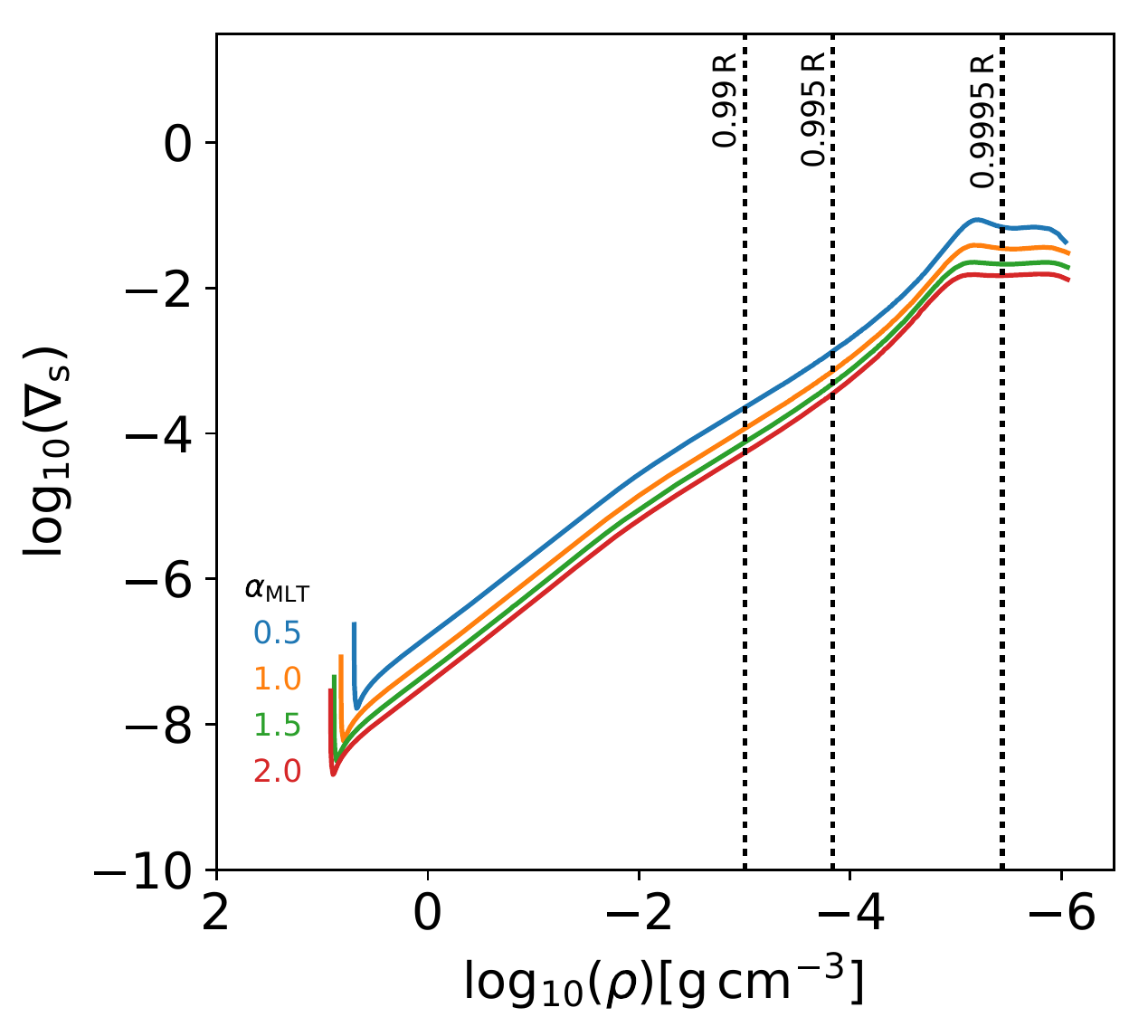}{0.5\textwidth}{(a)}
          \fig{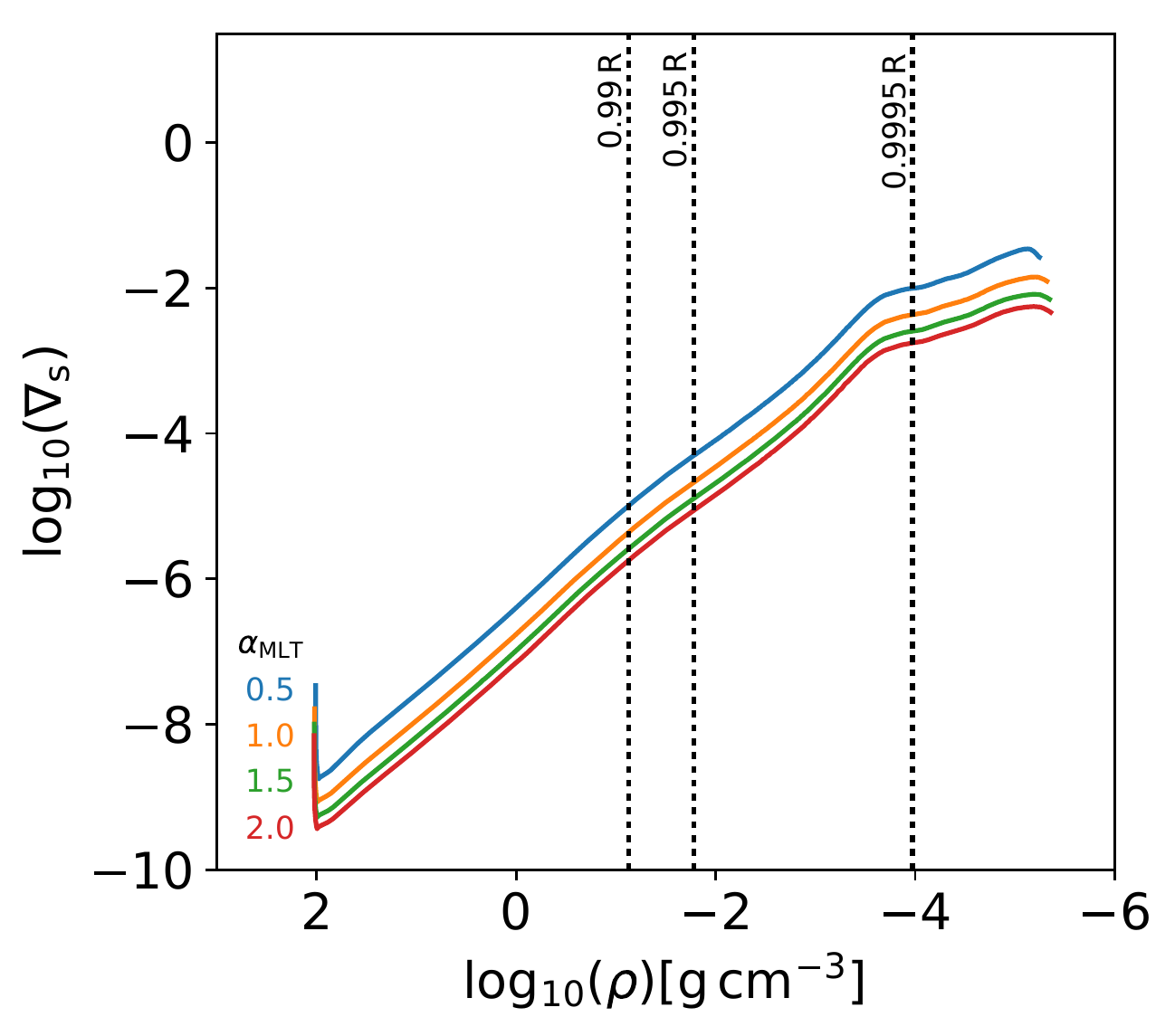}{0.513\textwidth}{(b)}
          }
\caption{$\log_{10}{(\nabla_{\text{s}})}$ as a function of $\log_{10}{(\rho)}$, for $0.3 \, \text{M}_\odot$, (a) $10 \, \text{Myr}$ (b) $1 \, \text{Gyr}$ stellar models at $\alpha_{\text{MLT}} = 0.5 - 2.0$ ($\Delta 0.5$). As $\alpha_{\text{MLT}}$ decreases, {the superadiabaticity} $\nabla_{\text{s}}$ increases throughout the stellar interior, but $\nabla_{\text{s}}$ is inherently lower in main-sequence models.\label{fig:logsupergrad_v_logRho}}
\end{figure*}

\subsection{Role of specific entropy in standard MLT} \label{subsec:role_entropy}

Heat transport, entropy, and the stellar structure are tightly linked in fully-convective objects. Here, we briefly review these links, outlining how changes in the convective efficiency of classical MLT modify the internal entropy structure and hence the stellar radius. The material in this section largely duplicates standard results found elsewhere \citep[see e.g.,][]{2004sipp.book.....H}, but we include it here as background for our studies in \S~\ref{sec:rot_inhibition_conv} - \S~\ref{sec:depth_dep_alpha}.

For an ideal gas without radiation pressure, the specific entropy (i.e., the entropy per unit mass) $s$ is

\begin{equation}\label{eq:closed_form_s}
s \simeq s_0 + \frac{N_{\text{A}} k_{\text{B}}}{\mu} \ln{\left(\frac{T^{1/(\gamma - 1)}}{\rho} \right)},
\end{equation}
where $s_0$ is {a constant}, $N_{\text{A}}$ is Avogadro's constant, $k_{\text{B}}$ is Boltzmann's constant, $\mu$ is the mean molecular weight, $T$ is temperature, $\rho$ is density, and $\gamma$ is the adiabatic exponent.

To examine how the specific entropy changes in response to variations in the convective efficiency, we first constructed a series of standard 1D stellar structure models using MESA. Here, we simply use the default setup provided by MESAstar: the MLT prescription is that of \citet{Cox:1968aa}; the atmospheric boundary conditions are MESA's ``simple" option, in which the photosphere is located at optical depth $\tau = 2/3$, the surface temperature is given by the Eddington $T(\tau)$ relation, and the opacity is calculated in an iterative fashion (see \citet{Paxton:2011aa} for details); the metallicity is fixed at $Z = 0.02$. We model stars only at a fixed mass of $0.3 \, \text{M}_\odot$, evolving each model from the pre-main-sequence up to an age of $4 \, \text{Gyr}$. Models of this mass are convective throughout their interiors. We vary the mixing length parameter $\alpha_{\text{MLT}}$ to vary the convective efficiency, effectively reducing the distance traveled by convective elements \citep{Trampedach:2014aa}.  Taken together, these choices imply that our models are somewhat more idealized depictions of a $0.3 \, \text{M}_\odot$ star than the most sophisticated ones in use today \citep[e.g.,][]{2015A&A...577A..42B}.  For example, in reality (and in more complete models) convection extends well into the optically thin regime, mainly because the formation of H$_2$ decreases the adiabatic gradient, favoring convection \citep{1997A&A...327.1039C}.  Values of the effective temperature in models including this effect will generally differ from those reported here (which simply assume the Eddington $T(\tau)$ relation).  We choose this simpler boundary condition partly because it allows us to compare more directly with analytical theory below, and because we are interested mainly in \emph{changes} between models with differing $\alpha_{\text{MLT}}$ rather than in the absolute values of $T_{\text{eff}}$, $R$, etc.  

We turn first to consideration of the superadiabatic gradient $\nabla_{\text{s}} \equiv (\nabla - \nabla_{\text{ad}})$, which is a dimensionless measure of the entropy gradient. In Figure~\ref{fig:logsupergrad_v_logRho}, we plot $\log_{10}{(\nabla_{\text{s}})}$ as a function of logarithmic density $\log_{10}{(\rho)}$ for $0.3 \, \text{M}_\odot$, $10 \, \text{Myr}$ pre-main-sequence and $1 \, \text{Gyr}$ main-sequence stellar models with $\alpha_{\text{MLT}} = 0.5 - 2.0$ ($\Delta 0.5$). Vertical dotted lines in this figure onwards indicate average radial positions in the region of the surface layer. A few key features are readily apparent: first, in the bulk of the convection zone, $\nabla_{\text{s}}$ reaches negligible values due to highly efficient convective transport, where the temperature gradient is nearly adiabatic. Nearer the surface, $\nabla_{\text{s}}$ increases, driven by the continuous decline in the density and temperature of the plasma. Convection carries nearly all the flux until radii of greater than $0.995 \, \text{R}$, where $\text{R}$ is the radius of a given model, and is highly efficient over most of that region; radiative diffusion begins to carry a non-negligible amount of flux only above $0.9995 \, \text{R}$. Comparing the left and right panels of Figure~\ref{fig:logsupergrad_v_logRho}, we see that $\nabla_{\text{s}}$ is somewhat lower in the main-sequence models (right panel) than on the pre-main-sequence. In both sets of models, at all depths $\nabla_{\text{s}}$ depends on the convective efficiency: less efficient convection, which in these models corresponds simply to a smaller value of $\alpha_{\text{MLT}}$, means that a higher $\nabla_{\text{s}}$ is required to carry the same heat flux.

To quantify how changing $\alpha_{\text{MLT}}$ influences the run of $\nabla_{\text{s}}$, and so explain the trends visible in Figure~\ref{fig:logsupergrad_v_logRho}, we consider the convective flux $F_{\text{conv}}$ as defined in the classic MLT prescription of \citet{1958ZA.....46..108B}, as implemented in MESA:

\begin{equation} \label{eq:F_conv_general}
	F_{\text{conv}} = \frac{1}{4 \sqrt{2}} c_p (p \rho Q)^{1/2} T (\nabla - \nabla')^{3/2} \alpha_{\text{MLT}}^2,
\end{equation}
where $c_p$ is the specific heat capacity (at constant pressure), $p$ is pressure, $Q = -(\partial \ln{\rho} / \partial \ln{T})_p$ is the isobaric expansion coefficient, and {$\nabla' = (d \ln{T} / d \ln{p})'$} is the temperature gradient of the rising element \citep{Cox:1968aa}. Following \citet{Cox:1968aa}, we can solve for the convective efficiency $\Gamma = A (\nabla - \nabla')^{1/2}$, which is the ratio of energy successfully transported and that which is lost by a convective element, in terms of $\nabla_{\text{s}}$, and express $\nabla - \nabla'$ as a function of $\nabla_{\text{s}}$:

\begin{equation} \label{eq:nabla_terms_of_nabla_s}
	\nabla - \nabla' = \left(\frac{\Gamma}{A}\right)^2 = \frac{1}{4A^2} \left(\sqrt{1 + 4 A^2 \nabla_{\text{s}}} - 1\right)^2,
\end{equation}
where

\begin{equation} \label{eq:A_mlt}
	A = \frac{Q^{1/2} c_p \kappa g \rho^{5/2} H_p^2}{12 \sqrt{2} a c p^{1/2} T^3} \alpha_{\text{MLT}}^2 \equiv A_{\text{other}} \alpha_{\text{MLT}}^2
\end{equation}
is the ratio of convective and radiative conductivities, where $\kappa$ is opacity, $g$ is gravitational acceleration, $a$ is the radiation constant, and $c$ is the speed of light. 

Using equations~(\ref{eq:F_conv_general}) and~(\ref{eq:nabla_terms_of_nabla_s}), we express $\nabla_{\text{s}}$ as a function of $\alpha_{\text{MLT}}$:

\begin{multline} \label{eq:nabla_alpha_general_relation}
	\nabla_{\text{s}} = \left(\frac{4 \sqrt{2} F_{\text{conv}}}{c_p (p \rho Q)^{1/2} T}\right)^{2/3} \alpha_{\text{MLT}}^{-4/3} \\ 
	+ \frac{1}{A_{\text{other}}} \left(\frac{4 \sqrt{2} F_{\text{conv}}}{c_p (p \rho Q)^{1/2} T}\right)^{1/3} \alpha_{\text{MLT}}^{-8/3}.
\end{multline}
Equation~(\ref{eq:nabla_alpha_general_relation}) reflects the fact that there are two regimes of convective efficiency $\Gamma \sim A \nabla_{\text{s}}^{1/2}$. As noted by \citet{Gough:1976aa}, stellar convection theories tend asymptotically toward two regimes: high ($\Gamma \gg 1$, left term) and low ($\Gamma \ll 1$, right term) convective efficiency. The connection between these two asymptotic limits is very thin, so the structure of this transition is typically not significant in the astrophysical context.

\begin{figure*}
\gridline{\fig{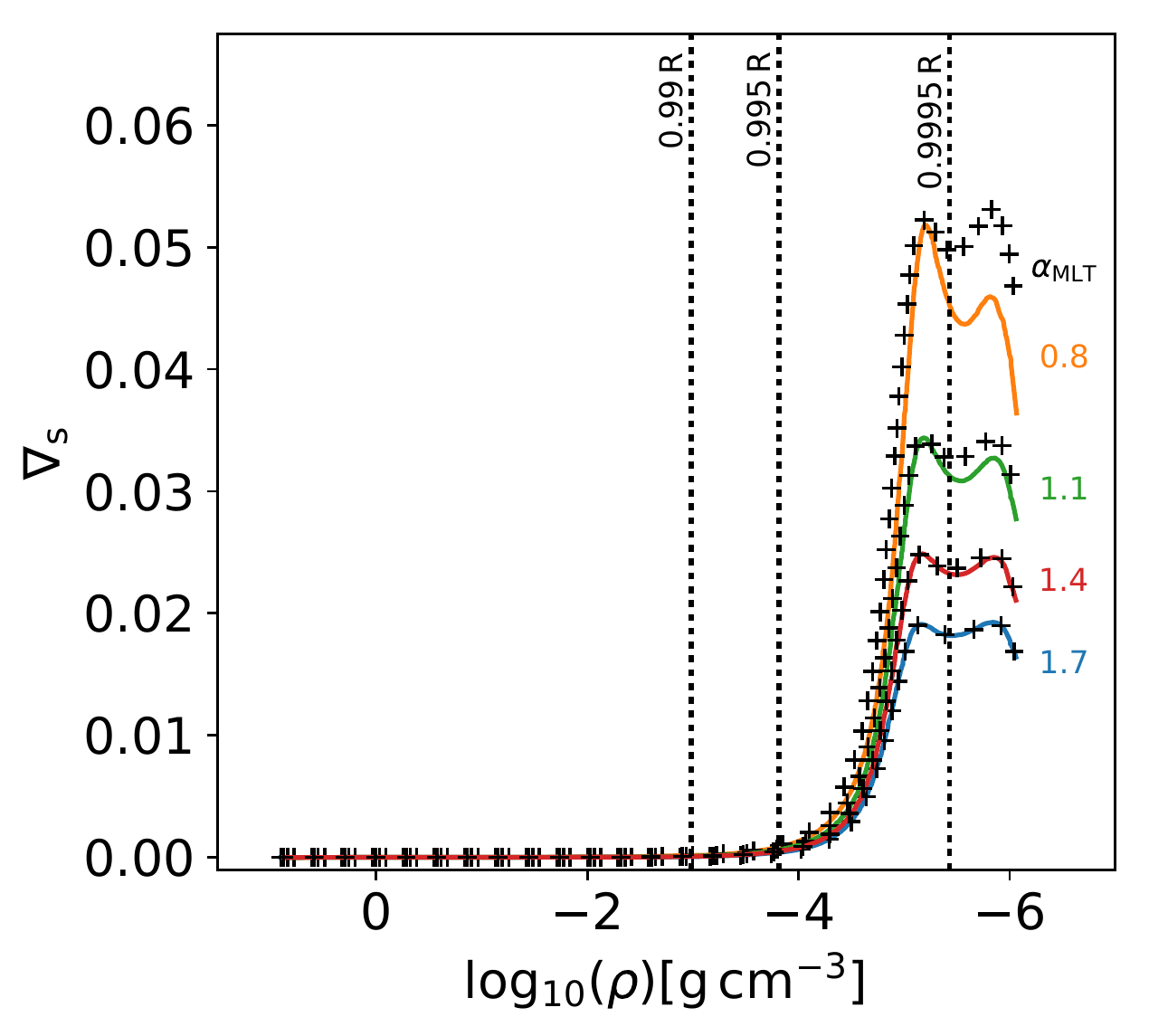}{0.5\textwidth}{(a)}
          \fig{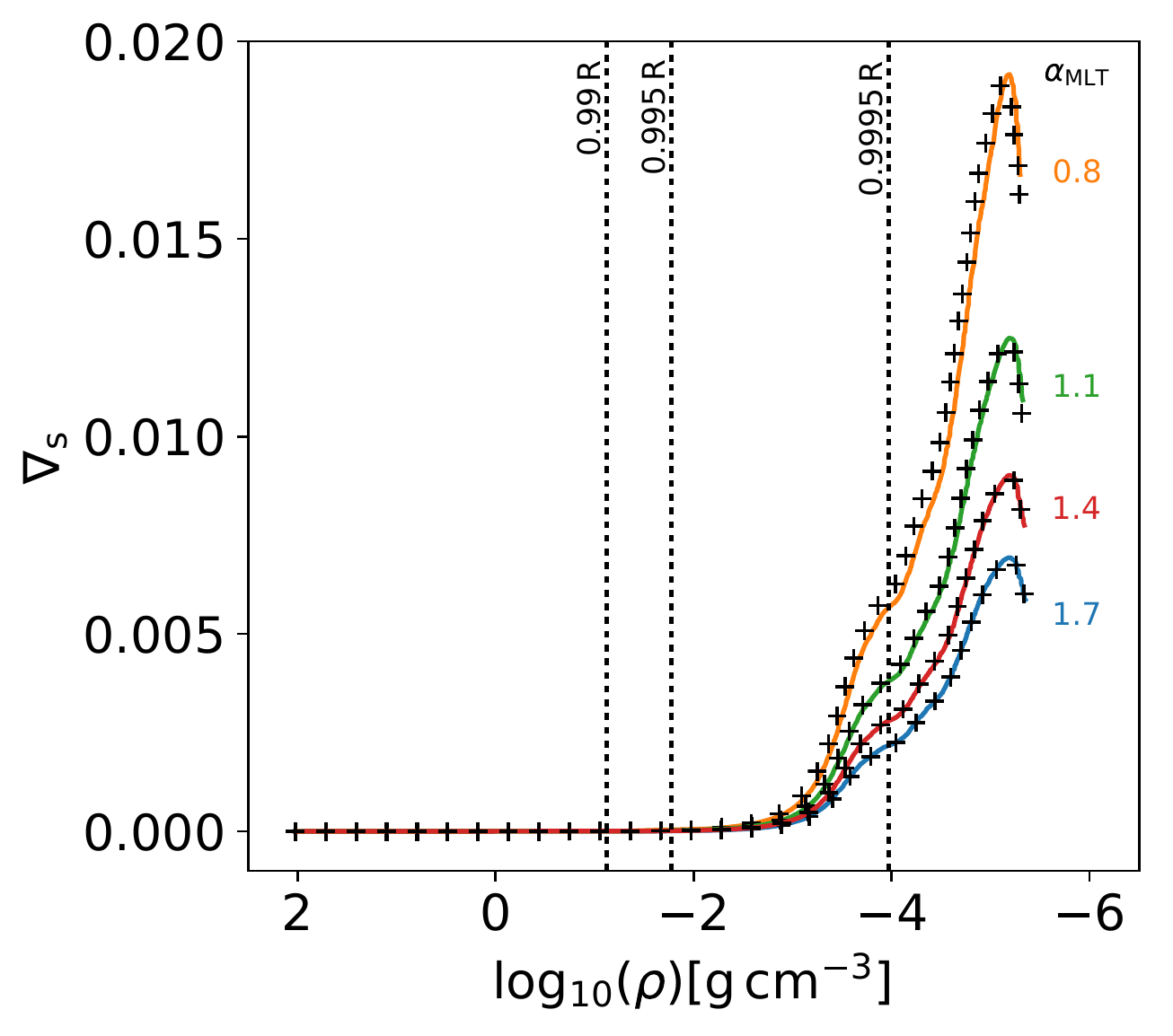}{0.515\textwidth}{(b)}
          }
\caption{$\nabla_{\text{s}}$ as a function of $\log_{10}{(\rho)}$, comparing the outputted values and those reproduced using equation~(\ref{eq:nabla_alpha_efficient_full}) (plus markers), for a selection of $0.3 \, \text{M}_\odot$, (a) $10 \, \text{Myr}$ (b) $1 \, \text{Gyr}$ stellar models at $\alpha_{\text{MLT}} = 0.8 - 1.7$ ($\Delta 0.3$). $\alpha_{\text{MLT}} = 1.7$ is chosen to be the ``unperturbed" model. As $\alpha_{\text{MLT}}$ decreases, the $10 \, \text{Myr}$ model's reproduced $\nabla_{\text{s}}$ increasingly diverges right at the photosphere, due to the non-negligible low efficiency regime. \label{fig:supergrad_interpol_v_logRho}}
\end{figure*}

\begin{figure*}
\gridline{\fig{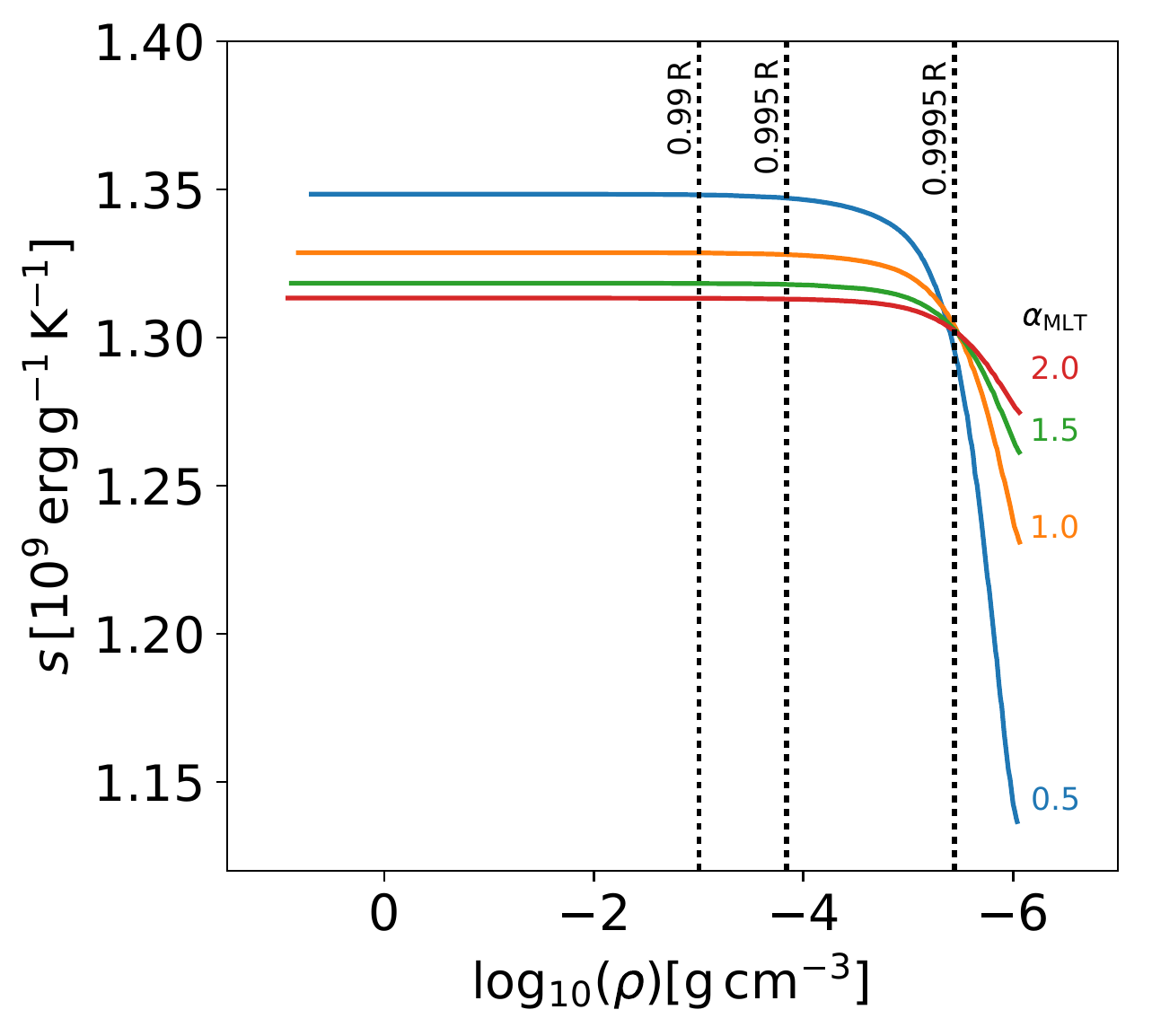}{0.5\textwidth}{(a)}
          \fig{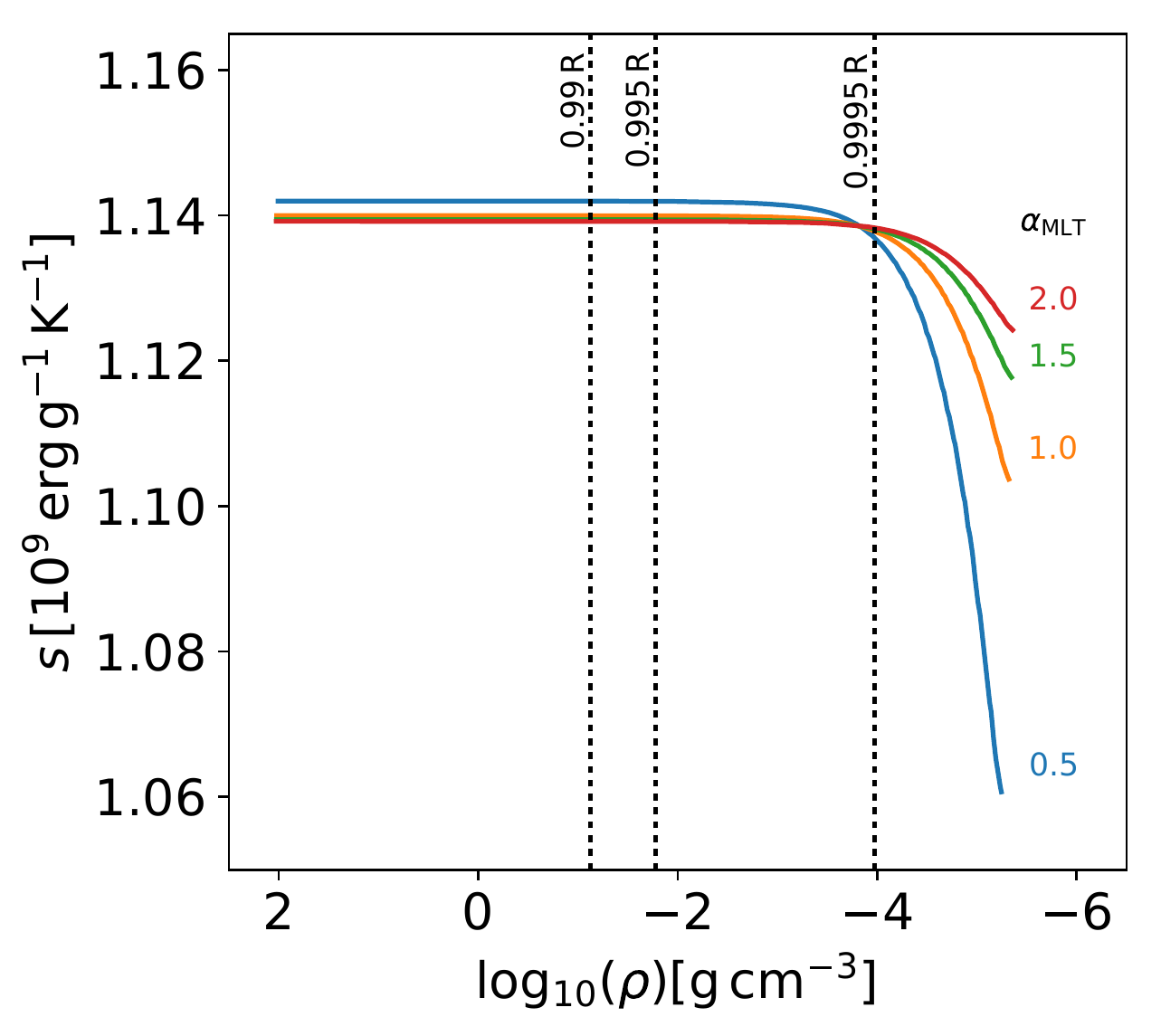}{0.5\textwidth}{(b)}
          }
\caption{$s$ as a function of $\log_{10}{(\rho)}$ for $0.3 \, \text{M}_\odot$, (a) $10 \, \text{Myr}$ (b) $1 \, \text{Gyr}$ stellar models at $\alpha_{\text{MLT}} = 0.5 - 2.0$ ($\Delta 0.5$). Decreasing $\alpha_{\text{MLT}}$ increases $s_{\text{ad}}$, i.e., the asymptotic value of specific entropy in the bulk of the convection zone, but to a lesser extent for main-sequence models.\label{fig:entropy_v_logRho}}
\end{figure*}

For homologous stellar models of highly efficient convection, where luminosity (hence convective flux) is fixed throughout the radial distribution in the bulk of the stellar interior,

\begin{equation} \label{eq:nabla_alpha_efficient}
	\nabla_{\text{s}} \propto \alpha_{\text{MLT}}^{-4/3},
\end{equation}
demonstrating that in this regime a decrease in $\alpha_{\text{MLT}}$ corresponds to an monotonic increase of $\nabla_{\text{s}}$ in the bulk of the convection zone \citep[e.g.,][]{Christensen-Dalsgaard:1997aa}. 

From equation~(\ref{eq:nabla_alpha_efficient}), it is possible to reproduce a majority of a model's $\nabla_{\text{s}}$ profile using the model's $\alpha_{\text{MLT}}$, and an unperturbed, or reference, model's $\alpha_{\text{MLT}}$ and $\nabla_{\text{s}}$, via

\begin{equation} \label{eq:nabla_alpha_efficient_full}
	\nabla_{\text{s}} \simeq \nabla_{\text{s}_0} \left(\frac{\alpha_{\text{MLT}}}{\alpha_{\text{MLT}_0}}\right)^{-4/3},
\end{equation}
where zero subscripts denote values from the unperturbed model. This is valid only for models where the convective flux remains roughly the same as in our fiducial model. In Figure~\ref{fig:supergrad_interpol_v_logRho}, we plot the outputted $\nabla_{\text{s}}$ and those reproduced using equation~({\ref{eq:nabla_alpha_efficient_full}) as a function of $\log_{10}{(\rho)}$ for $0.3 \, \text{M}_\odot$, $10 \, \text{Myr}$ and $1 \, \text{Gyr}$ stellar models with $\alpha_{\text{MLT}} = 0.8 - 1.7$ ($\Delta 0.3$). We choose $\alpha_{\text{MLT}} = 1.7$ to be our unperturbed model and the lower limit $\alpha_{\text{MLT}} = 0.8$ corresponds to the lowest $\alpha_{\text{MLT}}$ for which the convective flux is similar to the unperturbed model. We plot $\nabla_{\text{s}}$ linearly to show the surface layers more clearly. Small deviations are increasingly evident right near the photosphere in the $10 \, \text{Myr}$ models with decreasing $\alpha_{\text{MLT}}$, as the ``low efficiency" regime (ignored in equation~(\ref{eq:nabla_alpha_efficient_full})) begins to come into play. However, the approximation of equation~(\ref{eq:nabla_alpha_efficient_full}) captures the behavior of $\nabla_{\text{s}}$ up to $\approx 0.9995 \, \text{R}$.

We turn next to an analysis of the specific entropy in the same models. In Figure~\ref{fig:entropy_v_logRho}, we plot $s$ as a function of $\log_{10}{(\rho)}$ for these models. We obtain $s$ as a function of the radial distribution $r$ in our stellar models by taking the outputted central specific entropy $s_{\text{c}}$ and integrating the specific entropy gradient $ds/dr$ up to a radial point $r'$:

\begin{equation} \label{eq:s_s_ad_int_ds_dr}
s(r') = s_{\text{c}} + \int_0^{r'} \frac{ds}{dr} \, dr.
\end{equation}
$ds / dr$ is related to the superadiabaticity $\nabla_{\text{s}}$ through the first and second law of thermodynamics:

\begin{equation} \label{eq:ds_dr_relation}
\frac{ds}{dr} = - \frac{c_p}{H_p} \nabla_{\text{s}}.
\end{equation}

In the bulk of the convection zone, specific entropy asymptotically converges with depth towards a nearly-constant specific entropy value $s_{\text{ad}}$. The value of $s_{\text{ad}}$ largely determines the stellar structure, including the stellar radius. As noted by \citet{Gough:1976aa}, a perfect theory of convection would specify this adiabat (i.e., fix $s_{\text{ad}}$), but in practise it must be calibrated via observations. To be specific, note that for a fully-convective isentropic star (with $\gamma = 5/3$), we would have $s \propto \ln{(T^{3/2}/\rho)} = const$. In this case, properties at the center `c' and the photosphere `ph' would be directly linked, with $(T_{\text{c}}^{3/2}/\rho_{\text{c}}) = (T_{\text{ph}}^{3/2}/\rho_{\text{ph}})$, where $T_{\text{ph}} \equiv T_{\text{eff}}$. Specifying the surface properties and the adiabat would in this case clearly suffice to determine the properties of the star everywhere in its interior.

However, standard stellar structure models are not perfectly isentropic. Ascending into the surface layers, specific entropy decreases: although $\nabla_{\text{s}}$ is nearly constant there (Figure~\ref{fig:logsupergrad_v_logRho}), the entropy gradient (equation~(\ref{eq:ds_dr_relation})) is increasingly negative. This arises because although $c_p$ remains high even near the surface (in fact, in these models it is higher at $0.9995 \, \text{R}$ than at $0.99 \, \text{R}$), $H_p$ declines monotonically, implying that $ds/dr$ increases in magnitude near the surface. This non-zero $ds/dr$ implies that there is an entropy jump $\Delta s$ between the interior adiabat and the surface value. If this is the only region where $ds/dr$ is non-zero, then the ratio of the central and photospheric properties, from the logarithmic argument of equation~(\ref{eq:closed_form_s}), is now a function of $\Delta s$:

\begin{equation}\label{eq:delta_s_ratio}
\frac{T_{\text{c}}^{1/(\gamma - 1)}/\rho_{\text{c}}}{T_{\text{ph}}^{1/(\gamma - 1)}/\rho_{\text{ph}}} = \exp{\left(\frac{\mu \Delta s}{N_{\text{A}} k_{\text{B}}}\right)},
\end{equation}
demonstrating explicitly how noticeable values of $\Delta s$ may influence the stellar properties of fully-convective models. 

Examining the variation of specific entropy in Figure~\ref{fig:entropy_v_logRho}, a few key trends are clear. At both $10 \, \text{Myr}$ and $1 \, \text{Gyr}$, models with lower $\alpha_{\text{MLT}}$ always have a larger contrast $\Delta s$ between the photosphere and the deep interior. The lower-$\alpha_{\text{MLT}}$ models also have a lower specific entropy at the photosphere $s_{\text{ph}}$. In the pre-main-sequence models, models at lower $\alpha_{\text{MLT}}$ also possess a higher internal entropy $s_{\text{ad}}$, but by an age of $1 \, \text{Gyr}$ this variation has largely vanished, with only the very lowest-$\alpha_{\text{MLT}}$ model here ($\alpha_{\text{MLT}} = 0.5$) possessing a noticeably higher $s_{\text{ad}}$. These features can be understood as discussed below.

First, consider the overall entropy contrast $\Delta s$ in the near-surface layers. To quantify how the profile of specific entropy varies with $\alpha_{\text{MLT}}$, we first consider $\Delta s$ expressed in terms of $\nabla_{\text{s}}$ via equation~(\ref{eq:ds_dr_relation}):

\begin{equation} \label{eq:s_jump}
\Delta s = - \int_0^{R} \frac{ds}{dr} \, dr = \int_0^{R} \frac{c_p}{H_p} \nabla_{\text{s}} \, dr.
\end{equation}
Using equation~(\ref{eq:nabla_alpha_general_relation}), it can be shown that $\Delta s$ increases with decreasing $\alpha_{\text{MLT}}$:

\begin{multline} \label{eq:s_jump_alpha}
\Delta s = \alpha_{\text{MLT}}^{-4/3} \int_0^{R} \frac{c_p}{H_p} \left(\frac{4 \sqrt{2} F_{\text{conv}}}{c_p (p \rho Q)^{1/2} T}\right)^{2/3} \, dr \\ 
+ \alpha_{\text{MLT}}^{-8/3} \int_0^{R} \frac{c_p}{H_p} \frac{1}{A_{\text{other}}} \left(\frac{4 \sqrt{2} F_{\text{conv}}}{c_p (p \rho Q)^{1/2} T}\right)^{1/3} \, dr,
\end{multline}
where $\alpha_{\text{MLT}}$ is taken out of the integrands due to being depth-independent. As we are able to reproduce a majority of $\nabla_{\text{s}}$ via the high efficiency regime using equation~(\ref{eq:nabla_alpha_efficient_full}), it follows that for models where the convective flux remains roughly the same as in our unperturbed model that

\begin{equation}\label{eq:delta_s_prop_alpha_4_3}
\Delta s \propto \alpha_{\text{MLT}}^{-4/3}.
\end{equation}

Next, consider the photospheric entropy in the models. For an ideal gas with $\gamma=5/3$, 
\begin{equation} \label{eq:s_ph_T_eff}
s_{\text{ph}} \simeq \frac{N_{\text{A}} k_{\text{B}}}{\mu} \ln{\frac{T_{\text{eff}}^{5/2}}{p_{\text{ph}}}} \propto \frac{N_{\text{A}} k_{\text{B}}}{\mu} \ln{(T_{\text{eff}}^{23/2} \rho_{\text{ph}}^{1/2} R^2)},
\end{equation}
where $R$ is the stellar radius, and the proportionality assumes that the photosphere occurs at a pressure $p_{\text{ph}} \propto g/\kappa_{\text{ph}}$, with the surface opacity $\kappa_{\text{ph}}$ taken for simplicity to be dominated by H- opacity \citep{1988PASP..100.1474S}, which is proportional to $\rho_{\text{ph}}^{1/2}T_{\text{eff}}^{9}$. Note that in actuality, molecules also contribute substantially to the near-surface opacity in objects of this mass \citep{2005ApJ...623..585F}, and become more dominant at lower masses. The photospheric entropy is thus tightly linked to variations in the effective temperature, and this in turn is tightly constrained to lie within a narrow range: if the temperature were suddenly made much higher, for example, the opacity would sharply increase, increasing the optical depth at a given pressure level and hence driving the photosphere upwards (i.e., to lower pressure, and hence to lower temperatures). Conversely, much lower temperatures would lead to much lower opacities, requiring that the photosphere (at fixed optical depth) move inwards (to higher pressures, and higher temperatures). This behavior is well-known, and is essentially the basis for the ``forbidden region" of cool temperatures in pre-main-sequence evolution \citep{1961PASJ...13..450H}. In the present context, only modest variations in $T_{\text{eff}}$ are therefore allowed. Within this allowed range, models with lower $\alpha_{\text{MLT}}$ have a lower $T_{\text{eff}}$: for the same initial interior conditions, steeper entropy (and temperature) gradients are, per our discussion of $\Delta s$ above, required to carry out the same surface luminosity, and this leads to slightly lower surface temperatures (the subsequent evolution of $T_{\text{eff}}$ is somewhat more involved, as we will discuss more below, but the tendency to have lower $T_{\text{eff}}$ at lower $\alpha_{\text{MLT}}$ is robust). The strong dependence of $s_{\text{ph}}$ on $T_{\text{eff}}$ dominates over changes in $\rho_{\text{ph}}$ and stellar radii between models at a given age, implying (finally) that $s_{\text{ph}}$ is lower in models with lower $\alpha_{\text{MLT}}$.  

Finally, we turn to discussion of the nearly-constant specific entropy $s_{\text{ad}}$ in the deep interior of the models. This exhibits different behavior on the main-sequence than during the pre-main-sequence contraction phase. Recall that during this phase, stars descend along a Hayashi track at nearly constant $T_{\text{eff}}$; they contract because they are losing total energy (via radiative losses from the surface), so the contraction rate depends on the star's luminosity. From the virial theorem, the internal temperature of the star increases as its radius decreases ($T \propto R^{-1}$), but the increasing density ($\rho \propto R^{-3}$) results in a net loss of entropy. During this phase, it is clear from Figure~\ref{fig:entropy_v_logRho} that $s_{\text{ad}}$ is higher at a given age in models with lower $\alpha_{\text{MLT}}$. This mostly reflects the fact that these low-$\alpha_{\text{MLT}}$ models have had a slightly lower effective temperature during their contraction, and have ultimately lost somewhat less entropy at any fixed time; they therefore have a somewhat greater specific entropy at the time sampled in this figure. At these ages, the enhanced entropy contrast associated with lower $\alpha_{\text{MLT}}$ (per our discussion above) is thus not entirely confined to the near-surface layers: though the photospheric entropy is lower for low-$\alpha_{\text{MLT}}$ models, $s_{\text{ad}}$ is also higher.  

The pre-main-sequence contraction eventually ends because the interior temperature and density have increased enough for nuclear fusion in the core (rather than gravitational contraction) to provide the energy needed to offset the star's radiative losses at the surface. On the main-sequence, then, the value of $s_{\text{ad}}$ is not merely determined by the star's initial entropy and by its passive cooling (which was mediated by the near-surface layers): rather, it is bounded from below by the entropy production associated with nuclear fusion occurring in a steady state.  Of course this also is informed by the near-surface layers to some degree, but only insofar as these affect the entropy production rate by nuclear reactions. For the depth-independent $\alpha_{\text{MLT}}$ values probed here, these changes are modest, and so the deep interior entropy $s_{\text{ad}}$ is largely constant across models with varying $\alpha_{\text{MLT}}$ (at even smaller values of $\alpha_{\text{MLT}}$, $s_{\text{ad}}$ would be altered, as explored for example in \citealt{Chabrier:2007aa}). Thus in these models the higher $\Delta s$ associated with less efficient convection is almost entirely confined to the near-surface layers: the decrease in photospheric entropy with decreasing $\alpha_{\text{MLT}}$ compensates almost exactly for the increasing $\Delta s$.

\subsection{Scaling of stellar radius with $s_{\text{ad}}$ and $\alpha_{\text{MLT}}$} \label{subsec:radius_s_ad}

It has long been realized that a star's radius is sensitive to changes in its entropy \citep[see, e.g.][]{1988PASP..100.1474S,2004sipp.book.....H}. {For example, for a star with constant specific entropy, well-described by a polytropic model $p = K \rho^\gamma$, where $K$ is the polytropic constant, straightforward rearrangement gives}

\begin{equation}\label{eq:s_polytropic}
{s = \frac{N_{\text{A}} k_{\text{B}}}{\mu} \ln{(K)}}.
\end{equation}
{It can be shown that $K \propto M^{2-\gamma} R^{3 \gamma - 4}$ (see equation~(7.40) in \citealt{2004sipp.book.....H}), where $M$ is the stellar mass. By substituting this into equation~(\ref{eq:s_polytropic}) and integrating over the mass distribution of the stellar model, yielding the total entropy $S_{\text{tot}} \sim s M$ for a star of uniform composition, it can be shown that the stellar radius increases with the exponent of $S_{\text{tot}}$ for fixed mass:}

\begin{equation}\label{eq:R_exp_s_tot}
R \propto \exp{\left(\frac{\gamma - 1}{3 \gamma -4} \frac{\mu S_{\text{tot}}}{N_{\text{A}} k_{\text{B}} M}\right)},
\end{equation}
as noted for example in \citet{2004sipp.book.....H} (their equation~(7.150)). More precise relations between $R$, $S_{\text{tot}}$, and other variables can be derived in some specific cases, and these figure prominently in the classic theory of stellar structure \citep[e.g.,][]{1926ics..book.....E,1961PASJ...13..442H}. For example, for a star in hydrostatic equilibrium, the assumption of a perfectly isentropic interior allows relation of the central temperature, pressure, and density to the values of these quantities at the surface, following standard polytropic theory. If the nuclear energy generation $\epsilon$ is provided by fusion, it is further possible to solve for the radius of the star from first principles (by equating the luminosity produced by fusion, $L_{\text{fusion}} \propto R^3 \epsilon \propto R^3 \rho_{\text{c}}^2 T_{\text{c}}^{6}$ for the pp-chain, to the surface luminosity $L_{\text{surf}} = 4 \pi R^2 T_{\text{eff}}^4$, and adopting a closed-form expression for the surface opacity).  

However, the structure models calculated by MESA (or any other stellar structure code) are not isentropic. The level of departure from isentropy depends on details of the models, and in particular on the convective mixing length. In practice, as discussed in \S~\ref{subsec:role_entropy}, most of the entropy resides in the deep interior with nearly-constant specific entropy $s_{\text{ad}}$, so that $S_{\text{tot}} \simeq s_{\text{ad}} M$ and equation~(\ref{eq:R_exp_s_tot}) simplifies to
\begin{equation}\label{eq:R_exp_s_ad}
R \propto \exp{\left(\frac{\gamma - 1}{3 \gamma -4} \frac{\mu s_{\text{ad}}}{N_{\text{A}} k_{\text{B}}}\right)}.
\end{equation}
{Thus, we can relate the ratio of two stellar radii and the change in $s_{\text{ad}}$ between two fixed mass models:}

\begin{figure*}
\gridline{\fig{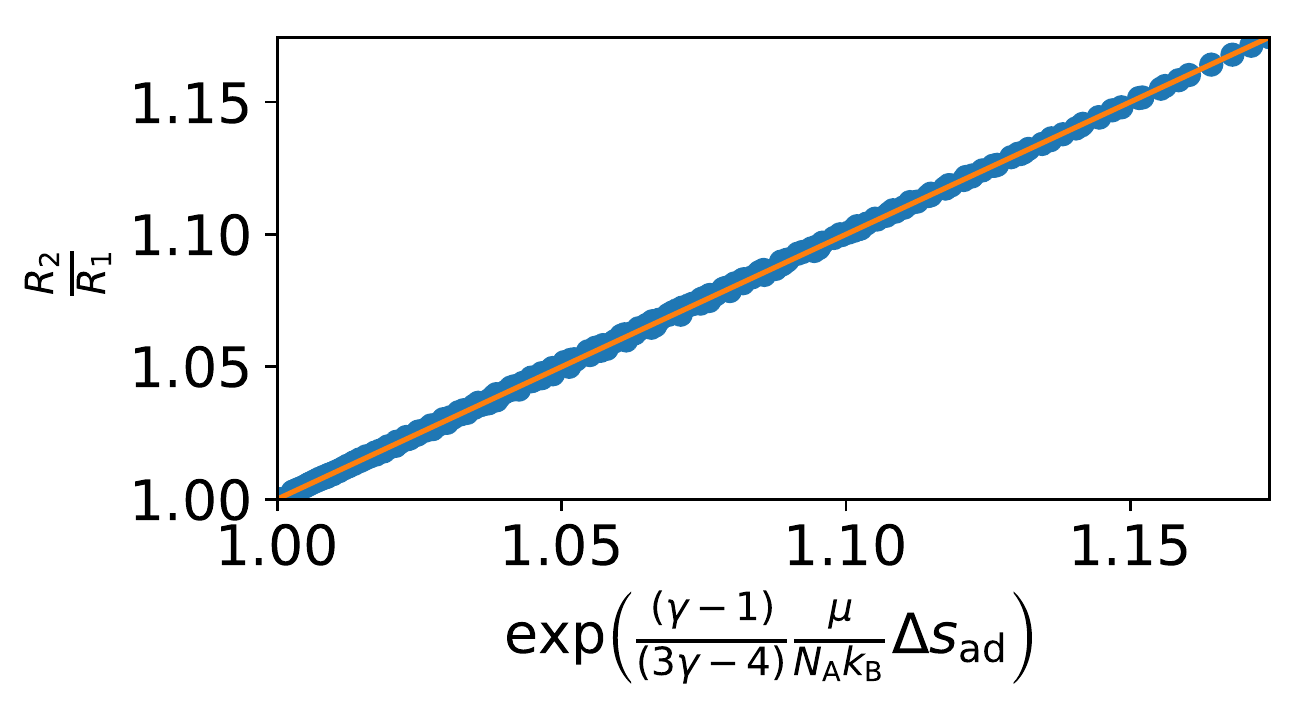}{0.5\textwidth}{(a)}
\fig{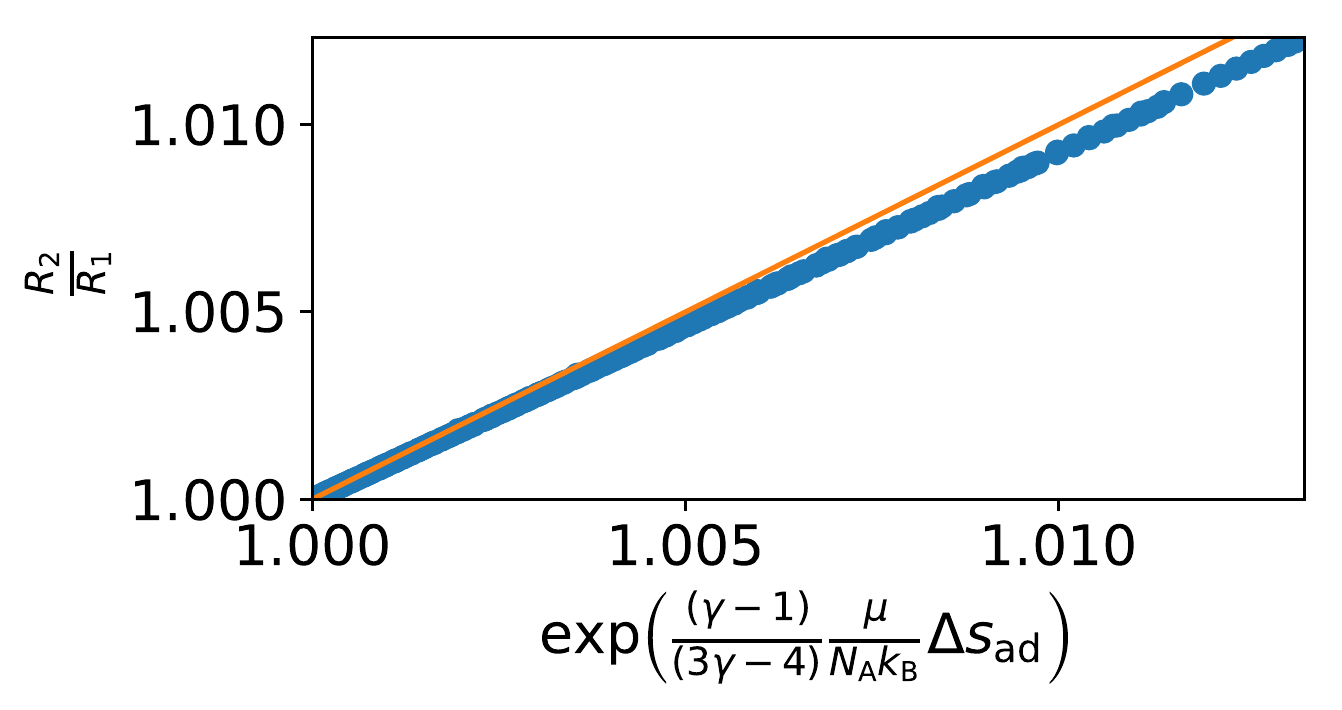}{0.51\textwidth}{(b)}
          }
\caption{The ratio of stellar radii $R_2/R_1$ as a function $\Delta s_{\text{ad}}$ via equation~(\ref{eq:radii_entropy_relation}), for $0.3 \, \text{M}_\odot$, (a) $10 \, \text{Myr}$ (b) $1 \, \text{Gyr}$ stellar models at $\alpha_{\text{MLT}} = 0.5 - 1.7$ ($\Delta 0.05$). We see a divergence at $1 \, \text{Gyr}$, which arise from deviations from the ideal equation of state. $y=x$ (orange) is plotted for ease of comparison.\label{fig:R2_R1_LHS_v_RHS_s_ad}}
\end{figure*}

\begin{equation}\label{eq:radii_entropy_relation}
\frac{R_2}{R_1} \simeq \exp{\left(\frac{\gamma - 1}{3\gamma - 4} \frac{\mu \Delta s_{\text{ad}}}{N_{\text{A}} k_{\text{B}}} \right)}.
\end{equation}
{where $R_1$, $R_2$ are the radii of the first and second model respectively, assuming a uniform $\gamma = 5/3$ for simplicity.} This illustrates how an increase in $s_{\text{ad}}$ ``inflates" the stellar radius of these stellar models. Choosing $\alpha_{\text{MLT}} = 1.7$ to be our unperturbed model, we determine an unperturbed stellar radius $R_0 = 0.683 \, \text{R}_\odot$ and $0.286 \, \text{R}_\odot$, for $10 \, \text{Myr}$ and $1 \, \text{Gyr}$ respectively. 

Models with different $\alpha_{\text{MLT}}$ have somewhat different radii. For example, at $10 \, \text{Myr}$, ``perturbing" our standard model by considering $\alpha_{\text{MLT}}$ in the range 0.5-1.7 results in radius inflation $\Delta R / R_0 \lesssim 17.5 \%$ ($R \lesssim 0.803 \, \text{R}_\odot$). However, for $1 \, \text{Gyr}$ models with the same range of $\alpha_{\text{MLT}}$, we only find $\Delta R / R_0 \lesssim 1.5 \%$ ($R \lesssim 0.289 \, \text{R}_\odot$); we analyze this important difference in the radius inflation between pre-main-sequence and main-sequence models in more detail below, but for now note that it stems partly from the lower superadiabaticity of these main-sequence models. This in turn implies that the properties of fixed mass fully-convective main-sequence stars are relatively insensitive to $\alpha_{\text{MLT}}$ in standard stellar structure models \citep[as noted previously by, e.g.,][]{Chabrier:2007aa,Feiden:2014aa}.

In Figure~\ref{fig:R2_R1_LHS_v_RHS_s_ad}, we examine the ratio of two outputted stellar radii as a function of $\Delta s_{\text{ad}}$ via equation~(\ref{eq:radii_entropy_relation}), for $0.3 \, \text{M}_\odot$ stellar models at both $10 \, \text{Myr}$ and $1 \, \text{Gyr}$ for all possible model comparisons between $\alpha_{\text{MLT}} = 0.5 - 1.7$ ($\Delta 0.05$). The line $y=x$, which would indicate perfect agreement with equation~(\ref{eq:radii_entropy_relation}), is over-plotted (orange line) for ease of comparison. At $10 \, \text{Myr}$ (left panel), the variations in stellar radii are captured extremely well by this expression; at $1 \, \text{Gyr}$ (right panel), they deviate from it slightly. The small deviations from equation~(\ref{eq:radii_entropy_relation}) arise partly from departures from the ideal equation of state assumed in our derivation of this equation. In particular, the central temperature for stars of this mass on the main sequence deviates slightly from the virial expectation that $T \propto M/R$ (owing partly to the fact that these interiors are somewhat degenerate). {Further deviations from equation~(\ref{eq:radii_entropy_relation}) arise due to our assumption of a uniform $\gamma = 5/3$ in deriving this expression; in our models, $\gamma$ is indeed roughly uniform (and $= 5/3$) in the interiors of our pre-main-sequence models, but deviates from this slightly on the main-sequence. (These deviations in turn arise partly from Coulomb interactions, which though small are not entirely negligible.)} Note, further, that the overall range in stellar radii across all models, and likewise the variation in $s_{\text{ad}}$ across these models, is much smaller than on the pre-main-sequence.  

\begin{figure*}
\gridline{\fig{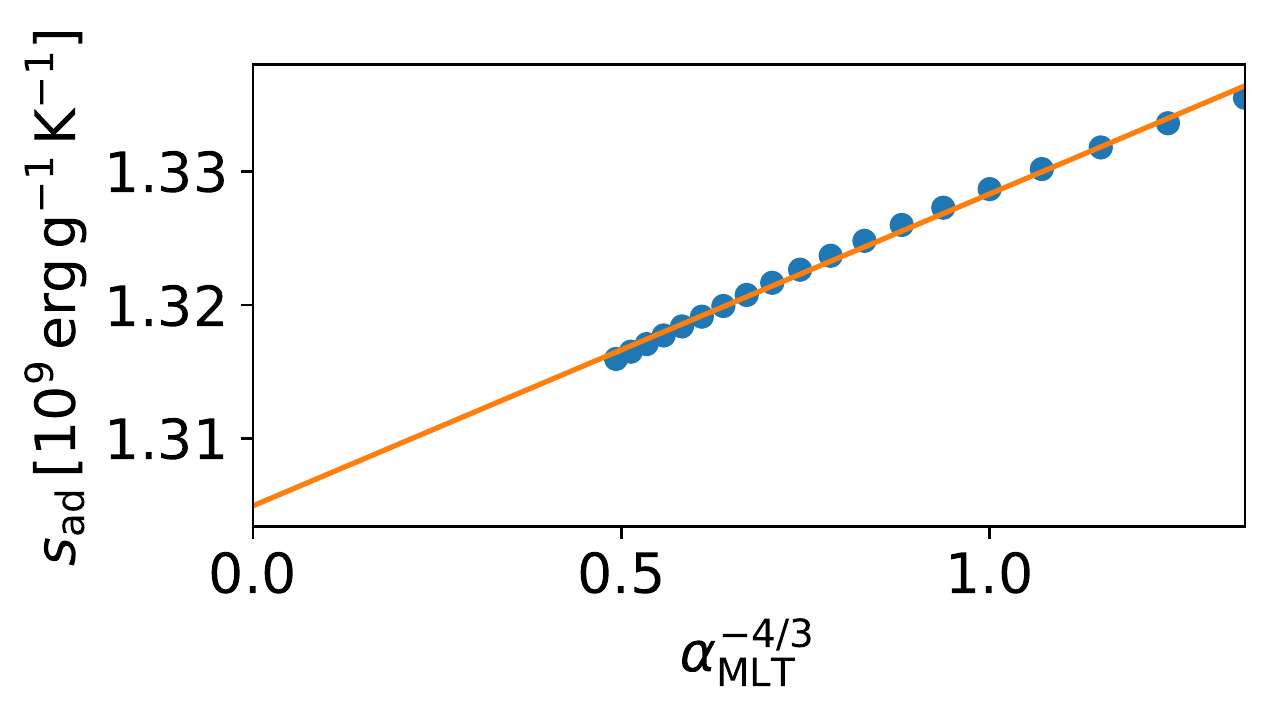}{0.5\textwidth}{(a)}
\fig{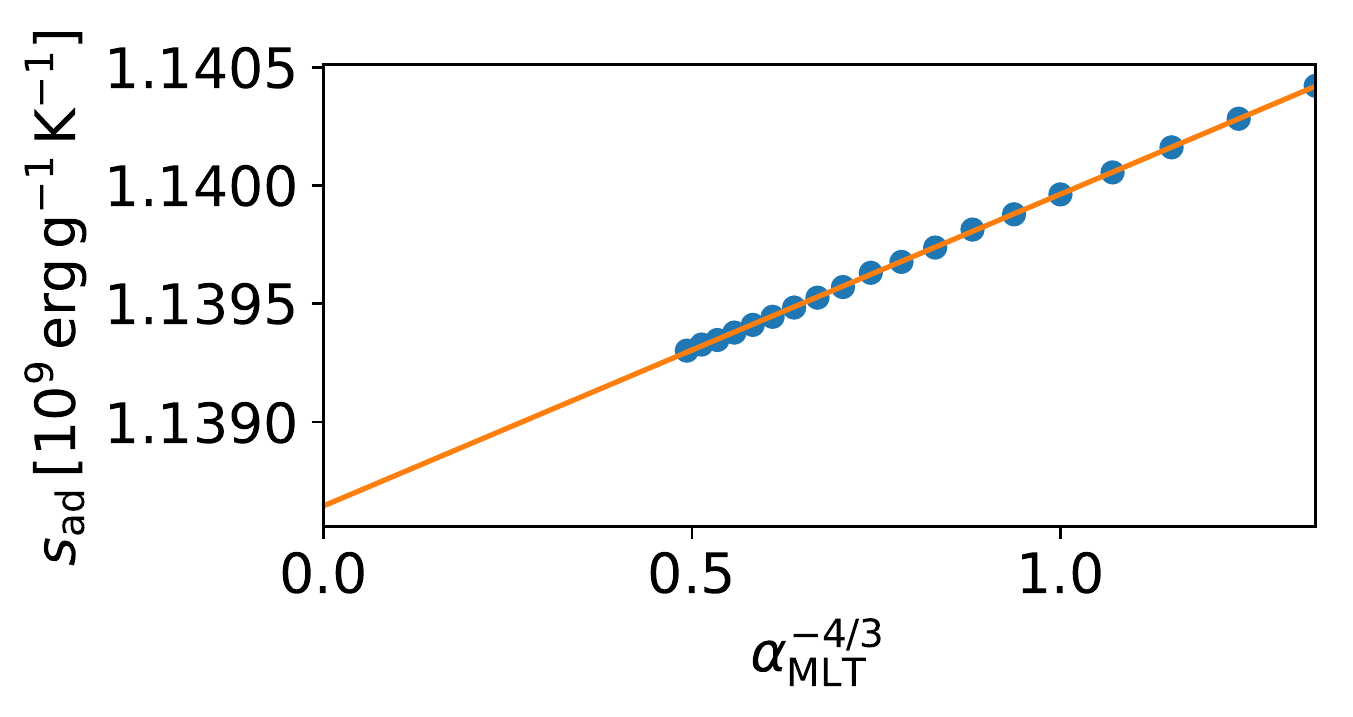}{0.525\textwidth}{(b)}
          }
\caption{$s_{\text{ad}}$ as a function of $\alpha_{\text{MLT}}^{-4/3}$ for $0.3 \, \text{M}_\odot$, (a) $10 \, \text{Myr}$ (b) $1 \, \text{Gyr}$ stellar models at $\alpha_{\text{MLT}} = 0.8 - 1.7$ ($\Delta 0.05$). We extrapolate to the isentropic value of $s_{\text{ad}}$ at that given age, which drops as a function of age in the pre-main-sequence, settling in the main-sequence. The trend between $s_{\text{ad}}$ and $\alpha_{\text{MLT}}^{-4/3}$ also decreases with age.} \label{fig:s_ad_v_alpha_4_3}
\end{figure*}

The changes in $s_{\text{ad}}$, and hence in the stellar radius, are linked to changes in $\alpha_{\text{MLT}}$. To examine this quantitatively, we must find how the value of the adiabat is linked to $\Delta s$. For example, if all the changes in $\Delta s$ between models were reflected simply in changes to the photospheric entropy $s_{\text{ph}}$, this would imply an $s_{\text{ad}}$ that is nearly uniform across models; meanwhile if $s_{\text{ph}}$ were instead somehow held constant across all models, changes in $\Delta s$ would translate directly to changes in $s_{\text{ad}}$. The true relation between $s_{\text{ph}}$ and $s_{\text{ad}}$ (and hence $\Delta s$) is more complex than either of these simple examples. Overall, though, as established previously, a decrease in $\alpha_{\text{MLT}}$ decreases $s_{\text{ph}}$ and (on the pre-main-sequence in particular) increases $s_{\text{ad}}$. 

To see roughly why this is so, note that in general the surface luminosity $L_{\text{surf}} \propto R^2 T_{\text{eff}}^4$, which (using equation~(\ref{eq:s_ph_T_eff})) can be written as 

\begin{equation}\label{eq:Lsurf_sph}
L_{\text{surf}} \propto \exp{\left(\frac{\mu s_{\text{ph}}}{N_{\text{A}} k_{\text{B}}} \right)} / (\rho_{\text{ph}}^{1/2} T_{\text{eff}}^{15/2}).
\end{equation}
On the pre-main-sequence, the luminosity is ultimately derived from gravitational contraction, with $L_{\text{surf}} \propto R^{-2} (dR/dt)$.  Equating the two, and noting how $R$ scales with $s_{\text{ad}}$ (equation~(\ref{eq:R_exp_s_ad})), implies that for contraction at nearly constant effective temperature, we must have

\begin{equation}\label{eq:Lsurf_sph2}
\exp{\left(-\frac{\gamma - 1}{3 \gamma -4} \frac{\mu s_{\text{ad}}}{N_{\text{A}} k_{\text{B}}} \right)} \propto \exp{\left(\frac{\mu s_{\text{ph}}}{N_{\text{A}} k_{\text{B}}} \right)} / (\rho_{\text{ph}}^{1/2} T_{\text{eff}}^{15/2}).
\end{equation}
This in turn implies that $s_{\text{ad}} \propto - s_{\text{ph}}$ on the pre-main-sequence (plus additional smaller terms). A similar proportionality holds on the main-sequence, where now the interior luminosity is generated by fusion, with $L \propto R^3 \epsilon \propto R^2 \rho_{\text{c}}^2 T_{\text{c}}^6 \propto R^{-9}$ for stars in virial equilibrium. This again implies $s_{\text{ad}} \propto - s_{\text{ph}}$, but with a different (and in fact significantly smaller) constant of proportionality. Thus in both cases, in comparing models of similar total convective flux ($\alpha_{\text{MLT}} = 0.8-1.7$), we have that $s_{\text{ad}} \propto \Delta s$, hence

\begin{equation}\label{eq:s_ad_prop_alpha_4_3}
s_{\text{ad}} \propto \alpha_{\text{MLT}}^{-4/3}.
\end{equation}
The constant of proportionality decreases with the age of the model---as discussed previously, the interior adiabat in pre-main-sequence models is more sensitive to variations in $\alpha_{\text{MLT}}$---but the proportionality holds true even for main-sequence models.

In Figure~\ref{fig:s_ad_v_alpha_4_3}, we examine $s_{\text{ad}}$ as a function of $\alpha_{\text{MLT}}^{-4/3}$ for $0.3 \, \text{M}_\odot$ stellar models at $\alpha_{\text{MLT}} = 0.8 - 1.7$ ($\Delta 0.05$) for both $10 \, \text{Myr}$ and $1 \, \text{Gyr}$, where the proportionality in equation~(\ref{eq:s_ad_prop_alpha_4_3}) holds for both ages here. We extrapolate to find $s_{\text{ad} (\alpha_{\text{MLT}} \rightarrow \infty)}$, i.e., the value corresponding to an isentropic model, which gives the constant of proportionality in equation~(\ref{eq:s_ad_prop_alpha_4_3}) as

\begin{equation}\label{eq:d_sad_s_alpha_4_3}
\frac{d s_{\text{ad}}}{d\alpha_{\text{MLT}}^{-4/3}} \approx \frac{s_{\text{ad}} - s_{\text{ad} (\alpha_{\text{MLT}} \rightarrow \infty)}}{\alpha_{\text{MLT}}^{-4/3}}.
\end{equation}
Thus, for fully-convective stellar models of similar total convective flux, one can predict the radius inflation between two models of known $\alpha_{\text{MLT}}$ without having to determine a perturbed model's $s_{\text{ad}}$, solely using the unperturbed model's $s_{\text{ad}}$, and $s_{\text{ad} (\alpha_{\text{MLT}} \rightarrow \infty)}$ at a given age:

\begin{equation} \label{eq:R2_R1_delta_alpha_4_3_relation}
{\frac{R_2}{R_1} \approx \exp{\left(\frac{\gamma - 1}{3\gamma - 4} \frac{\mu}{N_{\text{A}} k_{\text{B}}} \frac{s_{\text{ad}_1} - s_{\text{ad} (\alpha_{\text{MLT}} \rightarrow \infty)}}{\alpha_{\text{MLT}_1}^{-4/3}} \Delta \left(\alpha_{\text{MLT}}^{-4/3}\right)\right)}}.
\end{equation}

\begin{figure}
\includegraphics[width=0.5\textwidth]{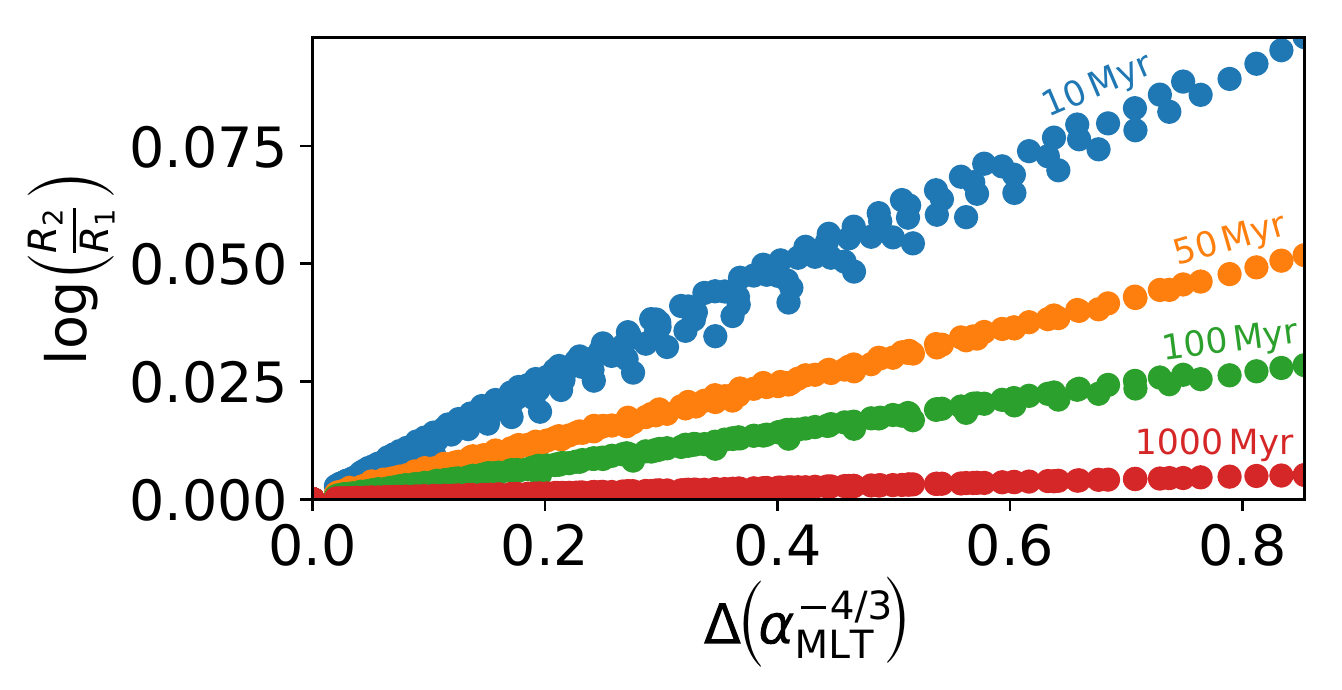}
\caption{$\log{(R_2/R_1)}$ as a function of $\Delta \left(\alpha_{\text{MLT}}^{-4/3}\right)$ at various ages for $0.3 \, \text{M}_\odot$ stellar models at $\alpha_{\text{MLT}} = 0.8 - 1.7$ ($\Delta 0.05$). During the pre-main-sequence, the trend decreases with age (see equation~(\ref{eq:radius_inflation_time})), eventually reaching levels of negligible radius inflation in the main-sequence.} \label{fig:ln_R_2_R_1_v_delta_alpha_4_3}
\end{figure}

In Figure~\ref{fig:ln_R_2_R_1_v_delta_alpha_4_3}, we examine the ratio of two outputted stellar radii as a function of $\Delta \left(\alpha_{\text{MLT}}^{-4/3}\right)$ at different ages between $10 \, \text{Myr}$ and $1 \, \text{Gyr}$. Models at all $\alpha_{\text{MLT}}$ contract on the pre-main-sequence, with $dR/dt \propto R^4$, implying in turn that $R \propto t^{-1/3}$ if the effective temperature remains constant. In our models, the radius inflation between two models of differing $\alpha_{\text{MLT}}$ decreases with age, becoming almost negligible during the main-sequence ($1 \, \text{Gyr}$). This time dependence ultimately reflects the fact that (as shown in Figure~\ref{fig:s_ad_v_alpha_4_3}) $d s_{\text{ad}}/d \alpha_{\text{MLT}}^{-4/3}$ (equation~(\ref{eq:d_sad_s_alpha_4_3})) changes with time, becoming much shallower on the main sequence; per equation~(\ref{eq:R2_R1_delta_alpha_4_3_relation}), this means a less pronounced radius inflation for a given change in $\alpha_{\text{MLT}}$. Empirically, we find that }  

\begin{equation} \label{eq:radius_inflation_time}
\frac{R_2}{R_1} \propto t^{-0.03 \Delta \left(\alpha_{\text{MLT}}^{-4/3}\right)},
\end{equation}
demonstrating that a larger change in $\alpha_{\text{MLT}}$ between two models does indeed result in the model contracting more rapidly with time.

As previously discussed, during the main-sequence, $s_{\text{ad}}$ is predominantly bounded by the entropy production via nuclear fusion, thus any changes in $s_{\text{ph}}$ in our models fail to produce noticeable changes in $s_{\text{ad}}$. Hence, for our range of main-sequence models, where $\Delta s_{\text{ph}} \lesssim 10^7 \, \text{erg} \, \text{g}^{-1} \, \text{K}^{-1}$, we find that $R_2 \approx R_1$, as demonstrated by the trend at $1 \, \text{Gyr}$ in Figure~{\ref{fig:ln_R_2_R_1_v_delta_alpha_4_3}}. 

Note that our lowest-efficiency models in Figures~\ref{fig:logsupergrad_v_logRho} and~\ref{fig:entropy_v_logRho} have $\alpha_{\text{MLT}} = 0.5$; at even lower values, radius inflation is possible even on the main sequence, as demonstrated for example by \citet{Chabrier:2007aa}. In this regime, however, the convective flux is not the same as at higher values of $\alpha_{\text{MLT}}$ (that is, the nuclear energy generation in the interior is affected), breaking the assumptions made in our analysis. Indeed, \citet{Chabrier:2007aa} show that at $\alpha_{\text{MLT}} \approx 0.05$ a radiative (stable) core begins to form in the interior, violating our assumption that the star is fully-convective.  We defer analysis of such cases to other work.

\section{Rotational inhibition of convection: Stevenson (1979) formulation} \label{sec:rot_inhibition_conv}

\subsection{Theory: rotational modification to MLT} \label{subsec:theory_rot}

As noted in \S~\ref{sec:intro}, rotation generally acts to inhibit convection. In linear theory, this inhibition manifests as an increase in the critical Rayleigh number required to drive convection \citep{1961hhs..book.....C}. The effects of rotation in the non-linear regime are more difficult to gauge, but many authors have argued that ultimately the temperature gradient required to transport a given heat flux by convection must increase somewhat if the rotation is sufficiently rapid. \citet{Stevenson:1979aa} (\citetalias{Stevenson:1979aa}, hereafter), for example, derived a mixing length prescription for rotating convection through consideration of the growth of linear, Boussinesq convective modes, constructing a finite amplitude theory by assuming that non-linearities, such as shear instabilities, limit the amplitude of the flow. Following \citet{1954RSPSA.225..196M}, \citetalias{Stevenson:1979aa} argued that the convective flow is dominated by the modes that transport the most heat. \citetalias{Stevenson:1979aa} use this model to relate $\nabla_{\text{s}}$ in a ``perturbed" model (at rotation rate $\Omega$) to the unperturbed (non-rotating) model's:

\begin{equation} \label{eq:general_rotation}
	\left(\frac{\nabla_{\text{s}}}{\nabla_{\text{s}_0}}\right)^{5/2} - \frac{\nabla_{\text{s}}}{\nabla_{\text{s}_0}} = \frac{1}{41} \text{Ro}^{-2} \equiv \frac{4}{41} \tau_{\text{c}_0}^2 \Omega^2,
\end{equation}
where $\tau_{\text{c}_0}$ is the convective turnover time of the unperturbed model, and $\text{Ro} \equiv (2 \tau_{\text{c}_0} \Omega)^{-1}$ is the Rossby number.

In the slow regime, i.e., $\text{Ro} \gg 1$,

\begin{equation} \label{eq:slow_rotation}
	\nabla_{\text{s}} \simeq \nabla_{\text{s}_0} \left(1 + \frac{1}{62} \text{Ro}^{-2}\right) \equiv \nabla_{\text{s}_0} \left(1 + \frac{4}{62} \tau_{\text{c}_0}^2 \Omega^2 \right),
\end{equation}
converging towards the non-rotating model. In the rapid regime, i.e., $\text{Ro} \ll 1$,

\begin{equation} \label{eq:rapid_rotation}
	\nabla_{\text{s}} \simeq 0.23 \nabla_{\text{s}_0} \text{Ro}^{-4/5} \equiv 0.92 \nabla_{\text{s}_0} \tau_{\text{c}_0}^{4/5} \Omega^{4/5}.
\end{equation}
As $\nabla_{\text{s}} \propto ds/dr$, this mechanism modifies the gradient of the specific entropy, i.e., specific entropy asymptotically converges to a different adiabat in the presence of rotation.

We are motivated to explore this reformulation of MLT partly because more recent investigations have suggested similar scalings for the temperature gradient and/or velocity in rapidly-rotating convection. For example, as noted in \S~\ref{sec:intro}, \citet{Barker:2014aa} derive a rotating MLT equivalent to that of \citetalias{Stevenson:1979aa} via simplified physical arguments, achieving the same scaling between $\nabla_{\text{s}}$ ($dT/dz$ in their case) and $\Omega$ when in the rapidly-rotating regime. To test their relationship, they take an average of $dT/dz$ from the middle third of the convection zone in a series of 3D hydrodynamical simulations of Boussinesq convection in a Cartesian box. They find that $dT/dz$ in the simulations does indeed scale with $\Omega$ as dictated by equation~(\ref{eq:rapid_rotation}), and likewise that the typical velocities and spatial structures amidst the flow also scale with $\Omega$ in the manner predicted by the theory. Previously, \citet{2012PhRvL.109y4503J} also examined the transport in rapidly-rotating convection using a set of asymptotically reduced equations. They likewise find that heat transport in the rapidly-rotating regime is ``throttled" by convection in the bulk of the domain---in marked contrast to the non-rotating case, which is controlled mainly by the boundary layers.  Overall, their theoretical model yields scalings of $dT/dz$ as a function of $\Omega$ that are arguably compatible with those in \citetalias{Stevenson:1979aa} and \citet{Barker:2014aa}. The broad concordance between these different theoretical models suggest that the MLT formulation adopted in \citetalias{Stevenson:1979aa}, though undoubtedly a simplified description of the complex flows occurring in actual stars, may nonetheless adequately capture how the primary quantity of interest for stellar convection---namely the temperature or entropy gradient as a function of the flux---varies with rotation rate.

We therefore incorporate rotational effects into our 1D stellar structure models by implementing the modified MLT formulation of \citetalias{Stevenson:1979aa} into MESA. Observations and simulations of fully-convective stars have indicated that they are likely to rotate mostly as solid bodies, supporting our choice of using a fixed $\Omega$ to model rotation inhibition. \citet{Barnes:2005aa} shows surface differential rotation diminishes with increasing convective depth in low-mass stellar observations, and magnetohydrodynamical (MHD) simulations performed by, e.g., \citet{Browning:2008aa} and \citet{2015ApJ...813L..31Y,2016GeoJI.204.1120Y}, suggest that magnetic fields react strongly on flows, helping to enforce solid-body rotation. 

\subsection{Radius inflation: S79 models} \label{subsec:s79_models}

\begin{figure*}
\gridline{\fig{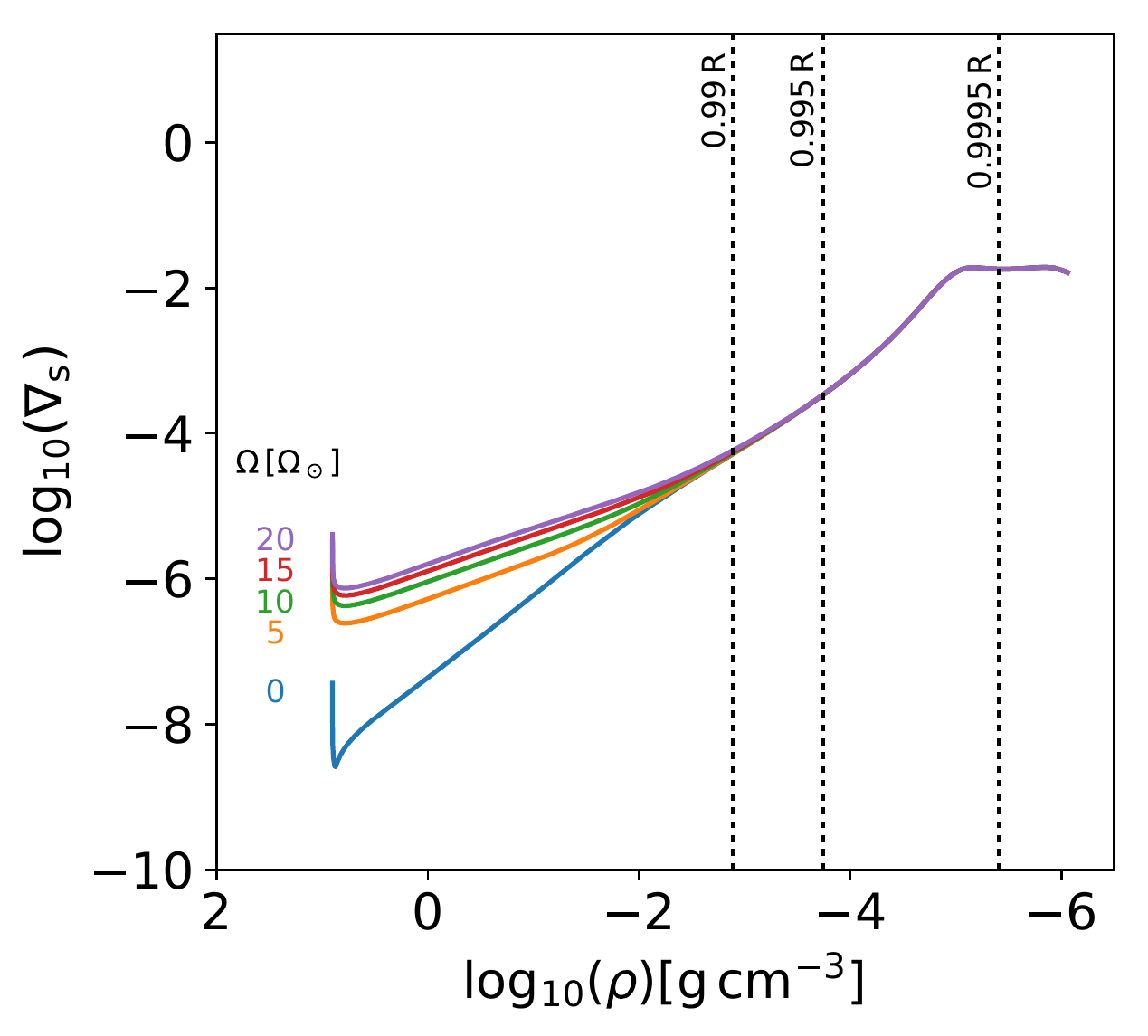}{0.5\textwidth}{(a)}
          \fig{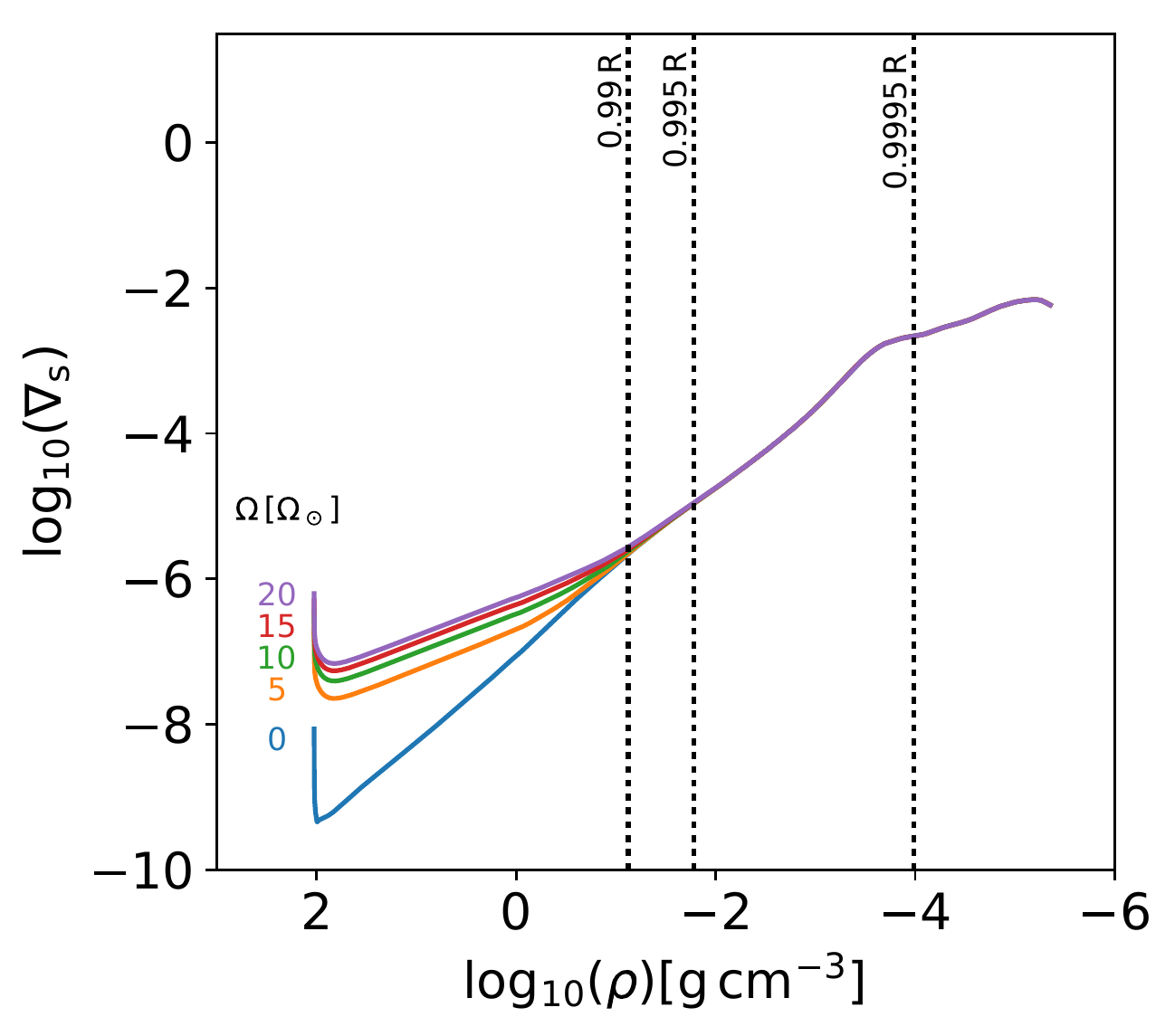}{0.51\textwidth}{(b)}
          }
\caption{$\log_{10}{(\nabla_{\text{s}})}$ as a function of $\log_{10}{(\rho)}$, for $0.3 \, \text{M}_\odot$, (a) $10 \, \text{Myr}$ (b) $1 \, \text{Gyr}$ stellar models at $\Omega = 0 - 20 \, \Omega_\odot$ ($\Delta 5 \, \Omega_\odot$). As $\Omega$ increases, $\nabla_{\text{s}}$ increases throughout the bulk of the stellar interior ($\text{Ro} \ll 1$), but becomes comparable to the unperturbed model in the near-surface layers ($\text{Ro} \gg 1$).\label{fig:logsupergrad_v_logRho_rot}}
\end{figure*}

Some active low-mass stars are fast rotators, with rotation velocities $v_{\text{rot}} \gtrsim 10 \, \text{km s}^{-1}$ \citep[e.g.,][]{Reid:2002aa,Mohanty:2003aa} in some cases. We test $\Omega = 5 - 20 \, \Omega_\odot$ ($\Delta 5 \, \Omega_\odot$), which is $\lesssim 10\%$ of the break-up velocity at $10 \, \text{Myr}$, and $\lesssim 3\%$ at $1 \, \text{Gyr}$, of our unperturbed $0.3 \, \text{M}_\odot$, $\alpha_{\text{MLT}} = 1.7$ stellar model. These produce typical rotation velocities of $v_{\text{rot}} \simeq 7-27 \, \text{km s}^{-1}$ and $v_{\text{rot}} \simeq 3-11 \, \text{km s}^{-1}$ for $10 \, \text{Myr}$ and $1 \, \text{Gyr}$ respectively. We have not attempted to account for changes in the effective gravity as $\Omega$ increases; since the angular velocity in all cases is only a small fraction of the breakup velocity, these effects probably play only a minor role. At each $\Omega$, {we calculate a new value of $\nabla_{\text{s}}$ at each point in the mass distribution, by modifying the non-rotating} $\nabla_{\text{s}}$ according to equation~(\ref{eq:general_rotation}), representing a ``rotating" version of the 1D stellar structure model. The depth-dependence of $\nabla_{\text{s}}$ is then determined by the profile of $\text{Ro}$, which in turn depends on the convective overturning time at every depth in the model. Here, we take this overturning time simply to be $\tau_{\text{c}_{\text{0}}}$ from the \emph{unperturbed} model---that is, we neglect the small changes in overturning time associated with changes in the convective velocity at rapid rotation. This simplification has the consequence that our models slightly underestimate the influence of rotation at any fixed $\Omega$ (compared to a fully self-consistent model), but we will see in a moment that this effect is utterly negligible for the overall structure.

In Figure~\ref{fig:logsupergrad_v_logRho_rot}, we plot $\log_{10}{(\nabla_{\text{s}})}$ as a function of $\log_{10}{(\rho)}$ for $0.3 \, \text{M}_\odot$, $\alpha_{\text{MLT}} = 1.7$ stellar models, at both $10 \, \text{Myr}$ and $1 \, \text{Gyr}$, for $\Omega = 0 - 20 \, \Omega_\odot$ ($\Delta 5 \, \Omega_\odot$). In the bulk of the convection zone, convective velocities are low, i.e., $\text{Ro} \ll 1$, hence this region becomes more superadiabatic than the unperturbed model by a few orders of magnitude. However, due to the already near-adiabaticity in this region, this perturbation in $\nabla_{\text{s}}$ does not influence the stellar structure noticeably. In the surface layers, where convective velocities become increasingly rapid, $\text{Ro}$ remains $\gg 1$ at all the $\Omega$ values sampled, resulting in negligible changes in the superadiabaticity there, i.e., $\nabla_{\text{s}} \simeq \nabla_{\text{s}_0}$.

The near equivalence of $\nabla_{\text{s}}$ in the surface layers of all these models, and their near-adiabaticity in the bulk of the convection zone, together imply that there are negligible differences between the specific entropy profiles of models at varying rotation rates. As discussed in \S~\ref{sec:entropy}, the radius of the star is determined primarily by the interior adiabat (i.e., $s_{\text{ad}}$), which in turn is largely established by the near-surface layers. Because the near-surface layers have $\text{Ro} \gg 1$ and thus are almost totally uninfluenced by convection, the entropy jump in all our rotating models is nearly identical to that in the non-rotating case. This in turn means that the specific entropy of the deep interior $s_{\text{ad}}$ is unchanged by rotation, even though the deep layers of the star are strongly influenced by Coriolis forces ($\text{Ro} \ll 1$), and $\nabla_{\text{s}}$ varies considerably between models there. This, following the discussion in \S~\ref{subsec:radius_s_ad}, finally implies that rotation will lead only to negligible changes in the overall structure and radius of the star.  

This expectation is confirmed in our models. We measured $\Delta R / R_0 \sim 10^{-2} \%$ and $\sim 10^{-4} \%$ for $\Omega = 5 - 20 \, \Omega_\odot$ models at $10 \, \text{Myr}$ and $1 \, \text{Gyr}$ respectively. Thus, implementing rotational inhibition of convection using this modified formulation of MLT does not produce noticeable changes in the stellar radius. This is, again, due mainly to the depth-dependence of the convective velocities and hence of the Rossby number: if the star were instead well-characterized by a single depth-independent Rossby number, radius inflation would be much more noticeable (for low enough values of $\text{Ro}$).

\section{Magnetic inhibition of convection: MacDonald \& Mullan (2014) formulation} \label{sec:mag_inhibition_conv}

\subsection{Theory: magnetic modification to MLT} \label{subsec:theory_mag}

It is not clear how best to encapsulate the influence of magnetism on convection in 1D stellar structure models. Clearly magnetic fields can inhibit flows via the Lorentz force. However, in the presence of rotation, the effects of magnetism can be more complex, with magnetized rotating fluids sometimes \emph{more} amenable to convection than their non-magnetic equivalents (see \S~\ref{sec:intro}). As with rotation, the impact of magnetism in the non-linear regime is much less clear. Various authors have turned to different prescriptions for encapsulating these effects in 1D models, motivated by physical arguments and results from linear theory, as summarized also in \S~\ref{sec:intro}. Here, we have chosen to focus our attention on one such model, namely that proposed by \citet{MacDonald:2014aa} (\citetalias{MacDonald:2014aa}, hereafter), which is a slightly modified form of \citealt{Mullan:2001ab}); we have chosen this model not because it is necessarily superior to others (e.g., \citealt{Feiden:2012aa}, or the reduced-$\alpha_{\text{MLT}}$ models of \citealt{Chabrier:2007aa}), but because its physical motivation is clear, it has been employed in a series of follow-on papers \citep[see, e.g.,][]{2015MNRAS.448.2019M,MacDonald:2017aa,MacDonald:2017ab}, and it is straightforward to implement in a 1D stellar evolution code. In this section, we briefly describe this prescription, its physical motivation, and then discuss its implementation into MESA models. We aim here to examine whether the mechanism by which radii are inflated in these ``magnetic" models is substantially the same as in the non-magnetic cases discussed in \S~\ref{sec:entropy} and \S~\ref{sec:rot_inhibition_conv}; that is, we examine how the radii, specific entropy, and adopted magnetic prescription are linked.  We show that radius inflation in the \citetalias{MacDonald:2014aa} models is, as in their non-magnetic cousins, associated with changes in the specific entropy of the deep interior, which in turn is linked to the entropy contrast in the near-surface layers.

The models of \citetalias{MacDonald:2014aa} are based partly on the linear stability work of \citet{Gough:1966aa}, who derived a criterion for convective instability onset due to a magnetic field in certain circumstances. In non-magnetic models, the criterion of convective onset is purely local; magnetic fields connect parcels of fluid at different levels, so such a criterion is not generally obtainable \citep{Gough:1966aa}. However, simple local stability criteria exist for particularly elementary magnetic field configurations. In practice, \citetalias{MacDonald:2014aa} modify the Schwarzschild criterion due to the presence of a magnetic field:

\begin{equation} \label{eq:schwarz_crit_mag}
	\nabla_{\text{rad}} > \nabla_{\text{ad}} + \frac{\delta}{Q},
\end{equation}
where

\begin{equation} \label{eq:delta}
	\delta = \frac{B_{\text{v}}^2}{B_{\text{v}}^2 + 4 \pi \gamma P_{\text{g}}}
\end{equation}
is a magnetic inhibition parameter, and $Q = -(\partial \ln{\rho} / \partial \ln{T})_p$ is the isobaric expansion coefficient. In this expression, $P_{\text{g}}$ is the gas pressure and $B_{\text{v}}$ is taken by \citetalias{MacDonald:2014aa} to represent the vertical component of the magnetic field, on the grounds that this component figures prominently in the linear stability analysis of \citet{Gough:1966aa}. More generally, we might take $B_{\text{v}}$ as a crude proxy encompassing both the strength of the field at a point and some aspects of its spatial morphology. This parameter ($\delta/Q$) is added to every instance of $\nabla_{\text{ad}}$ in the MLT prescription, in order to determine the perturbed temperature gradient at a given convective energy flux (or vice versa). Physically, this amounts to asserting that the dimensionless temperature gradient in non-linear convection tends not towards $\nabla_{\text{ad}}$, as it would for efficient non-magnetized, non-rotating convection at sufficiently high Rayleigh number, but to $\nabla_{\text{ad}} + \delta/Q$. We have not attempted to take into consideration other effects arising from the presence of a magnetic field (e.g., magnetic pressure). At each time step, the model evolves self-consistently using the perturbed structure. The criterion expressed in equation~(\ref{eq:schwarz_crit_mag}) differs from that used in \citet{Mullan:2001ab} by a factor $Q$, which was adopted in \citetalias{MacDonald:2014aa} onwards to account for non-ideal thermodynamic behavior.

\begin{figure*}
\gridline{\fig{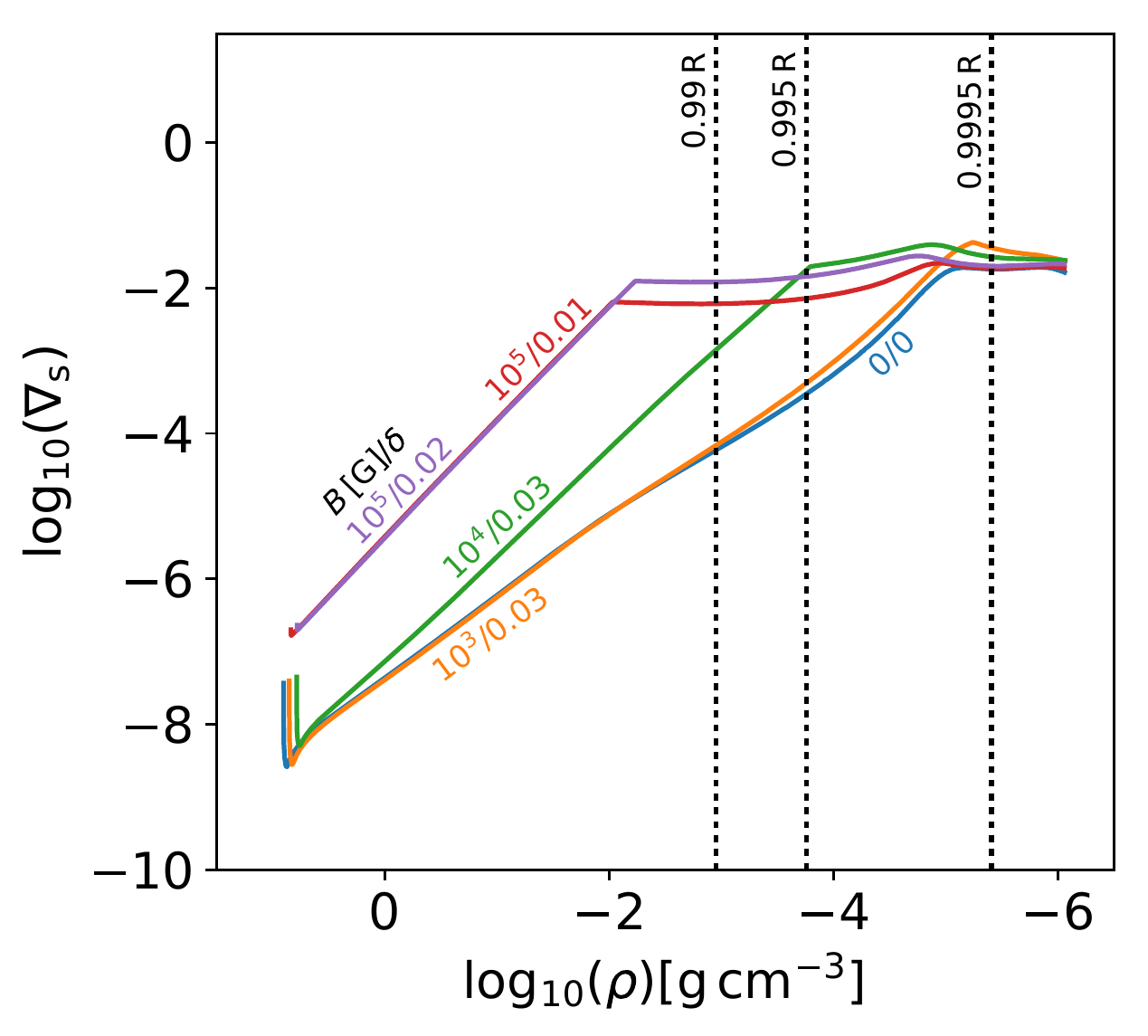}{0.5\textwidth}{(a)}
          \fig{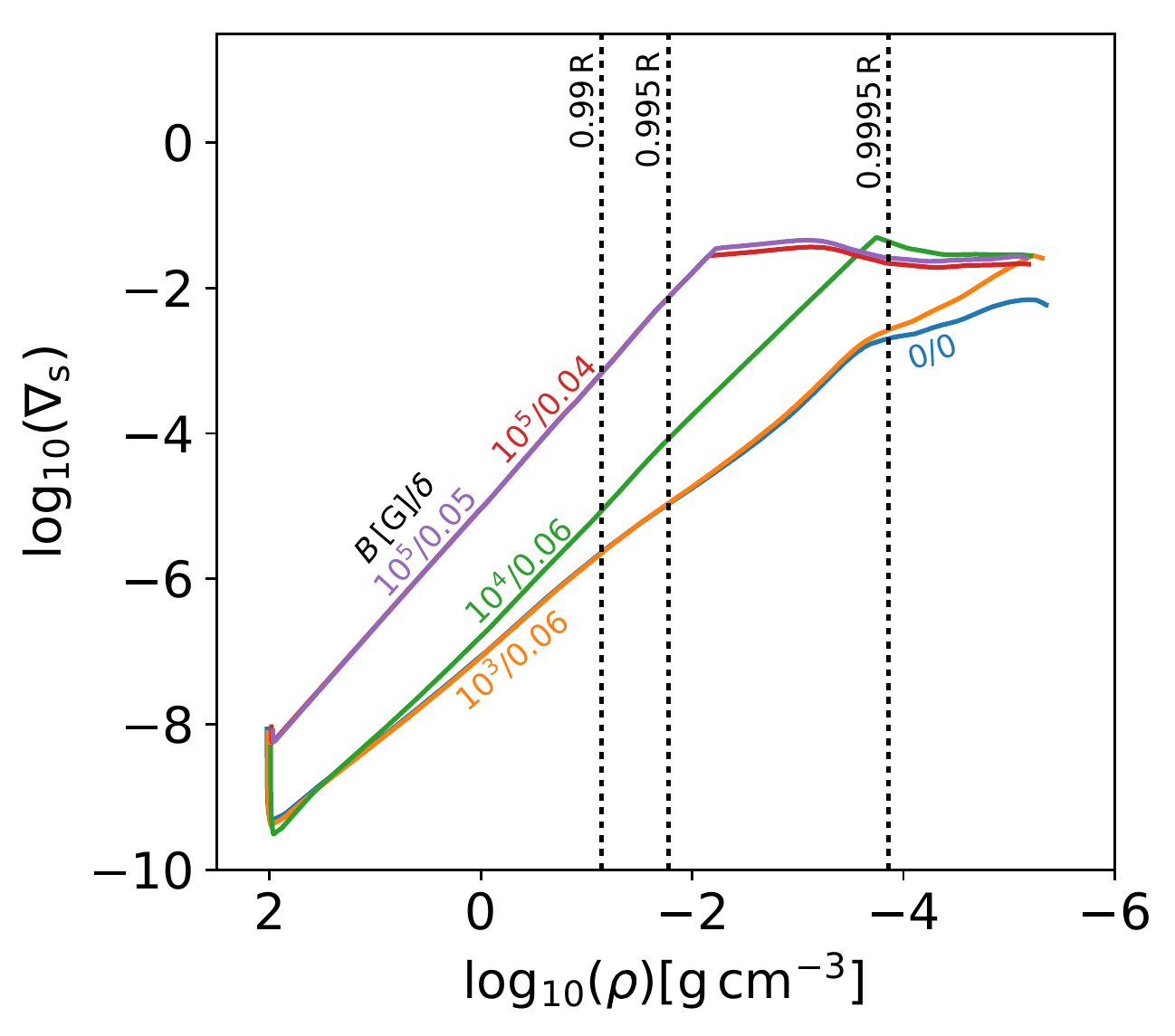}{0.51\textwidth}{(b)}
          }
\caption{$\log_{10}{(\nabla_{\text{s}})}$ as a function of $\log_{10}{(\rho)}$, for $0.3 \, \text{M}_\odot$, (a) $10 \, \text{Myr}$ (b) $1 \, \text{Gyr}$, $\alpha_{\text{MLT}} = 1.7$ stellar models at some combinations of $B_{\text{v-max}} = 10^3-10^5 \, \text{G}$ and (a) $\delta = 0.01 - 0.03$ (b) $\delta = 0.04 - 0.06$, including the unperturbed model. Increasing $\delta$ noticeably increases $\nabla_{\text{s}}$ where $B_{\text{v}} < B_{\text{v-max}}$, and increasing $B_{\text{v-max}}$ increases the depth at which $\delta$ noticeably increases $\nabla_{\text{s}}$.\label{fig:logsupergrad_v_logRho_mag}}
\end{figure*}

Higher values of $B_{\text{v}}$ inhibit convection, requiring a steeper temperature gradient to transport an equivalent heat flux; hence, increasing $\delta$ will increase the superadiabaticity of the stellar interior. The choice of radial profile for $\delta$ is, in these models, somewhat arbitrary. \citetalias{MacDonald:2014aa} choose $\delta = const$ from the surface downwards to some radius $r_{\text{max}}$, where $B_{\text{v}}$ reaches its critical strength $B_{\text{v-max}}$; thus, $\delta$ rapidly decreases with depth for $r<r_{\text{max}}$. Best-fit magneto-convection models performed by this reformulation of MLT are more sensitive to $\delta$ than to the chosen $B_{\text{v-max}}$. The range of vertical surface magnetic field strengths $B_{\text{v-surf}}$ in the models is not dictated by the large range of uncertainty in $B_{\text{v-max}}$, i.e., deep interior field strengths, but rather to the range of $\delta$ considered.

\subsection{Radius inflation: MM14 models} \label{subsec:MM14_models}

We implement this magnetic inhibition of convection into MESA, producing ``magnetic" $0.3 \, \text{M}_\odot$, $\alpha_{\text{MLT}} = 1.7$ stellar models at both $10 \, \text{Myr}$ and $1 \, \text{Gyr}$. For ease of comparison with prior work, we adopt the same strategy as \citetalias{MacDonald:2014aa} by assuming $\delta$ is constant down to some radius $r_{\text{max}}$ at which $B=B_{\text{v-max}}$; below this point, $\delta$ decreases rapidly in accord with the rising gas pressure. It must be noted at the outset that this assumption amounts to asserting that the magnetic pressure remains a constant fraction of the gas pressure at depths between the surface and $r_{\text{max}}$. In non-linear 3D simulations of turbulent stellar dynamos, the field strength is typically not directly related to the gas pressure at any given depth, but is set by the dynamics of the convection coupled to rotation and shear \citep[e.g.,][]{2006ApJ...638..336D,Browning:2008aa,2016GeoJI.204.1120Y}. But once this choice of $\delta$ profile is made, the model is specified fully by the choice of surface $\delta$ and by the value of $B_{\text{v-max}}$.

The total gas pressure increases rapidly with depth, so if no $B_{\text{v-max}}$ is specified, the magnetic field strengths implied by a $\delta=const$ profile would quickly become enormous. Some of the first studies along these lines, for example, allowed for fields of sufficient strength that the formation of a radiative core would result \citep[e.g.,][]{Mullan:2001ab}. Some later models adopted a maximum field strength of order $1 \, \text{MG}$, \citep{MacDonald:2012aa,Mullan:2015aa}. Recently, \citet{Browning:2016aa} suggested $B_{\text{v-max}} \sim 10^5 \, \text{G}$ to be an extreme upper limit for the maximum field strengths found in these fully-convective low-mass stars. They argue that at a given magnetic field strength, large-scale field configurations are subject to the constraints of magnetic buoyancy instabilities, whilst Ohmic dissipation associated with small-scale field configurations was enough to exceed the stellar luminosity in some cases. Combining these constraints produced an upper limit on the maximum field strength of $B_{\text{v-max}} \leq 800 \, \text{kG}$, for models of particularly simple magnetic field spatial structure. Additional, stronger constraints come again from 3D simulations of dynamo action in these objects. For example, \citet{2015ApJ...813L..31Y} found $B_{\text{v-max}} \approx 14 \, \text{kG}$ for a fully-convective M dwarf with a rotation period of 20 days, and likewise the simulations of \citet{Browning:2008aa} found fields of order the equipartition strength (with the turbulent convective energy density). Broadly, we think models in which the field does not greatly exceed values of order $10^4 \, \text{G}$ are most realistic (as also studied recently, for example, by \citealt{MacDonald:2017ab}). Note that as $B_{\text{v-max}}$ approaches the value of the surface field, the profile assumed for $\delta$ becomes increasingly irrelevant; in that limit, the field strength throughout the interior is just the constant $B_{\text{v-max}} \approx B_{\text{surf}}$.  

Motivated by these considerations, we test $B_{\text{v-max}} = 10^3 - 10^5 \, \text{G}$ ($\Delta 1 \, \log_{10}{(\text{G})}$) at both ages. Note that we include $10^5 \, \text{G}$ for comparison with prior work and to demonstrate the utility of our mechanism even in the extreme field cases, even though we think, as noted above, that $10^4 \, \text{G}$ is a reasonable upper limit. We use $\delta = 0.01-0.03$ ($\Delta 0.005$) for our $10 \, \text{Myr}$ models, giving $B_{\text{v-surf}} \lesssim 0.3 \, \text{kG}$. We use an extended range of $\delta = 0.01 - 0.06$ ($\Delta 0.005$) for our $1 \, \text{Gyr}$ models, to counteract the suppression of radius inflation in main-sequence models, producing $B_{\text{v-surf}} \lesssim 0.9 \, \text{kG}$.

In Figure~\ref{fig:logsupergrad_v_logRho_mag}, we plot $\log_{10}{(\nabla_{\text{s}})}$ as a function of $\log_{10}{(\rho)}$ for $0.3 \, \text{M}_\odot$, $\alpha_{\text{MLT}} = 1.7$ stellar models, for some combinations of $B_{\text{v-max}} = 10^3-10^5 \, \text{G}$, with $\delta = 0.01 - 0.03$ for $10 \, \text{Myr}$ models and $\delta = 0.04 - 0.06$ for $1 \, \text{Gyr}$ models, which we compare with the unperturbed model. In accord with equation~(\ref{eq:schwarz_crit_mag}), $\nabla_{\text{s}} \simeq \nabla_{\text{s}_0} + \delta / Q_0$ at all depths. Changes in $Q$ are negligible between models, hence we used the unperturbed value. In the bulk of the convection zone, where $B_{\text{v}} = B_{\text{v-max}}$, $\nabla_{\text{s}} \sim B_{\text{v-max}}^2 / Q_0 \gamma P_{\text{gas}} \gg \nabla_{\text{s}_0}$, thus a factor of ten increase in $B_{\text{v-max}}$ results in a factor of $\sim 100$ increase in superadiabaticity. As $\delta$ increases, $\nabla_{\text{s}}$ increases in the surface layers. The point at which $\nabla_s$ transitions---from a nearly-constant value near the surface to a steeply declining profile in the interior---is mediated by the point at which the vertical surface magnetic field (here set by $\delta$) reaches $B_{\text{v-max}}$, because interior to that point the gas pressure begins to exceed the magnetic pressure by an increasingly large amount.  

\begin{figure*}
\gridline{\fig{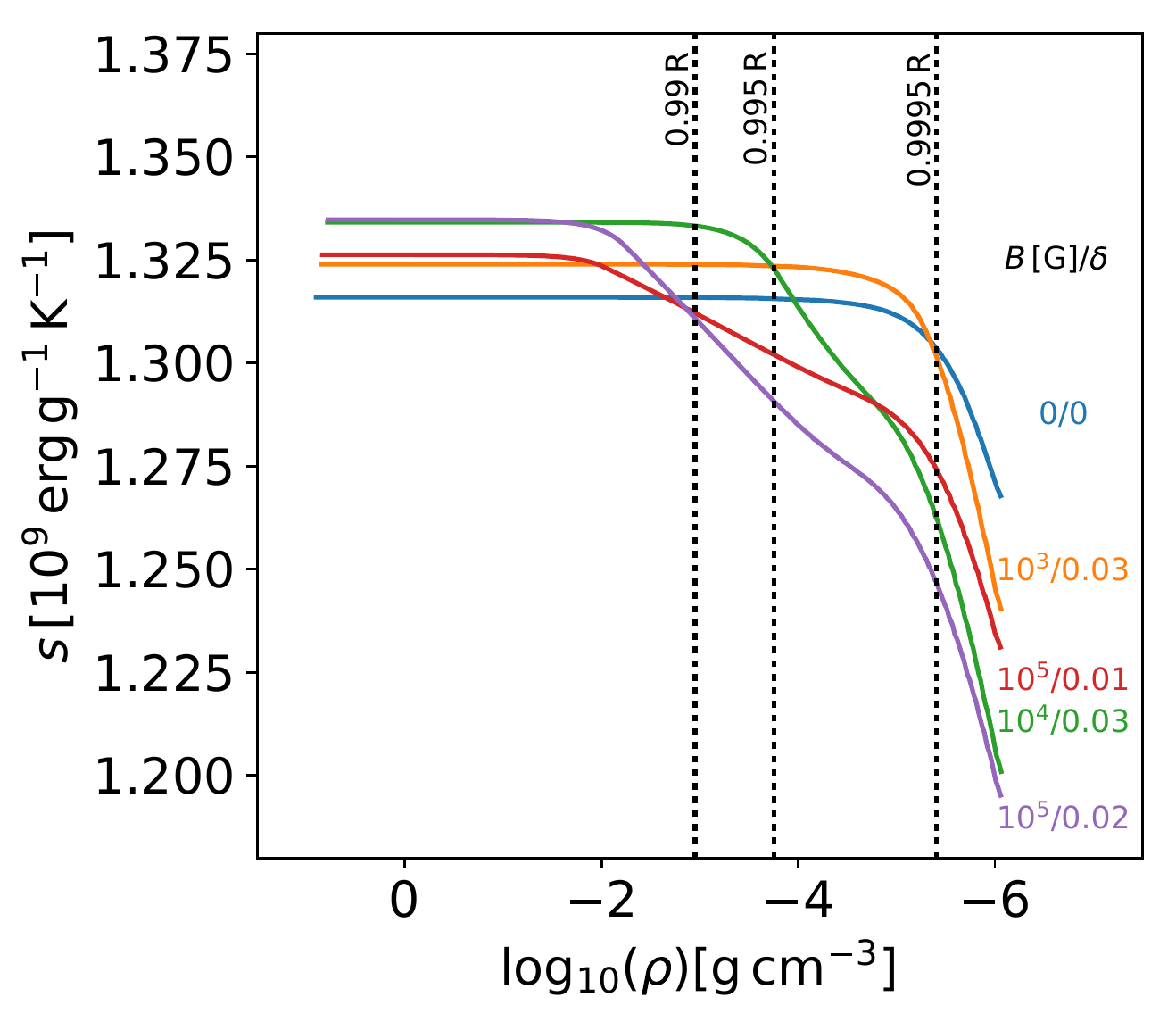}{0.515\textwidth}{(a)}
          \fig{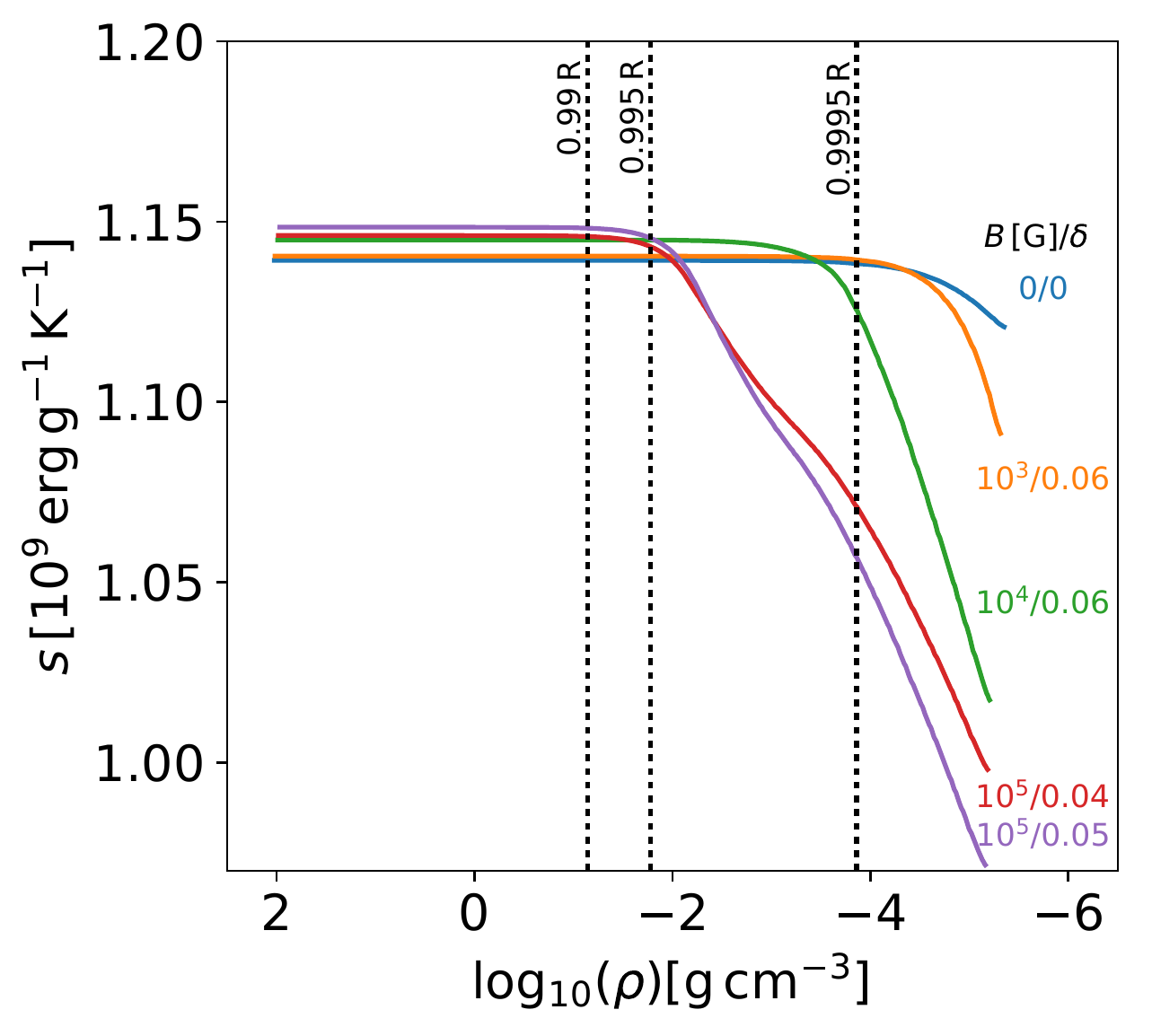}{0.5\textwidth}{(b)}
          }
\caption{$s$ as a function of $\log_{10}{(\rho)}$, for $0.3 \, \text{M}_\odot$, (a) $10 \, \text{Myr}$ (b) 1 Gyr, $\alpha_{\text{MLT}} = 1.7$ stellar models at some combinations of $B_{\text{v-max}} = 10^3-10^5 \, \text{G}$ and (a) $\delta = 0.01 - 0.03$ (b) $\delta = 0.04 - 0.06$, including the unperturbed model.\label{fig:entropy_v_logRho_mag}}
\end{figure*}

\begin{figure*}
\gridline{\fig{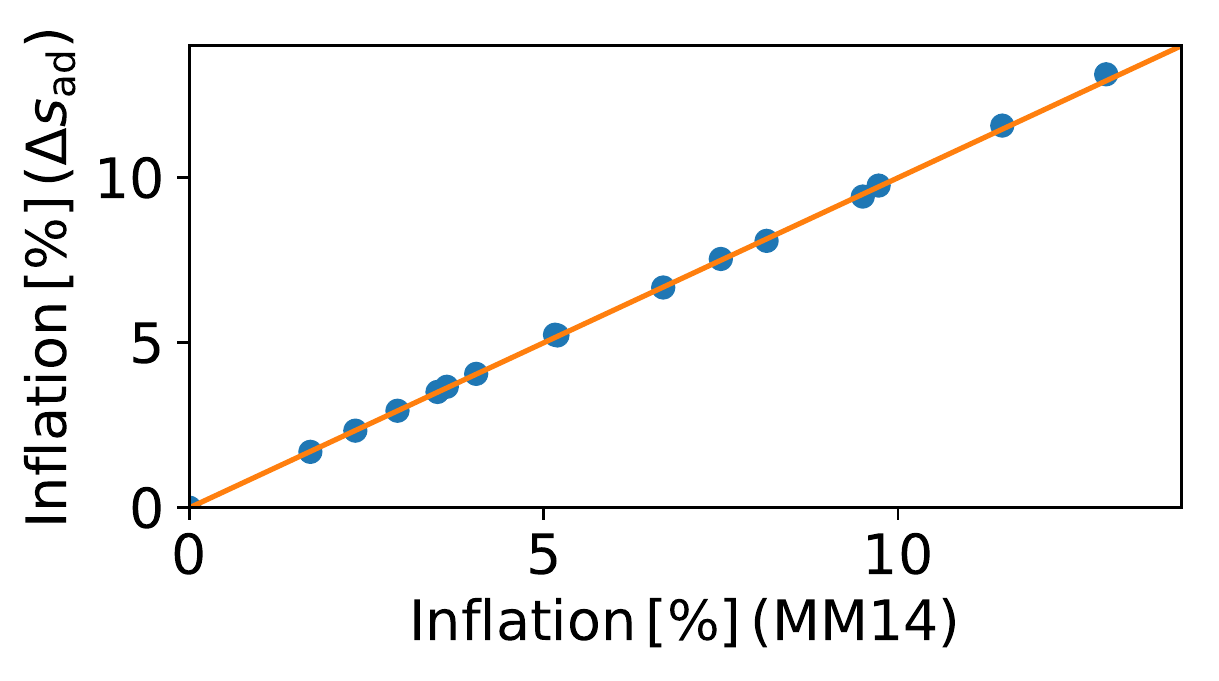}{0.515\textwidth}{(a)}
		  \fig{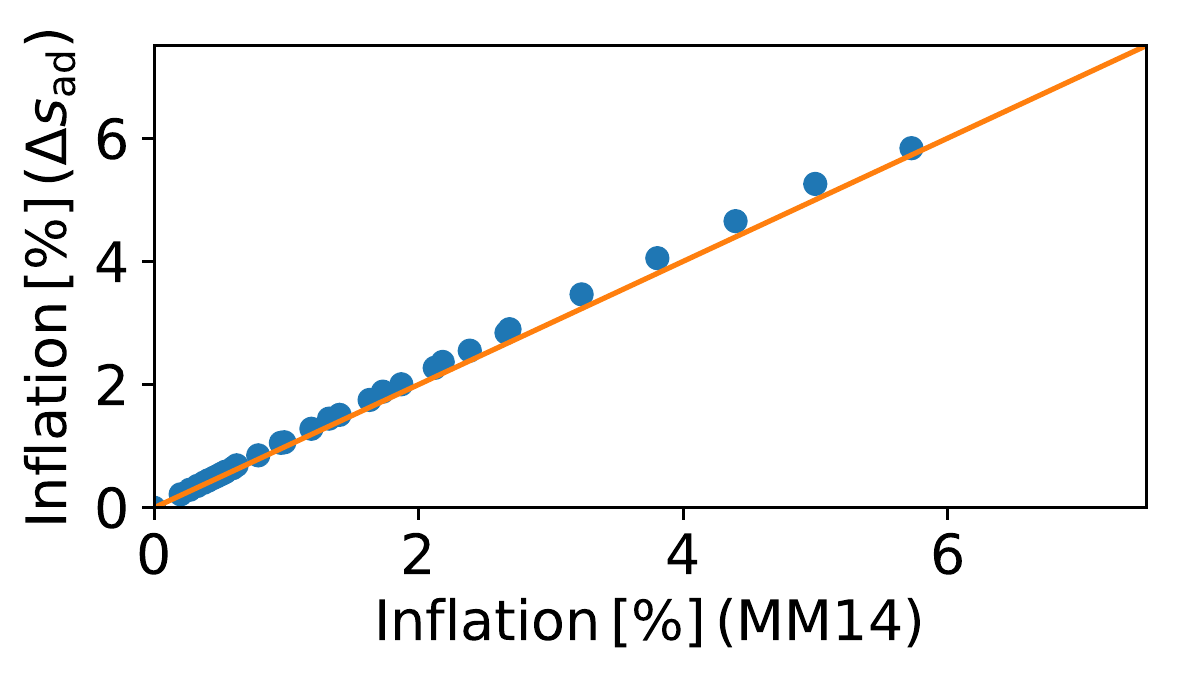}{0.5\textwidth}{(b)}
        }
\caption{Radius inflation determined from $\Delta s_{\text{ad}}$ via equation~(\ref{eq:radii_entropy_relation}) as a function of the outputted radius inflation from the \citetalias{MacDonald:2014aa} models for $0.3 \, \text{M}_\odot$, (a) $10 \, \text{Myr}$ (b) $1 \, \text{Gyr}$, $\alpha_{\text{MLT}} = 1.7$ stellar models at all combinations of ${B_{\text{v-max}}} = 10^3-10^5 \, {\text{G}}$ ($\Delta 1 \, \log_{10}{(\text{G})}$) and (a) $\delta = 0.01 - 0.03$ ($\Delta 0.005$) (b) $\delta = 0.01 - 0.06$ ($\Delta 0.005$). $y=x$ (orange) is plotted for ease of comparison.\label{fig:R2_v_R2_s_ad}}
\end{figure*}

In Figure~\ref{fig:entropy_v_logRho_mag}, we plot $s$ as a function of $\log_{10}{(\rho)}$ for the same stellar models. At fixed $\delta$, the photospheric entropy $s_{\text{ph}}$ decreases monotonically with increasing $B_{\text{v-max}}$; likewise at fixed $B_{\text{v-max}}$, increasing $\delta$ decreases $s_{\text{ph}}$. In turn, $s_{\text{ad}}$ is shown to increase strongly with $\delta$, and to a lesser extent $B_{\text{v-max}}$. Pre-main-sequence stars with lower $s_{\text{ph}}$ have higher $s_{\text{ad}}$ for the reasons discussed in \S~\ref{sec:entropy}; hence, stars with higher $B_{\text{v-max}}$ and $\delta$ tend to have a higher $s_{\text{ad}}$. On the main-sequence, variations in $s_{\text{ad}}$ are smaller, due to the self-regulation of the star through nuclear fusion. However, the differences in $s_{\text{ph}}$ induced by changes in $\delta$ or $B_{\text{v-max}}$ are larger than in our fixed-$\alpha_{\text{MLT}}$ models. A larger entropy contrast, as a result of higher superadiabaticity in the surface layers, produces small but noticeable changes in $s_{\text{ad}}$. As in \S~\ref{sec:entropy} and \S~\ref{sec:rot_inhibition_conv}, stellar structure is largely insensitive to the increasing $\nabla_{\text{s}}$ in the deep interior; it responds more readily to an increased $\nabla_{\text{s}}$ in the surface layers. 

\begin{figure*}
\gridline{\fig{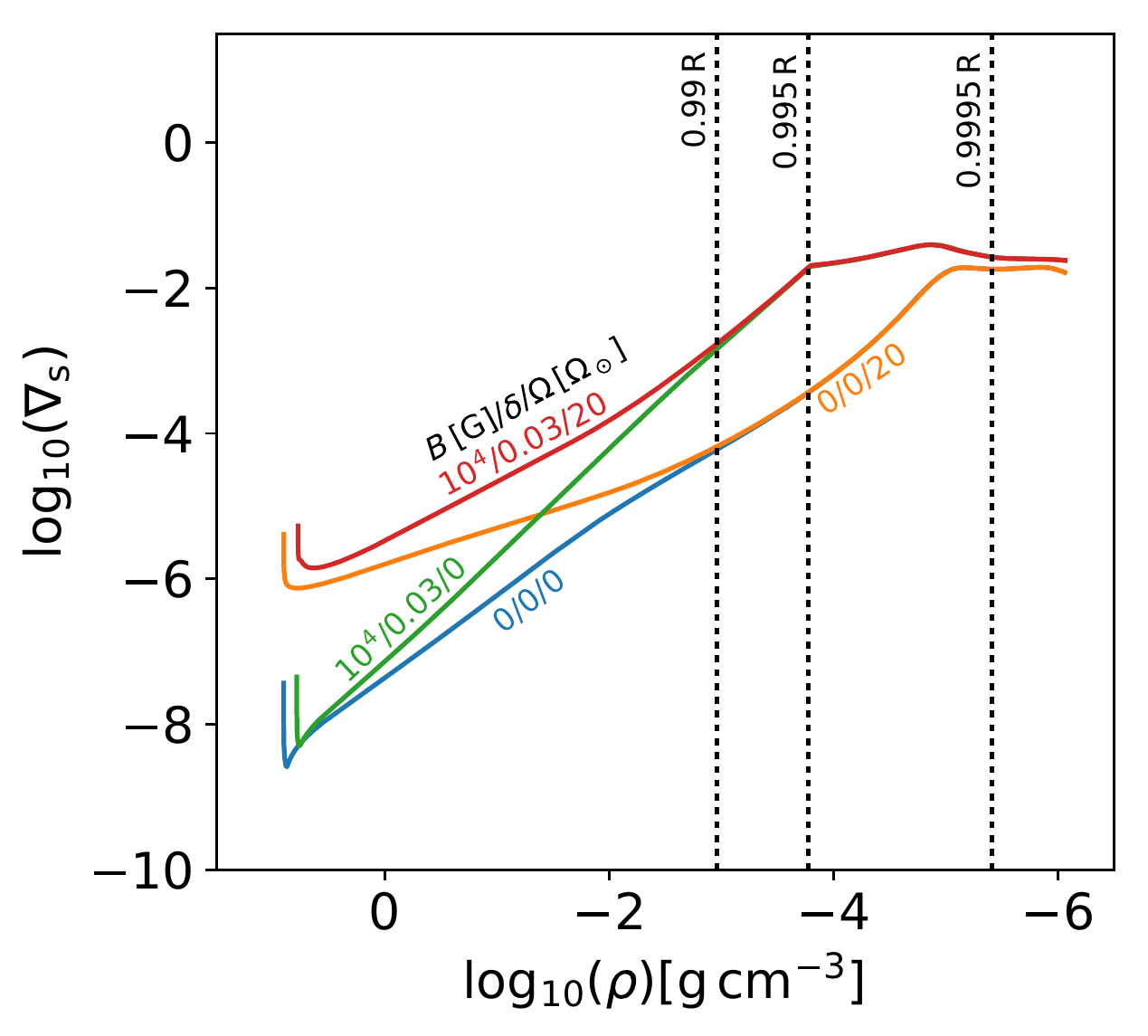}{0.5\textwidth}{(a)}
          \fig{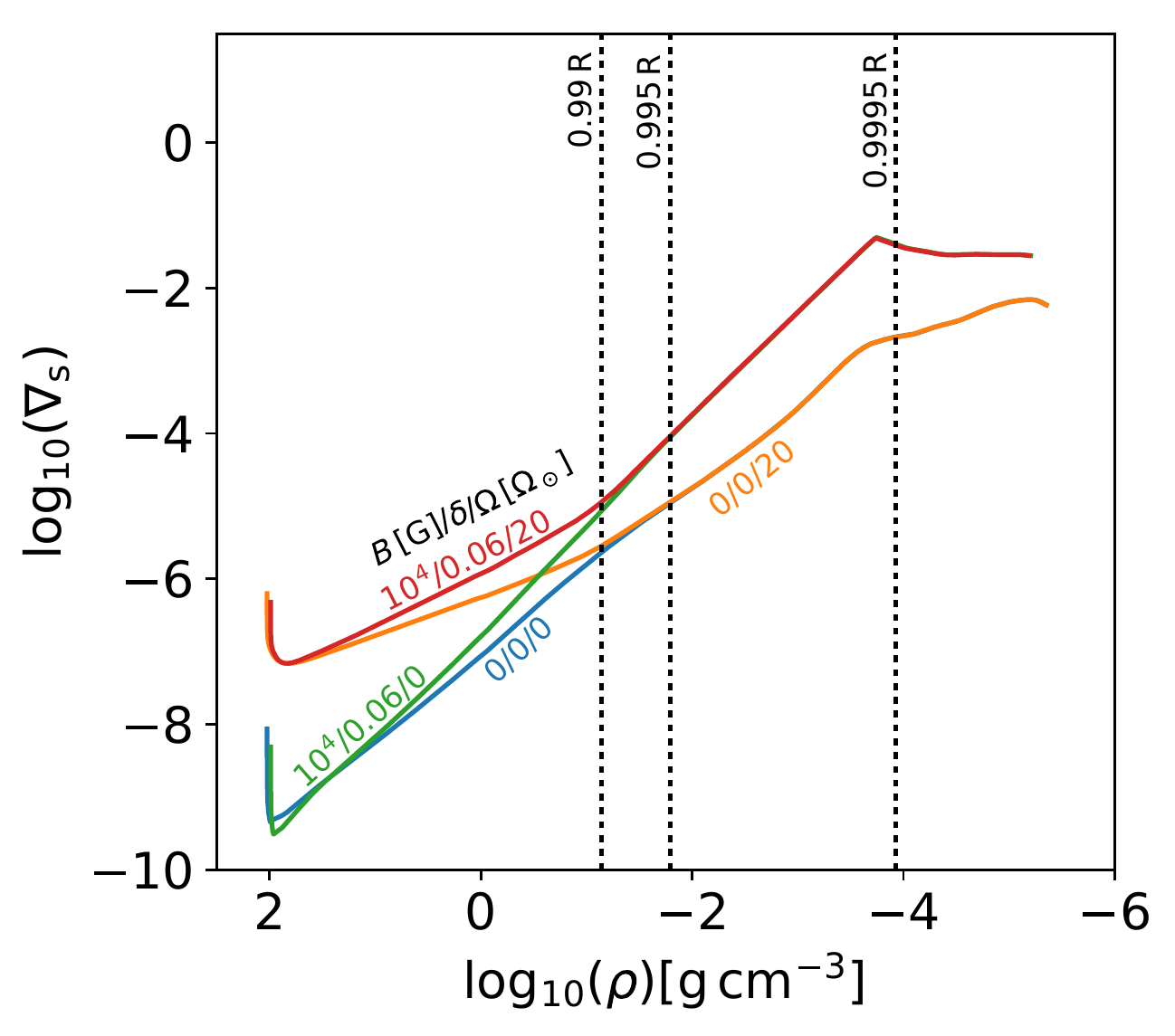}{0.51\textwidth}{(b)}
          }
\caption{$\log_{10}{(\nabla_{\text{s}})}$ as a function of $\log_{10}{(\rho)}$, for a $0.3 \, \text{M}_\odot$, (a) $10 \, \text{Myr}$ (b) $1 \, \text{Gyr}$, $\alpha_{\text{MLT}} = 1.7$ stellar model at $\Omega = 20 \, \Omega_\odot$, $B_{\text{v-max}} = 10^4 \, \text{G}$, and (a) $\delta = 0.03$ (b) $\delta = 0.06$, including the rotating-only, magnetic-only, and unperturbed models.\label{fig:logsupergrad_v_logRho_rot_mag}}
\end{figure*}

In Figure~\ref{fig:R2_v_R2_s_ad}, we examine radius inflation calculated via $\Delta s_{\text{ad}}$ (equation~(\ref{eq:radii_entropy_relation})) as a function of the outputted radius inflation, showing good agreement for these \citetalias{MacDonald:2014aa} models. For $\delta = 0.01 - 0.03$ models, we find $\Delta R / R_0 \lesssim 13 \%$ ($R \lesssim 0.771 \, \text{R}_\odot$) for our range of perturbed models at $10 \, \text{Myr}$, and $\Delta R / R_0 \lesssim 2 \%$ ($R \lesssim 0.292 \, \text{R}_\odot$) at $1 \, \text{Gyr}$. For $\delta = 0.04 - 0.06$ models at $1 \, \text{Gyr}$, we find $\Delta R / R_0 \lesssim 6 \%$ ($R \lesssim 0.302 \, \text{R}_\odot$). Overall, we find greater changes in $s_{\text{ph}}$ in these models than in the fixed-$\alpha_{\text{MLT}}$ main-sequence models in \S~\ref{subsec:radius_s_ad}, which is enough to slightly perturb $s_{\text{ad}}$ from the value predominantly determined via nuclear fusion, producing small, yet noticeable radius inflation. There is a slight divergence for our most-inhibited fully-convective models, due to the increasing effective depth of the magnetic inhibition of convection. For those models, the asymptotic increase towards $s_{\text{ad}}$ is reached at ever-increasing depth, thus our approximation $S_{\text{tot}} \simeq s_{\text{ad}} M$ becomes increasingly less accurate. Therefore, with increasing levels of radius inflation, the accuracy of using $s_{\text{ad}}$ alone to determine the stellar radius decreases.

\section{Combined inhibition of convection by rotation and magnetism}\label{sec:rot_and_mag}

Both the \citetalias{Stevenson:1979aa} rotational and \citetalias{MacDonald:2014aa} magnetic reformulations of MLT  modify the superadiabaticity of a model. In the ``magnetic" case, the superadiabaticity in the surface layers is noticeably increased between 0.99-0.995 R, and slightly increased from 0.995 R up to the photosphere (see Figure~\ref{fig:logsupergrad_v_logRho_mag}). In the ``rotating" case, there is a small difference in $\nabla_{\text{s}}$ in the 0.99-0.995 R region, but negligible difference after this point up towards the photosphere (see Figure~\ref{fig:logsupergrad_v_logRho_rot}). Here, we briefly examine whether the \emph{combination} of rotation and magnetism using these prescriptions could increase the radius of a model even further.  To do so, we first modify the criterion for convection using the \citetalias{MacDonald:2014aa} formulation, as in \S~\ref{sec:mag_inhibition_conv}; the resulting model is then used as the ``unperturbed" model for an application of the rotational formulation described in \S~\ref{sec:rot_inhibition_conv}. Hence, the enhanced superadiabaticity near the surface in the magnetic models may be further increased by the rotation, with possible impacts on the structure. Of course, this is a very crude approximation; as noted in \S~\ref{sec:intro}, the combined effects of rotation and magnetism may be considerably more complex than either simply rotation or magnetism acting alone, and these effects may not be additive (and indeed, in the case of the linear onset of convection, are not). Nonetheless we adopt it here as a first attempt at the problem.

In Figure~\ref{fig:logsupergrad_v_logRho_rot_mag}, we plot $\log_{10}{(\nabla_{\text{s}})}$ as a function of $\log_{10}{(\rho)}$, comparing a ``magnetic rotating" $0.3 \, \text{M}_\odot$, $\alpha_{\text{MLT}} = 1.7$ stellar model at $\Omega = 20 \, \Omega_\odot$, $B_{\text{v-max}} = 10^4 \, \text{G}$, with $\delta = 0.03$ for $10 \, \text{Myr}$ and $\delta = 0.06$ for $1 \, \text{Gyr}$, compared with the rotating-only case, the magnetic-only case, and the unperturbed model. We choose the most-perturbed model at each age to be $10^4 \, \text{G}$ in order to investigate the highest possible radius inflation attained by the addition of ``rotational" effects at what we think is a realistic maximum field strength. At both ages, the superadiabaticity of our ``magnetic rotating" model is higher than in the magnetic-only case by orders of magnitude within the deep convection zone, where convective velocities are low (i.e. $\text{Ro} \ll 1$).  Closer to the surface, this difference diminishes (because $\text{Ro}$ increases there).

\begin{figure}
\includegraphics[width=0.5\textwidth]{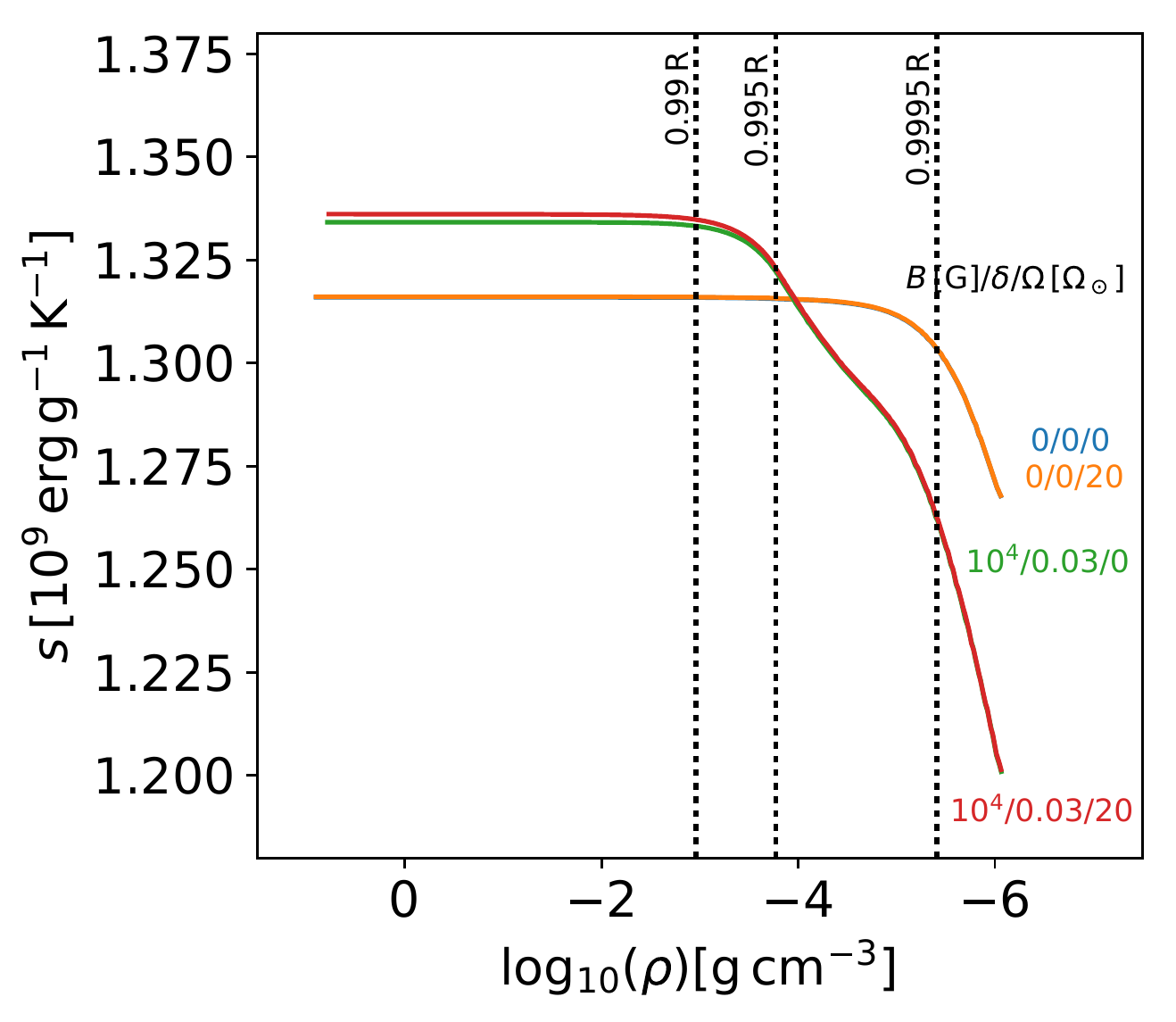}
\caption{$s$ as a function of $\log_{10}{(\rho)}$, for a $0.3 \, \text{M}_\odot$, $10 \, \text{Myr}$, $\alpha_{\text{MLT}} = 1.7$ stellar model at $\Omega = 20 \, \Omega_\odot$, $B_{\text{v-max}} = 10^4 \, \text{G}$, and $\delta = 0.03$, including the rotating-only, magnetic-only, and unperturbed models. The rotating-only case is near-identical to the unperturbed model. \label{fig:entropy_v_logRho_rot_mag}}
\end{figure}

We plot $s$ as a function of $\log_{10}{(\rho)}$ in Figure~\ref{fig:entropy_v_logRho_rot_mag} for our $10 \, \text{Myr}$ model. These changes in superadiabaticity are enough to produce a small change in $s_{\text{ad}}$ for our pre-MS model. As a result of this, our $10 \, \text{Myr}$ ``magnetic rotating" model is inflated by a further $1 \%$ compared to the magnetic-only case, giving $\Delta R / R_0 \simeq 10.5 \%$ ($R \simeq 0.755 \, \text{R}_\odot$). However, for our $1 \, \text{Gyr}$ model, there is negligible inflation, as the superadiabaticity is much lower throughout the surface layers compared to the pre-MS model, giving negligible changes in $s_{\text{ad}}$. These results suggest that the combination of rotation and magnetism may indeed further inflate the stellar radius, but the additional effect arising from rotation is only noticeable in the youngest models.

\section{Depth-dependent $\alpha_{\text{MLT}}$ as MLT proxies for rotation and magnetic fields}\label{sec:depth_dep_alpha}

The structure of a 1D stellar model constructed with a modified version of MLT, like the rotationally or magnetically-constrained versions described in \S~\ref{sec:rot_inhibition_conv} and \S~\ref{sec:mag_inhibition_conv}, cannot generally be duplicated by a model with a standard depth-independent $\alpha_{\text{MLT}}$. The reason for this is straightforward: in the standard 1D models, $\nabla_{\text{s}}$ throughout the stellar interior increases with decreasing $\alpha_{\text{MLT}}$, whereas for the \citetalias{Stevenson:1979aa} and \citetalias{MacDonald:2014aa} models the inhibition of convection depends on parameters that vary with depth---i.e., $\text{Ro}$ in the ``rotating" case and $\delta$ in the ``magnetic" case. It is not possible to mimic these effects with a standard depth-independent $\alpha_{\text{MLT}}$, no matter its value. They can, however, be captured by models that include a depth-\emph{dependent} $\alpha_{\text{MLT}}$ ($\alpha_{\text{MLT}}(r)$, hereafter), as described in this section.

Here, we provide explicit formulae linking a $\alpha_{\text{MLT}}(r)$ profile to the rotationally- and magnetically-inhibited convection formulae of \citetalias{Stevenson:1979aa} and \citetalias{MacDonald:2014aa} respectively. Our motivation for constructing such profiles is just that, in a given 1D stellar evolution code, it may be much more straightforward to input (or implement) a $\alpha_{\text{MLT}}(r)$ profile than to modify the whole underlying MLT formulation. Knowledge of the precise correspondence between $\alpha_{\text{MLT}}(r)$ and a particular depth-dependent theory of convective inhibition---arising from rotation, magnetism, or other effects---gives us the ability to model the non-standard 1D stellar structures arising from these effects without undue difficulty. 

Models constructed with modified MLT formulations of the type and magnitude considered here can be regarded as perturbations at each depth to a fiducial unperturbed model. We write the perturbed model's $\nabla_{\text{s}}$ as the unperturbed model's plus a given depth-dependent perturbation $\beta$:

\begin{equation} \label{eq:nabla_perturbation}
{\nabla_{\text{s}} = \nabla_{\text{s}_0} + \beta}.
\end{equation}
Thus any perturbation made to the superadiabaticity results in the modification of $ds/dr \propto \nabla_{\text{s}}$, implying that the specific entropy will asymptotically converge to a different adiabat.

In \S~\ref{subsec:role_entropy}, we found that a perturbed model's $\nabla_{\text{s}}$ could be reproduced using the unperturbed model's and each model's $\alpha_{\text{MLT}}$, i.e., equation~(\ref{eq:nabla_alpha_efficient_full}). By substituting equation~(\ref{eq:nabla_alpha_efficient_full}) into equation~(\ref{eq:nabla_perturbation}), we find an approximate expression for $\alpha_{\text{MLT}}(r)$ as a function of the unperturbed model's depth-\emph{independent} $\alpha_{\text{MLT}}$ and $\nabla_{\text{s}}$, and the perturbation $\beta$:

\begin{equation} \label{eq:depth_dep_alpha_general}
\alpha_{\text{MLT}}(r) \simeq \frac{\alpha_{\text{MLT}_0}}{\left(1 + \frac{\beta}{\nabla_{\text{s}_0}}\right)^{3/4}}.
\end{equation}

\begin{figure*}
\gridline{\fig{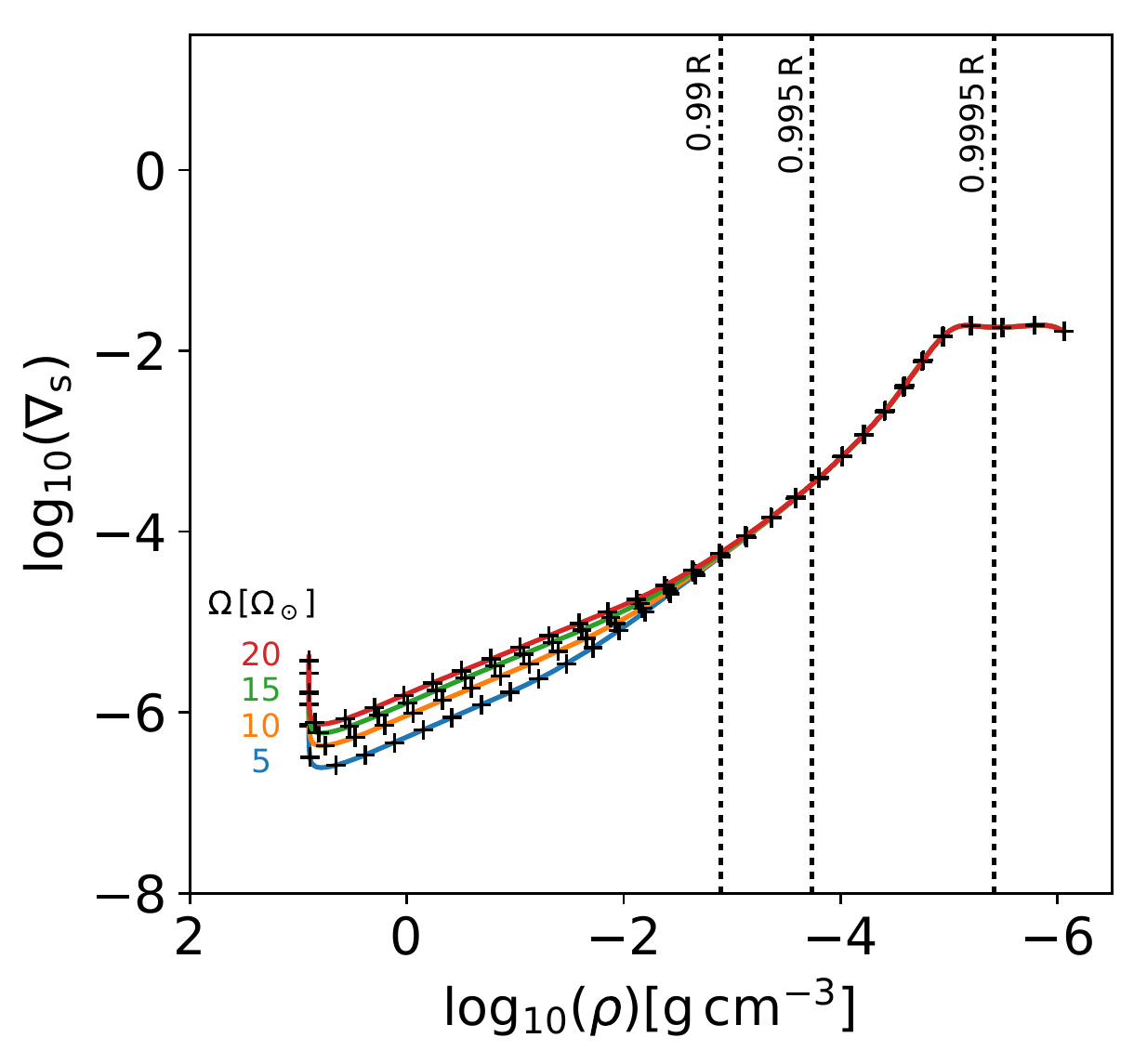}{0.5\textwidth}{(a)}
          \fig{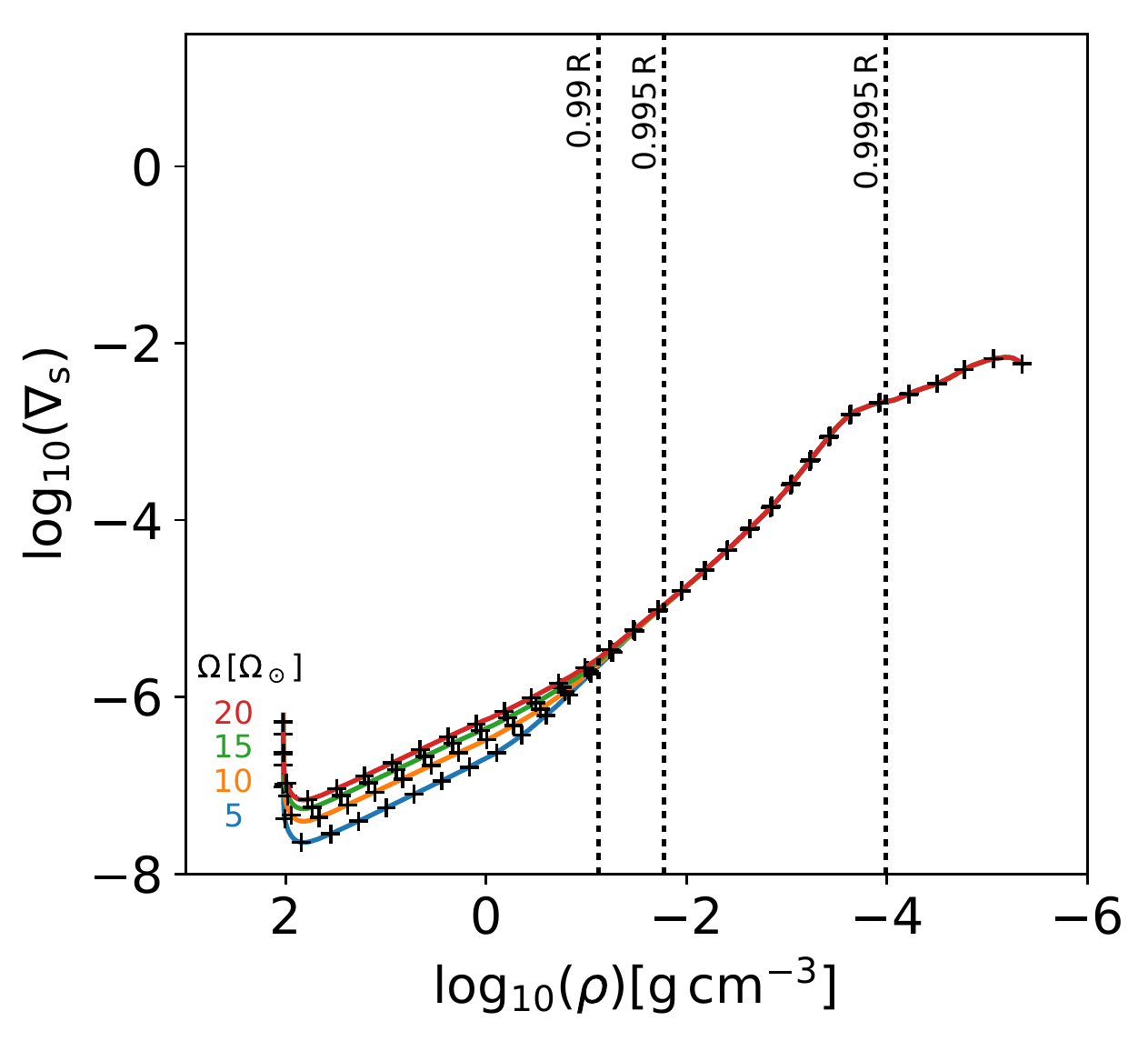}{0.51\textwidth}{(b)}
          }
\caption{$\log_{10}{(\nabla_{\text{s}})}$ as a function of $\log_{10}{(\rho)}$, comparing \citetalias{Stevenson:1979aa} ``rotating" models and our $\alpha_{\text{MLT}}(r)$ models (plus markers), for $0.3 \, \text{M}_\odot$, (a) $10 \, \text{Myr}$ (b) $1 \, \text{Gyr}$, $\alpha_{\text{MLT}} = 1.7$ stellar models at $\Omega = 5-20 \, \Omega_\odot$ ($\Delta 5 \, \Omega_\odot$).\label{fig:logsupergrad_v_logRho_depth_dep_rot}}
\end{figure*}

We find that a $\alpha_{\text{MLT}}(r)$ profile constructed using this expression allows us to reproduce virtually all of the radial variation of $\nabla_{\text{s}}$ in both our $10 \, \text{Myr}$ and $1 \, \text{Gyr}$ non-standard ``rotating" and ``magnetic" stellar structure models. First, consider  the case of the \citetalias{Stevenson:1979aa} ``rotating" MLT formulation. We express equation~({\ref{eq:general_rotation}}) in terms of $\alpha_{\text{MLT}}(r)$ and the unperturbed depth-independent $\alpha_{\text{MLT}}$ using equation~(\ref{eq:nabla_alpha_efficient}):

\begin{equation} \label{eq:general_rotation_alpha}
	\left(\frac{\alpha_{\text{MLT}}(r)}{\alpha_{\text{MLT}_0}}\right)^{-10/3} - \left(\frac{\alpha_{\text{MLT}}(r)}{\alpha_{\text{MLT}_0}}\right)^{-4/3} \simeq \frac{4}{41} \tau_{\text{c}_0}^2 \Omega^2.
\end{equation}
Therefore, in the case of the \citetalias{Stevenson:1979aa} models, the depth-dependent perturbation can be expressed as
\begin{equation} \label{eq:beta_rotation}
	\beta \simeq \nabla_{\text{s}_0} \left[\left(\frac{\alpha_{\text{MLT}}(r)}{\alpha_{\text{MLT}_0}}\right)^{-10/3} - \frac{4}{41} \tau_{\text{c}_0}^2 \Omega^2 - 1\right],
\end{equation}
giving

\begin{equation} \label{eq:depth_dep_alpha_rot}
\alpha_{\text{MLT}}(r) \simeq \frac{\alpha_{\text{MLT}_0}}{\left[\left(\frac{\alpha_{\text{MLT}}(r)}{\alpha_{\text{MLT}_0}}\right)^{-10/3} - \frac{4}{41} \tau_{\text{c}_0}^2 \Omega^2 \right]^{3/4}},
\end{equation}
which must be solved iteratively.

We can mimic the ``rotating" effects from the \citetalias{Stevenson:1979aa} MLT formulation in our 1D stellar structure models, solely using this $\alpha_{\text{MLT}}(r)$ profile. We modify MESA to input $\alpha_{\text{MLT}}(r)$ rather than the conventional fixed value and produce near-identical models to those produced using the \citetalias{Stevenson:1979aa} reformulation where we modified $\nabla_{\text{s}}$. To demonstrate this, in Figure~\ref{fig:logsupergrad_v_logRho_depth_dep_rot}, we plot $\log_{10}{(\nabla_{\text{s}})}$ as a function of $\log_{10}{(\rho)}$ for both our $\alpha_{\text{MLT}}(r)$ and \citetalias{Stevenson:1979aa} stellar models, at $10 \, \text{Myr}$ and $1 \, \text{Gyr}$; models constructed using the two techniques are indistinguishable here.

\begin{figure*}
\gridline{\fig{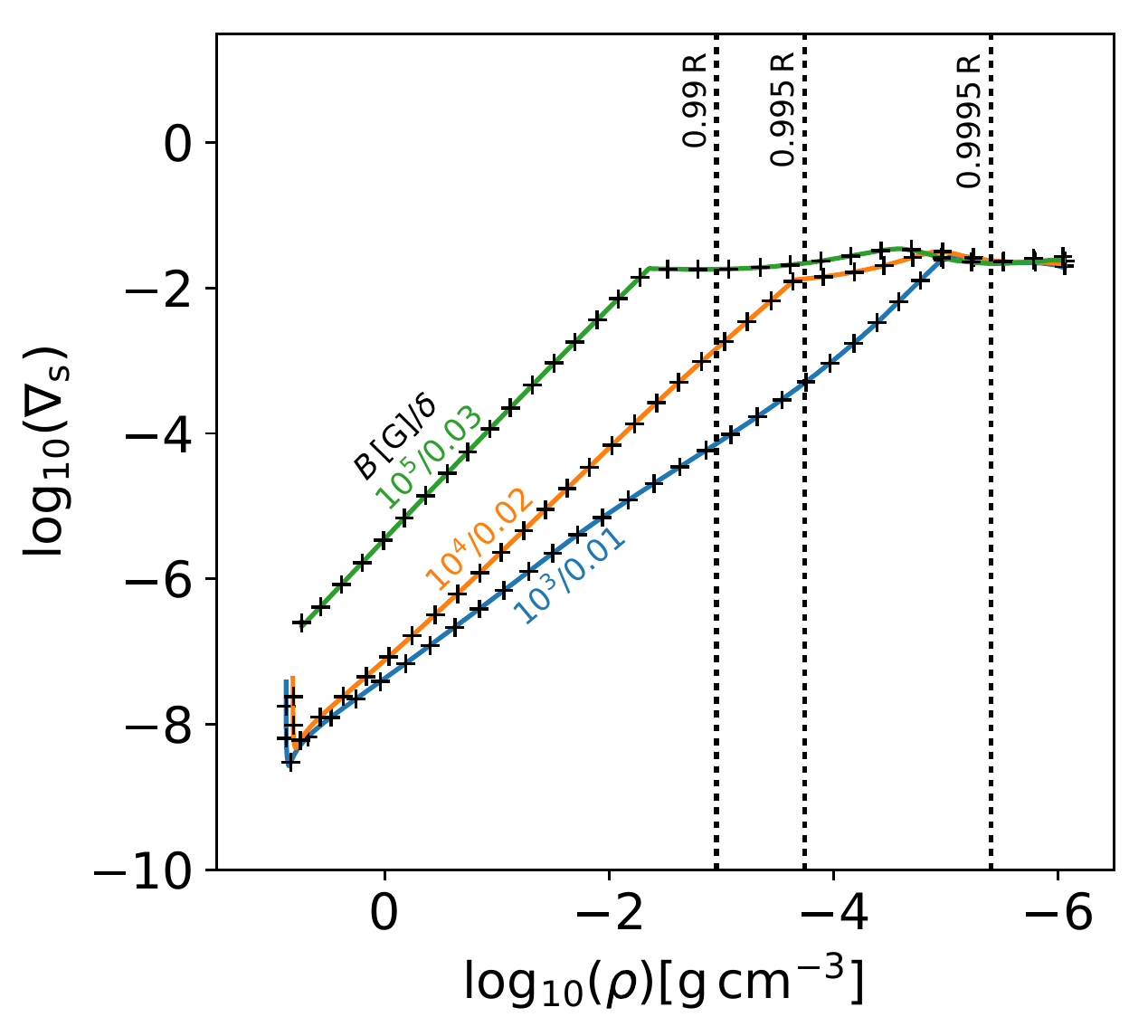}{0.5\textwidth}{(a)}
          \fig{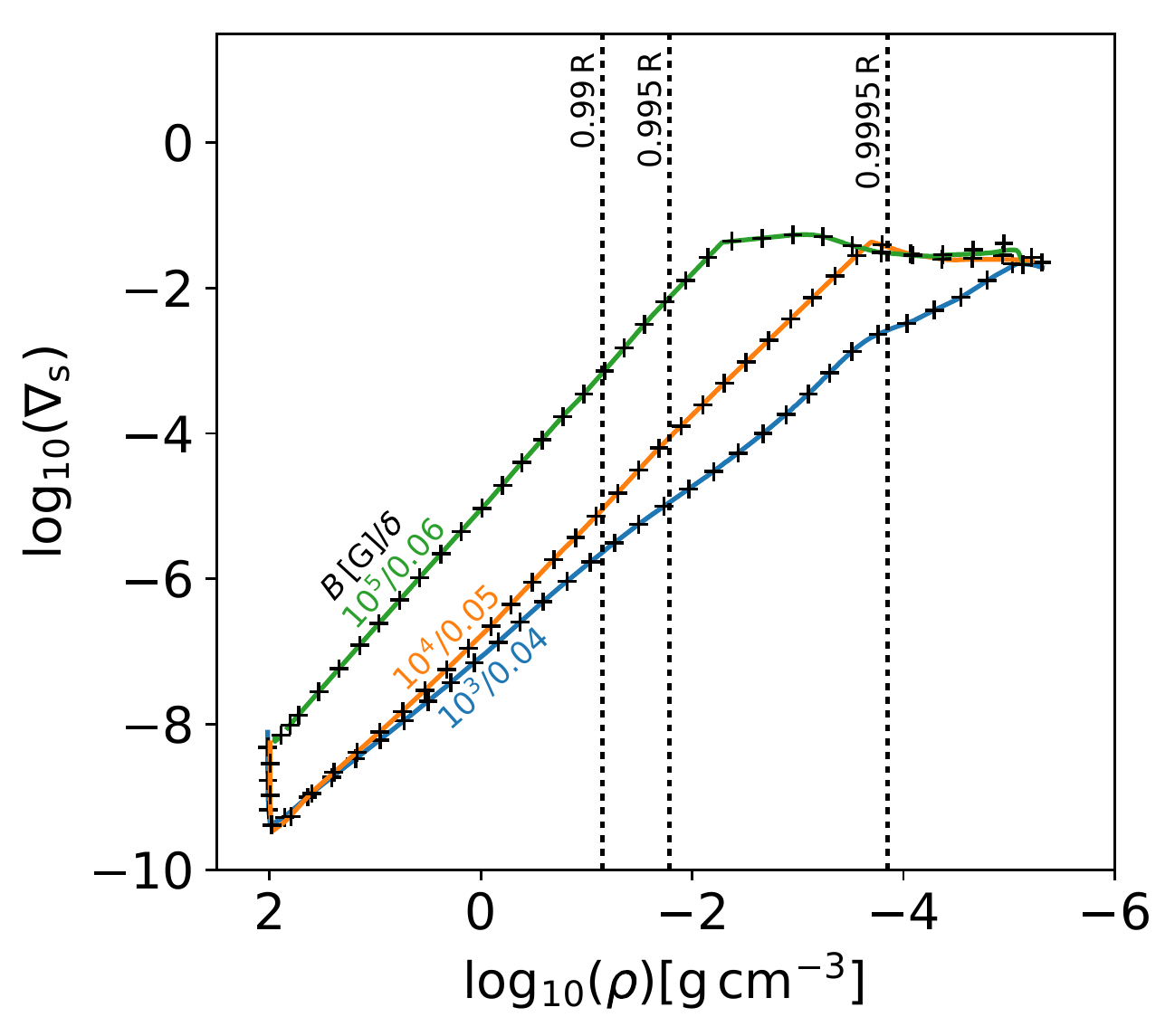}{0.51\textwidth}{(b)}
          }
\caption{$\log_{10}{(\nabla_{\text{s}})}$ as a function of $\log_{10}{(\rho)}$, comparing \citetalias{MacDonald:2014aa} ``magnetic" models and our $\alpha_{\text{MLT}}(r)$ models (plus markers), for $0.3 \, \text{M}_\odot$, (a) $10 \, \text{Myr}$ (b) $1 \, \text{Gyr}$, $\alpha_{\text{MLT}} = 1.7$ stellar models at some combinations of $B_{\text{v-max}} = 10^3-10^5 \, \text{G}$ and (a) $\delta = 0.01 - 0.03$ (b) $\delta = 0.04 - 0.06$.\label{fig:logsupergrad_v_logRho_depth_dep_mag}}
\end{figure*}

\begin{figure*}
\gridline{\fig{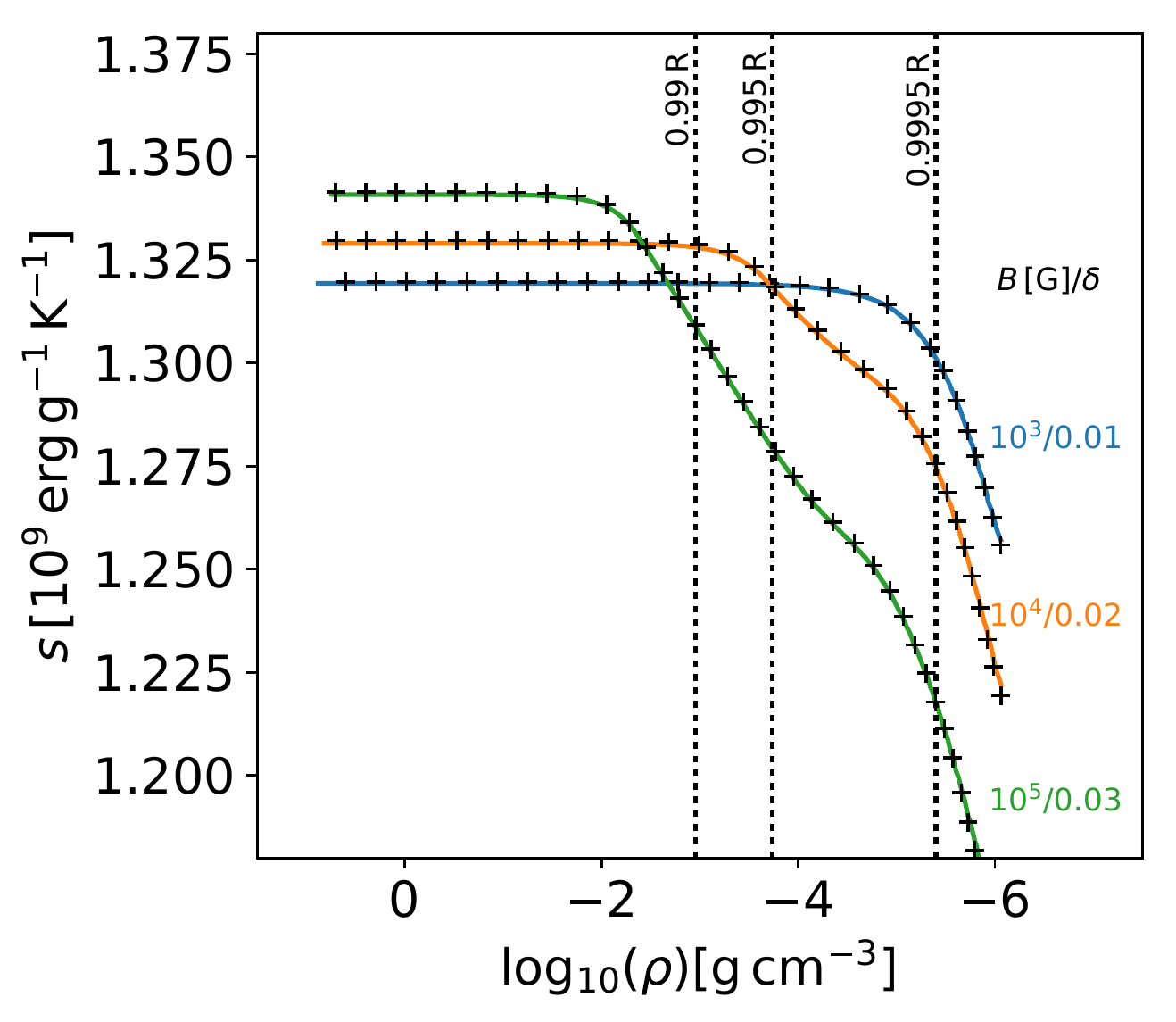}{0.51\textwidth}{(a)}
          \fig{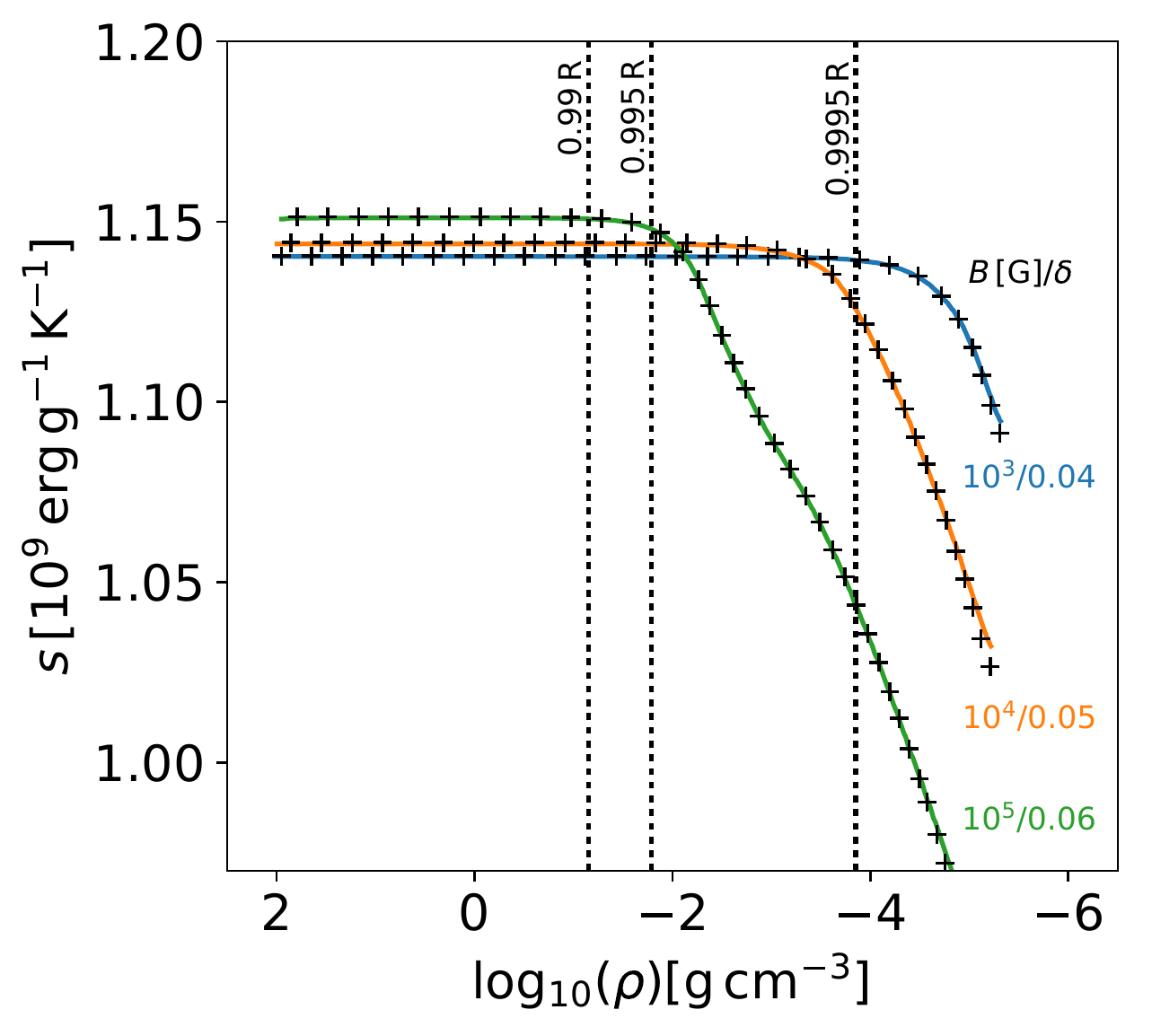}{0.5\textwidth}{(b)}
          }
\caption{$s$ as a function of $\log_{10}{(\rho)}$, comparing \citetalias{MacDonald:2014aa} ``magnetic" models and our $\alpha_{\text{MLT}}(r)$ models (plus markers), for $0.3 \, \text{M}_\odot$, (a) $10 \, \text{Myr}$ (b) $1 \, \text{Gyr}$, $\alpha_{\text{MLT}} = 1.7$ stellar models at some combinations of $B_{\text{v-max}} = 10^3-10^5 \, \text{G}$ and (a) $\delta = 0.01 - 0.03$ (b) $\delta = 0.04 - 0.06$. \label{fig:entropy_v_logRho_depth_dep_mag}}
\end{figure*}

We can apply the same technique to mimic the effects of ``magnetic" inhibition of convection via $\alpha_{\text{MLT}}(r)$. In the case of the \citetalias{MacDonald:2014aa} MLT formulation {in the high efficiency convective regime,} ${\beta \simeq \delta / Q_0}$, thus

\begin{equation} \label{eq:depth_dep_alpha_mag}
\alpha_{\text{MLT}}(r) \simeq \frac{\alpha_{\text{MLT}_0}}{\left(1 + \frac{\delta}{Q_0 \nabla_{\text{s}_0}}\right)^{3/4}}.
\end{equation}
We again input $\alpha_{\text{MLT}}(r)$ into MESA and reproduce near-identical models to those produced using the \citetalias{MacDonald:2014aa} reformulation. In Figures~\ref{fig:logsupergrad_v_logRho_depth_dep_mag} and~\ref{fig:entropy_v_logRho_depth_dep_mag}, we plot examples of $\log_{10}{(\nabla_{\text{s}})}$ and $s$ respectively as a function of $\log_{10}{(\rho)}$, produced by our $\alpha_{\text{MLT}}(r)$ models and our \citetalias{MacDonald:2014aa} models; excellent correspondence between the two model structures is evident.

\begin{figure*}
\gridline{\fig{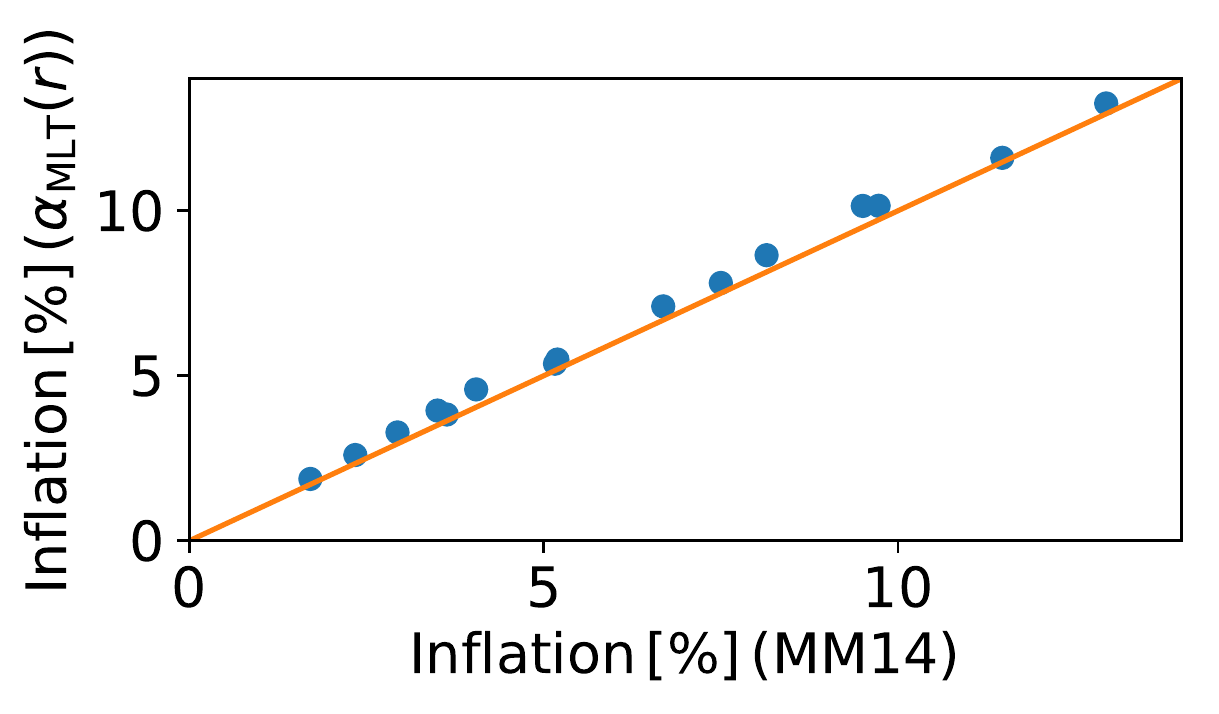}{0.51\textwidth}{(a)}
          \fig{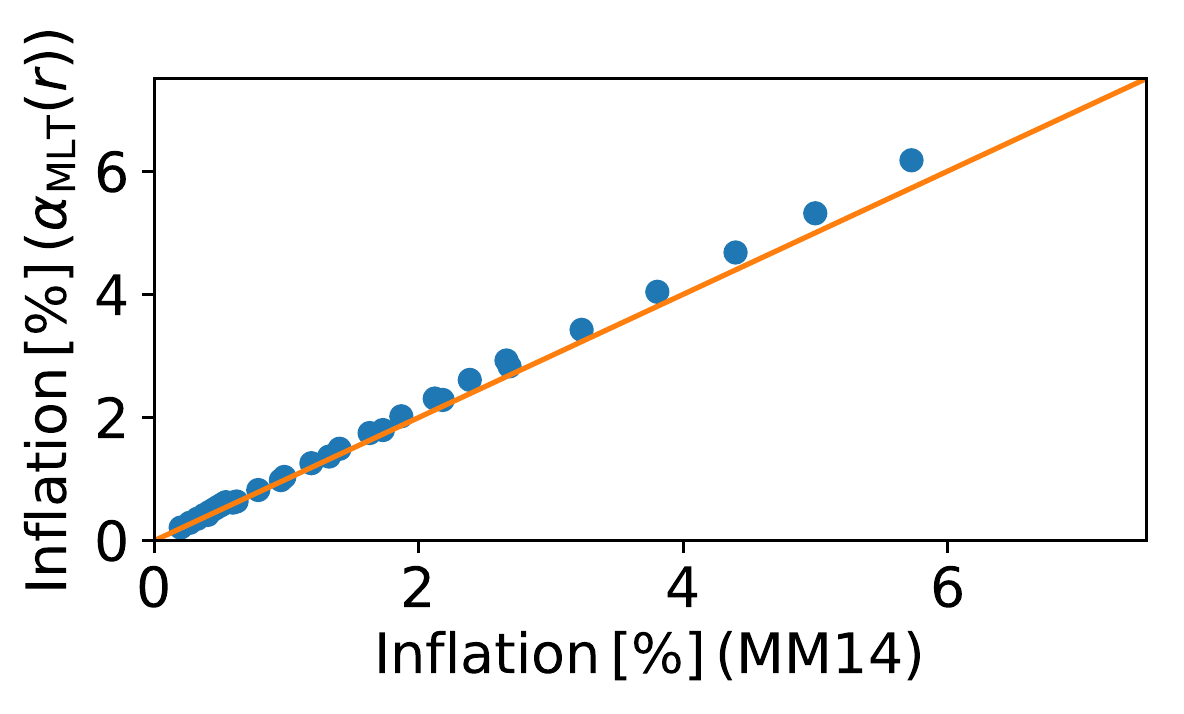}{0.5\textwidth}{(b)}
          }
\caption{Radius inflation from our $\alpha_{\text{MLT}}(r)$ models as a function of radius inflation from our \citetalias{MacDonald:2014aa} models, for $0.3 \, \text{M}_\odot$, (a) $10 \, \text{Myr}$ (b) $1 \, \text{Gyr}$, $\alpha_{\text{MLT}} = 1.7$ stellar models at all combinations of ${B_{\text{v-max}}} = 10^3-10^5 \, {\text{G}}$ ($\Delta 1 \, \log_{10}{(\text{G})}$) and (a) $\delta = 0.01 - 0.03$ ($\Delta 0.005$) (b) $\delta = 0.01 - 0.06$ ($\Delta 0.005$). $y=x$ (orange) is plotted for ease of comparison.\label{fig:R2_v_R2_depth_dep}}
\end{figure*}

In Figure~\ref{fig:R2_v_R2_depth_dep}, we examine radius inflation from our $\alpha_{\text{MLT}}(r)$ models as a function of the radius inflation from our \citetalias{MacDonald:2014aa} ``magnetic" models, at both $10 \, \text{Myr}$ and $1 \, \text{Gyr}$. They are in good agreement, with small divergences for our most-inhibited fully-convective models, as in \S~\ref{subsec:MM14_models}. As with the models discussed in \S~\ref{sec:rot_inhibition_conv} and \S~\ref{sec:mag_inhibition_conv}, this agreement is not just fortuitous: it stems from the fact that changes in the radii are linked to changes in $s_{\text{ad}}$, which are well-described by our $\alpha_{\text{MLT}}(r)$ profiles.  

\begin{figure*}
\gridline{\fig{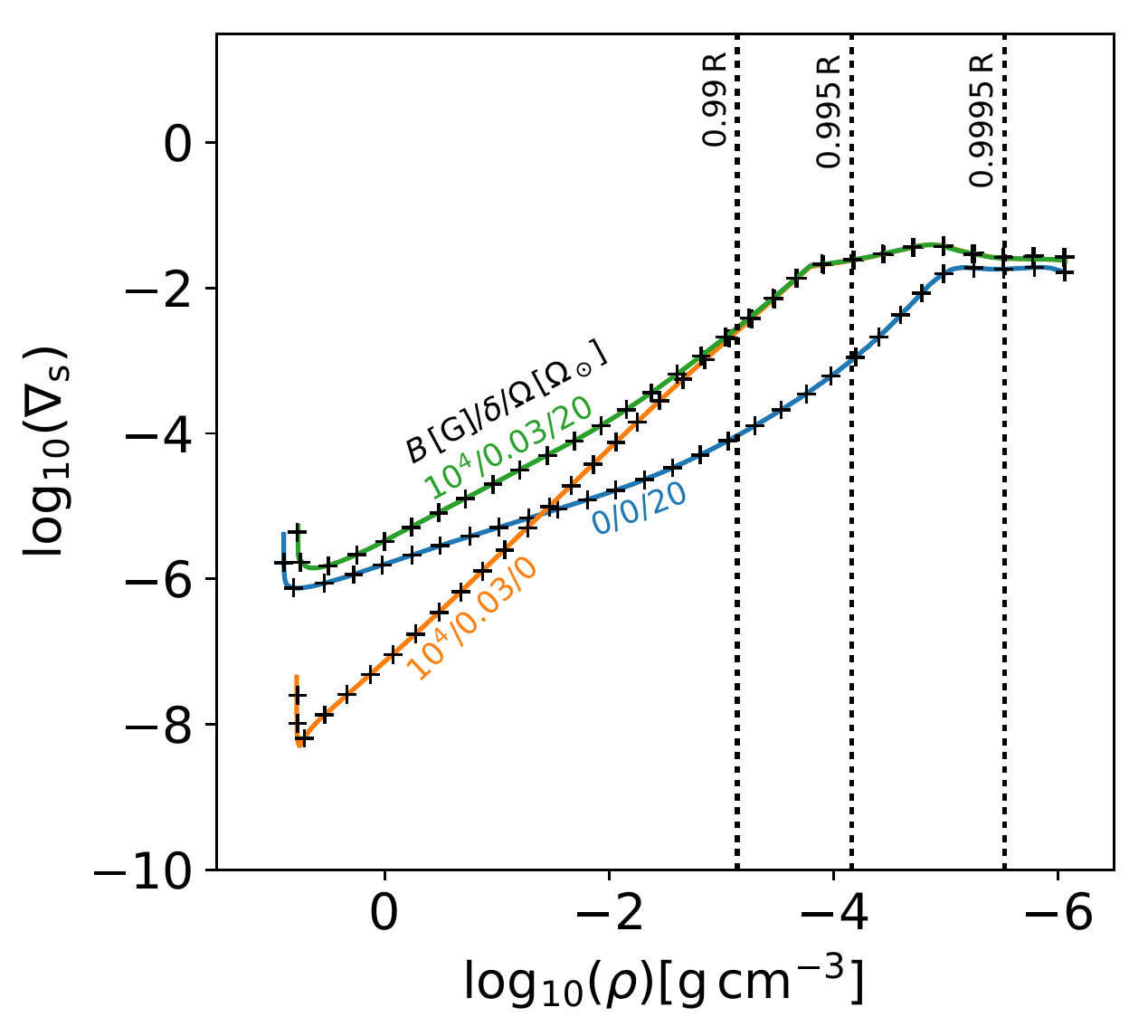}{0.5\textwidth}{(a)}
          \fig{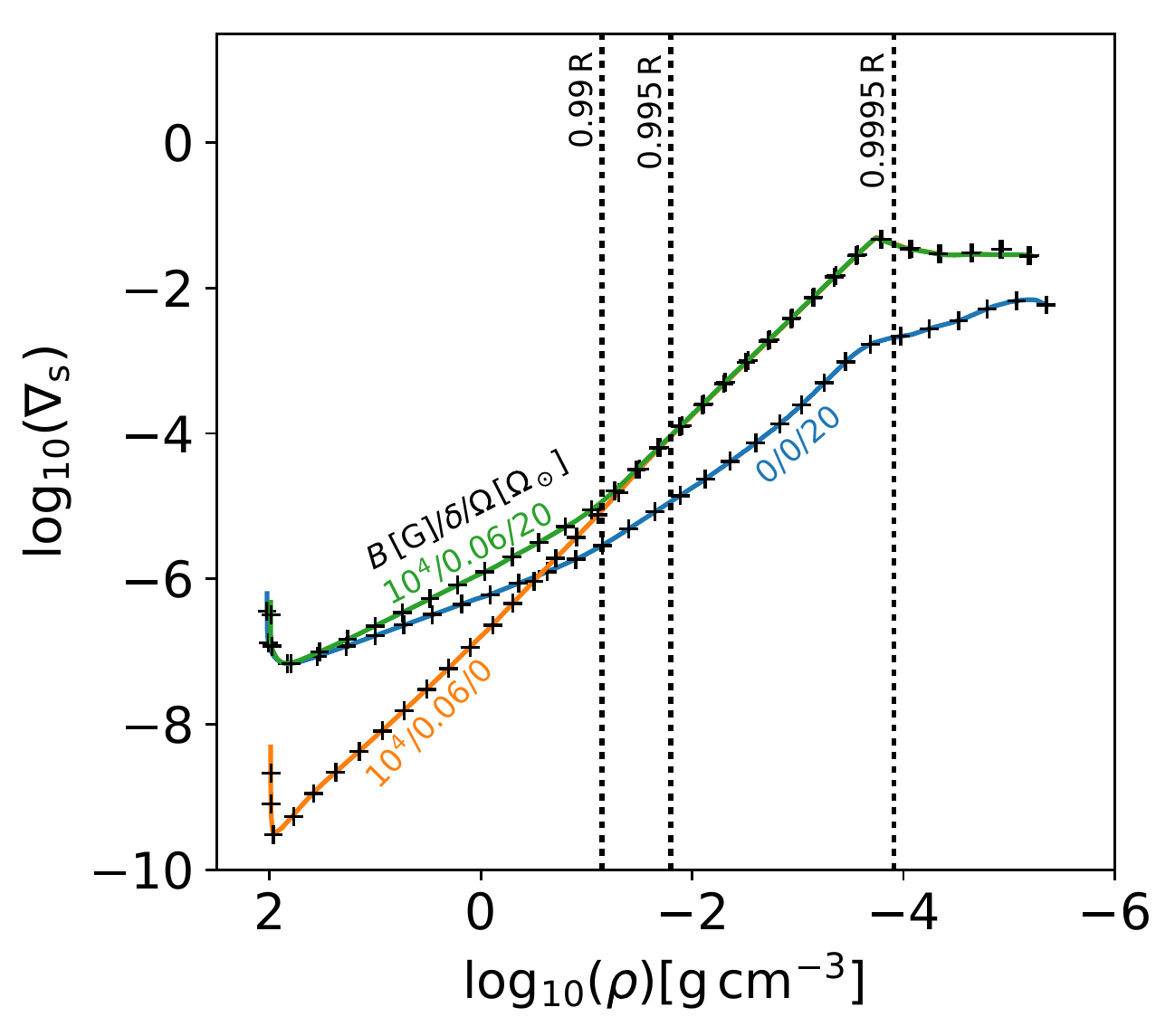}{0.51\textwidth}{(b)}
          }
\caption{$\log_{10}{(\nabla_{\text{s}})}$ as a function of $\log_{10}{(\rho)}$, comparing our $0.3 \, \text{M}_\odot$, (a) $10 \, \text{Myr}$ (b) $1 \, \text{Gyr}$, $\alpha_{\text{MLT}} = 1.7$ stellar model at $\Omega = 20 \, \Omega_\odot$, $B_{\text{v-max}} = 10^4 \, \text{G}$, and (a) $\delta = 0.03$ (b) $\delta = 0.06$, to our $\alpha_{\text{MLT}}(r)$ model (plus markers). We include the rotating-only and magnetic-only cases for further comparison.\label{fig:logsupergrad_v_logRho_depth_dep_rot_mag}}
\end{figure*}

\begin{figure}
\includegraphics[width=0.5\textwidth]{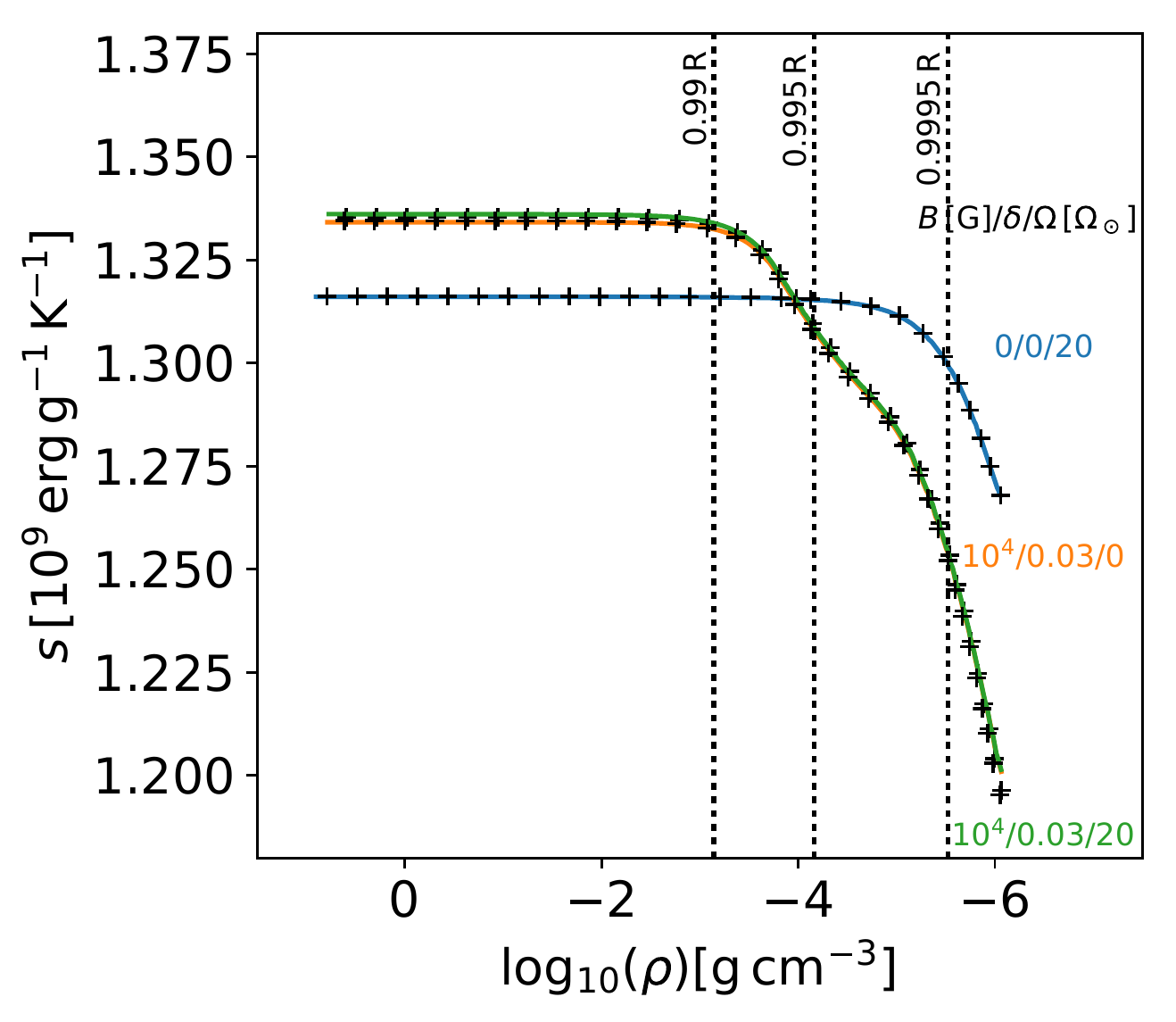}
\caption{$s$ as a function of $\log_{10}{(\rho)}$, comparing our $0.3 \, \text{M}_\odot$, $10 \, \text{Myr}$, $\alpha_{\text{MLT}} = 1.7$ stellar model at $\Omega = 20 \, \Omega_\odot$, $B_{\text{v-max}} = 10^4 \, \text{G}$, and $\delta = 0.03$, to our $\alpha_{\text{MLT}}(r)$ model (plus markers). We include the rotating-only and magnetic-only cases for further comparison.\label{fig:entropy_v_logRho_depth_dep_rot_mag}}
\end{figure}

We also create an $\alpha_{\text{MLT}}(r)$ expression for the combination of the magnetic and rotational reformulations of MLT (see \S~\ref{sec:rot_and_mag}), by treating $\alpha_{\text{MLT}_0}$ in equation~(\ref{eq:depth_dep_alpha_rot}) as the $\alpha_{\text{MLT}}(r)$ profile for the magnetic prescription in equation~(\ref{eq:depth_dep_alpha_mag}), which we will denote as $\alpha_{\text{MLT}}(r)_B$, producing

\begin{equation} \label{eq:depth_dep_alpha_rot_mag}
\alpha_{\text{MLT}}(r) \simeq \frac{\alpha_{\text{MLT}}(r)_B}{\left[\left(\frac{\alpha_{\text{MLT}}(r)}{\alpha_{\text{MLT}}(r)_B}\right)^{-10/3} - \frac{4}{41} \tau_{\text{c}_0}^2 \Omega^2 \right]^{3/4}},
\end{equation}
which must also be solved iteratively. In Figures~\ref{fig:logsupergrad_v_logRho_depth_dep_rot_mag} and~\ref{fig:entropy_v_logRho_depth_dep_rot_mag}, we plot $\log_{10}{(\nabla_{\text{s}})}$ for both ages and $s$ for $10 \, \text{Myr}$ respectively as a function of $\log_{10}{(\rho)}$, {produced by a particular ``rotating magnetic" model from \S~\ref{sec:rot_and_mag} and our $\alpha_{\text{MLT}}(r)$ model, including profiles from the equivalent rotating-only and magnetic-only cases}; again, we see excellent correspondence between the two model structures.

\begin{figure*}
\gridline{\fig{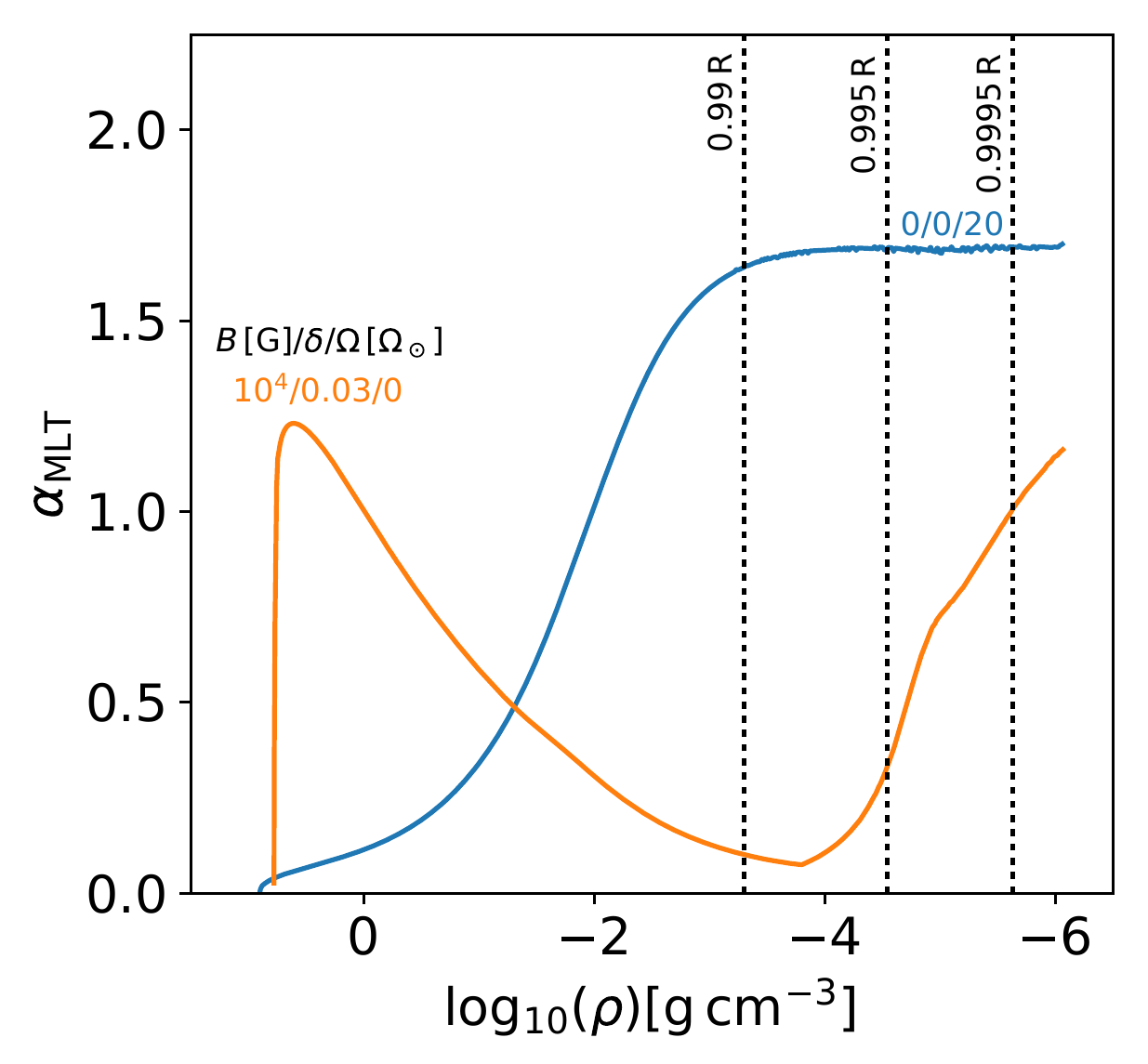}{0.5\textwidth}{(a)}
          \fig{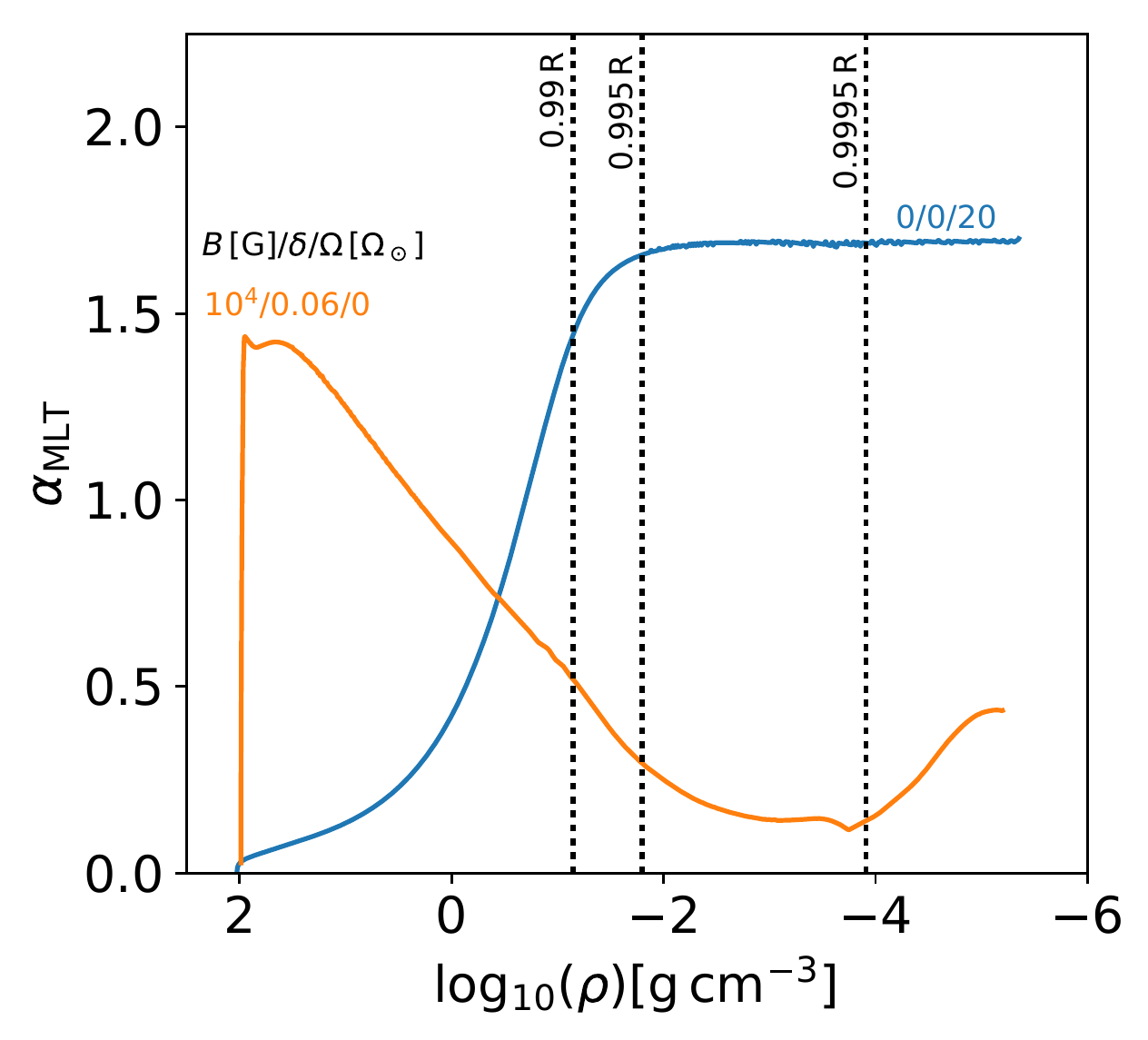}{0.51\textwidth}{(b)}
          }
\caption{{$\alpha_{\text{MLT}}(r)$ as a function of $\log_{10}{(\rho)}$, for a $0.3 \, \text{M}_\odot$, (a) $10 \, \text{Myr}$ (b) $1 \, \text{Gyr}$, $\alpha_{\text{MLT}} = 1.7$ stellar model in the rotating-only case ($\Omega = 20 \, \Omega_\odot$), and the magnetic-only case ($B_{\text{v-max}} = 10^4 \, \text{G}$, and (a) $\delta = 0.03$ (b) $\delta = 0.06$).}\label{fig:alpha_mlt_v_logRho}}
\end{figure*}

{In Figure~\ref{fig:alpha_mlt_v_logRho}, we plot $\alpha_{\text{MLT}}(r)$ as a function of $\log_{10}{(\rho)}$ at both ages for the same model, to show the differences in the depth-dependence of $\alpha_{\text{MLT}}(r)$ for the rotating-only and magnetic-only cases. For the rotating-only case, $\alpha_{\text{MLT}}(r)$ is constant and close to the unperturbed model value ($\alpha_{\text{MLT}} = 1.7$) across a majority of the surface layers (implying negligible changes in stellar structure), and drops rapidly with depth in the deep interior where $\text{Ro} \ll 1$. For the magnetic-only case, $\alpha_{\text{MLT}}(r)$ starts at a lower value at the photosphere, and drops sharply with depth in the surface layers (producing noticeable changes in stellar structure), rising again in the deep interior where $\delta$ drops rapidly.}

In \S~\ref{subsec:radius_s_ad}, we showed that it possible to determine an explicit relation between $s_{\text{ad}}$ and the depth-\emph{independent} $\alpha_{\text{MLT}}$ in standard 1D models. If this were possible in the depth-\emph{dependent} case as well, it would allow us to provide analytical estimates of how $s_{\text{ad}}$, and hence (via the formulae of \S~\ref{subsec:radius_s_ad}) the overall stellar radius, responds to changes in the depth-dependent convective inhibition parameters in any given theory (e.g., $\delta$ in the \citetalias{MacDonald:2014aa} formulation). Unfortunately, although we find that $s_{\text{ad}} \propto \Delta s$ in all of our $\alpha_{\text{MLT}}(r)$ models, it is no longer feasible to provide a simple analytical formula encapsulating the link between $s_{\text{ad}}$ and $\alpha_{\text{MLT}}(r)$. Essentially, this arises because we can no longer exclude $\alpha_{\text{MLT}}(r)$ from the integral producing $\Delta s$ in equation~(\ref{eq:s_jump_alpha}): in the fixed-$\alpha_{\text{MLT}}$ case for models with similar convective flux profiles, the integral associated with the high efficiency regime (excluding $\alpha_{\text{MLT}}$ due to its depth-independence) is near-homologous between models, allowing a direct proportionality between $\Delta s$ and $\alpha_{\text{MLT}}$ (equation~(\ref{eq:delta_s_prop_alpha_4_3})). However, in the $\alpha_{\text{MLT}}(r)$ case, this is not possible as the integral is now weighted by $\alpha_{\text{MLT}}(r)$ throughout the radial distribution; hence, in order to determine a change in $s_{\text{ad}}$ between two models of differing $\alpha_{\text{MLT}}(r)$, one must also have knowledge of all parameters in equation~(\ref{eq:s_jump_alpha}) for the perturbed model, rather than just $\alpha_{\text{MLT}}(r)$ and details of the unperturbed model.

\section{Discussion and conclusion} \label{sec:discussion_conclusion}

Rotation and magnetism both affect convection: the velocities, temperature gradients, and spatial structure that prevail in a magnetized, rotating flow are not generally the same as those that occur when rotation and magnetic fields are absent. In principle, the resulting changes in convective heat transport could affect the structure of stars or planets that host convection. Motivated by the observation that some low-mass stars appear to have larger radii than predicted by standard 1D stellar models, which parameterize the convective transport using MLT, several authors have suggested that rotation and/or magnetism may indeed be influencing the overall stellar structure. In this paper, we have examined this issue using 1D stellar models that attempt to incorporate both rotational and magnetic effects in a highly simplified way, and compared our results to models constructed using a standard version of MLT (modified here to allow for a mixing length parameter $\alpha_{\text{MLT}}$ that in some cases varies with depth). Below, we recapitulate our main findings and note some of their limitations.  

The structure of a star may be regarded as a function of its entropy, so assessing the structural impacts of rotation or magnetism amounts to determining the role these play in modifying the star's entropy. In \S~\ref{sec:entropy}, we reviewed the links between entropy, convective efficiency, and stellar radii in ``standard" 1D models, in which the mixing length parameter $\alpha_{\text{MLT}}$ is assigned a depth-independent value that must be calibrated by comparison with observations. In these models, reducing the convective efficiency via a decrease in $\alpha_{\text{MLT}}$ increases the temperature gradient required to carry an equivalent heat flux within the stellar interior. This translates into a larger entropy contrast between the photosphere and the deep interior for both pre-main-sequence and main-sequence models, which in turn influences the specific entropy attained in the deep interior (i.e., $s_{\text{ad}}$) in an age-dependent fashion. We explicitly determine the radius inflation of a given model from the difference in $s_{\text{ad}}$, with $\Delta \ln{R} \propto \Delta s_{\text{ad}}$. We also show how changes in the depth-independent $\alpha_{\text{MLT}}$ are directly related to changes in the stellar radius, in a manner similar to that described by \citet{Christensen-Dalsgaard:1997aa} for solar-like stars. 

One of our principal aims was to determine whether rotation alone could plausibly modify the convective transport enough to change a fully-convective star's radius by a noticeable amount. In \S~\ref{sec:rot_inhibition_conv}, we considered a rotationally-constrained version of MLT originally proposed by \citetalias{Stevenson:1979aa}, and given renewed vibrancy by the recent analyses and simulations of \citet{2012PhRvL.109y4503J} and \citet{Barker:2014aa}. By implementing this theory directly into our 1D MESA models, we find that rotation has a negligible impact on the star's overall radius. This is because the radius is determined primarily by the interior adiabat, which in turn is established largely by layers near the stellar surface. These layers are almost completely uninfluenced by rotation at any plausible rotational velocity---that is, $\text{Ro} \gg 1$ there because the convective velocity increases rapidly near the low-density photosphere---so rotation has little effect on $s_{\text{ad}}$ and hence on the stellar radius, even though flows in the deep interior of the star \emph{are} strongly affected by rotation. It is worth noting that if stars were instead well-characterized by a single depth-independent Rossby number, rotation would (in at least some stars) be important everywhere, and would have a much more significant impact on the radius; it is primarily the depth variation of convective velocities that makes this impossible.

In \S~\ref{sec:mag_inhibition_conv}, we argued that a particular prescription for incorporating the effects of magnetism into 1D stellar models, due to \citetalias{MacDonald:2014aa}, could be usefully analyzed using the same techniques developed in \S~\ref{sec:entropy}. In particular, we note that the effect of varying magnetic fields in this model is to vary the entropy content of the deep interior; once this is known, the stellar radius is also determined, via the same formula developed in \S~\ref{sec:entropy} (namely, equation~\ref{eq:radii_entropy_relation}) for standard MLT models. In accord with \citet{MacDonald:2017ab}, we find that \emph{if} magnetic fields indeed influence convective transport in the manner assumed here, fields of a plausible strength ($10^4 \, \text{G}$ or less) could noticeably ``inflate" the stellar radius. This inflation is larger (by about a factor of two) in models at $10 \, \text{Myr}$ than in those at an age of $1 \, \text{Gyr}$.

In \S~\ref{sec:rot_and_mag}, we showed that combining the rotational and magnetic reformulations of MLT, covered in \S~\ref{sec:rot_inhibition_conv} and \S~\ref{sec:mag_inhibition_conv} respectively, can indeed ``inflate" stellar radii by a further small amount. This demonstrates that the \citetalias{Stevenson:1979aa} rotation prescription is only effective at changing the stellar structure if the model is already ``perturbed" by magnetism. The superadiabaticity throughout the stellar interior increases with magnetic field strength; the effects of rotational inhibition can ``feed" on this, increasing the superadiabaticity somewhat further and producing small structural differences in some cases.  In our models, this additional effect is noticeable only on the pre-main-sequence. 

Finally, in \S~\ref{sec:depth_dep_alpha} we showed that both the rotationally- and magnetically-constrained versions of MLT explored in \S~\ref{sec:rot_inhibition_conv} and \S~\ref{sec:mag_inhibition_conv}, and the combination of these as shown in \S~\ref{sec:rot_and_mag}, can be duplicated by a ``standard" MLT model in which the mixing length parameter $\alpha_{\text{MLT}}$ is allowed to be depth-dependent. We provide explicit formulae linking the radially-variable $\alpha_{\text{MLT}}(r)$ to the rotational and magnetic formulations of \citetalias{Stevenson:1979aa} and \citetalias{MacDonald:2014aa} (equations~\ref{eq:depth_dep_alpha_rot} and~\ref{eq:depth_dep_alpha_mag} respectively), and we show that models constructed using these $\alpha_{\text{MLT}}(r)$ are indistinguishable from those directly employing the \citetalias{Stevenson:1979aa} or \citetalias{MacDonald:2014aa} models. These formulae enable the computation of ``magnetic" or ``rotating" models---within the assumptions of the \citetalias{Stevenson:1979aa} or \citetalias{MacDonald:2014aa} prescriptions---{without modification} of the mixing length formulation in a standard 1D stellar evolution code {(though they do require that codes be capable of modeling non-constant $\alpha_{\text{MLT}}$)}. We must caution, though, against taking these formulae as providing a \emph{quantitatively} correct assessment of how rotation and/or magnetism affect the heat transport (and hence the structure of the star) at every depth; this is in our opinion unlikely to be the case, since the formulations on which it is based (namely those of \citetalias{Stevenson:1979aa} and \citetalias{MacDonald:2014aa}) have many potential shortcomings, as detailed below.  We have derived and included these formulae mainly in order to illustrate \emph{how} rotation and magnetism (in these prescriptions) could affect the structure of the star---namely, by modifying its specific entropy, just as $\alpha_{\text{MLT}}(r)$ modifies the entropy in this depth-dependent MLT.  The trends deduced here (regarding the relative efficacy of these mechanisms, for example, in objects of different ages) may well be \emph{qualitatively} correct, even if the specific values of stellar radii, effective temperatures, etc., ultimately are not.    

A principal limitation of our work is its reliance throughout on particularly simple models of how the rotation or magnetism affect the convective transport. In considering the effects of rotation on the structure, we effectively assumed that only the variation of $ds/dr$ with $\Omega$ matters, and also that the rotationally-constrained MLT of \citetalias{Stevenson:1979aa} adequately captures this variation; both assumptions are questionable. For example, the simulations of \citet{Barker:2014aa}, which we cite as providing some numerical support for this scaling, effectively model only a single latitude near the pole (i.e., where rotation and the gravity vector are aligned); it is by no means clear that the same temperature scalings will hold at different latitudes.  In general rotation also introduces new anisotropy into the system (with motions increasingly aligned with the rotation axis in accord with the Taylor-Proudman constraint), implying that we might generally expect variations in the heat flux and/or entropy gradient with latitude.  It is unclear how these latitudinal variations could best be represented in a 1D stellar model, which intrinsically assumes spherical symmetry. Similarly, the scaling of temperature or entropy gradients with rotation rate may well depend on latitude; indeed, latitudinal variations in these quantities are often present in spherical shell simulations of rotating convection \citep[e.g.,][]{2004ApJ...601..512B,2018A&A...609A.124R}.

It must likewise be acknowledged that the effects of magnetism on the flow, and hence on the stellar structure, are still uncertain. In general they will depend on both the strength and the spatial morphology of the magnetic fields---which, in all the models quoted above and in our own work here, is not solved-for self-consistently as the outcome of a dynamo process, but instead must simply be imposed \emph{a priori}. Models making different assumptions about the interior field strengths have yielded substantially different results. {For example, the low-mass star models of \citet{Mullan:2001ab} explored fields of such strength ($\sim 100 \, \text{MG}$) that portions of the interior were rendered convective stable; this was motivated partly by the striking observational finding that the coronal heating efficiency of stars did not exhibit any clear break in behavior at around spectral types M3-M4, where stars are (in standard non-magnetic models) predicted to transition from being partially radiative to fully convective \citep[e.g.,][]{1993ApJ...410..387F}.} Many of the other models noted above, including \citet{MacDonald:2012aa} onwards, have considered much weaker fields, which are probably more realistic \citep[e.g.,][]{Browning:2016aa}. Meanwhile numerical simulations of the interiors of low-mass stars \citep{2006ApJ...638..336D,Browning:2008aa,2015ApJ...813L..31Y} suggest that in many cases dynamos in these objects may yield fields that are approximately in equipartition with the convective kinetic energy density, rising above this in the most rapidly rotating cases \citep[see, e.g., discussion in][]{2017arXiv170102582A}; the spatial structure of the fields is not yet certain, but is clearly influenced by the rotation rate (e.g., \citealt{2006GeoJI.166...97C,Browning:2008aa,2012A&A...546A..19G,2015ApJ...813L..31Y,Weber:2016aa,2017JFM...813..558A}; see also discussions in \citealt{2017LRSP...14....4B}).  The 1D models considered here (and for example in \citealt{MacDonald:2017ab}) are at least broadly consistent with these constraints on the overall field strengths, but we have made no effort to mimic the interior radial profile of the field, or to capture aspects of its actual spatial morphology---which, in any event, are still uncertain.

The effects of the magnetism on heat transport are also somewhat unclear, but note for example that \citet{2016GeoJI.204.1120Y} find that convective heat transport is actually \emph{enhanced} (relative to conductive transport) by the presence of magnetism in certain cases, in striking contrast to what is assumed in the \citetalias{MacDonald:2014aa} formulation (or likewise that of \citealt{Feiden:2014aa}, or in the reduced-$\alpha_{\text{MLT}}$ models discussed here). Of course the simulations operate in parameter regimes far removed from those in actual stellar interiors, but they are nonetheless indicative of the sometimes surprising dynamics that can occur when convection, rotation, and magnetism interact in spherical domains.  

More fundamentally, our models rely on the mixing length theory of convection, and on extremely simple atmospheric boundary conditions; both are crude approximations of the complex 3D transport occurring in these layers. Several authors have noted effects that are present in 3D convection but not easily captured in MLT \citep[e.g.,][]{1991ApJ...370..295C,2007ApJ...667..448M,2010ApJ...710.1619A,2017ApJ...845L..17C}. Likewise, the role of the near-surface layers, where 3D convection coupled to radiative transport ultimately helps set the stellar adiabat, has lately been studied using simulations and theory \citep[e.g.,][]{2014ApJ...785L..13T,Tanner:2016aa,Trampedach:2014aa,2015A&A...573A..89M}. It is beyond the scope of this paper to provide detailed comparison between the effects induced by magnetism or rotation and those arising from all other effects not included in our modeling.  However, it is worth noting that some of these effects must be clarified if a quantitative comparison between models and any specific observational data point is required. For example, variations in the surface atmospheric boundary condition and in metallicity, both fixed in our models, would modify the precise values of radius or effective temperature achieved at any given $\alpha_{\text{MLT}}$, whether depth-dependent or not \citep[see, e.g., discussions in][]{2014ApJ...785L..13T,Tanner:2016aa}.

Overall, our results suggest that rotation alone (if indeed it affects convection in the manner assumed here) cannot notably influence the overall structure of a fully-convective star, but magnetism might. To have a substantial influence, the magnetism (or indeed any other agent that modifies the heat transport) must impact layers relatively close to the stellar surface, which largely establish the star's overall adiabat and hence its radius. These effects can be duplicated using standard MLT, but at the cost of allowing a depth-dependent $\alpha_{\text{MLT}}(r)$ (intended to mimic the depth dependence of convective inhibition). In general, this may be difficult or impossible to calibrate using observations that probe the stellar surface alone. Further independent constraints on the form such depth-dependent convective inhibition must take---for example, by detailed comparison with 3D simulations incorporating rotation, magnetism, and radiative transport---may therefore be a prerequisite for truly predictive models of how magnetism affects the structure and evolution of these stars.

\acknowledgments

This research has been supported by the European Research Council under ERC grant agreements No. 337705 (CHASM), and by a Consolidated Grant from the UK STFC (ST/J001627/1). We have also benefited from access to the University of Exeter supercomputer, a DiRAC Facility jointly funded by STFC, the Large Facilities Capital Fund of BIS, and the University of Exeter. We also acknowledge PRACE for awarding us access to computational resources, namely Mare Nostrum based in Spain at the Barcelona Supercomputing Center, and Fermi and Marconi based at Cineca in Italy. We thank Isabelle Baraffe for helpful comments on a draft of the manuscript. {We also thank the referee for a thoughtful review that helped to improve the manuscript.}

%



\software{MESA \citep{Paxton:2011aa,Paxton:2013aa,Paxton:2015aa,2017arXiv171008424P}
          }

\bibliography{References}



\end{document}